\documentclass[pre, twocolumn, floatfix, superscriptaddress,nofootinbib,longbibliography]{revtex4-2}
\usepackage[utf8]{inputenc}
\usepackage{graphicx}
\usepackage{hyperref}
\usepackage{amssymb}
\usepackage{amsmath}
\usepackage{microtype}
\usepackage[table,dvipsnames]{xcolor}
\usepackage{braket}
\usepackage{dsfont}
\usepackage[export]{adjustbox}
\usepackage{soul}
\usepackage{gensymb}
\usepackage{setspace}
\usepackage{pifont}
\usepackage{booktabs}
\usepackage{array}

\hypersetup{
    colorlinks = true,
    linkcolor  = blue,
    citecolor  = blue,
    urlcolor   = blue,
}
\renewcommand{\thesection}{\arabic{section}}

\makeatletter
\renewcommand{\p@subsection}{}
\renewcommand{\p@subsubsection}{}
\makeatother

\renewcommand{\vec}[1]{\mathbf{#1}}

\def\U#1{\text{U}(#1)}
\def\Z#1{\mathbb{Z}_{#1}}

\DeclareMathOperator{\Li}{Li}
\newcommand{\mc}[1]{{\mathcal{#1}}}

\makeatletter
\newcounter{savesection}
\newcounter{apdxsection}
\renewcommand\appendix{\par
  \setcounter{savesection}{\value{section}}%
  \setcounter{section}{\value{apdxsection}}%
  \setcounter{subsection}{0}%
  \gdef\thesection{\@Alph\c@section}}
\newcommand\unappendix{\par
  \setcounter{apdxsection}{\value{section}}%
  \setcounter{section}{\value{savesection}}%
  \setcounter{subsection}{0}%
  \gdef\thesection{\@arabic\c@section}}
\makeatother

\begin{document}

\title{Hilbert space shattering and dynamical freezing in the quantum Ising model}
\author{Oliver Hart}
\email{oliver.hart-1@colorado.edu}
\affiliation{Department of Physics and Center for Theory of Quantum Matter, University of Colorado, Boulder, CO 80309 USA}
\author{Rahul Nandkishore}
\affiliation{Department of Physics and Center for Theory of Quantum Matter, University of Colorado, Boulder, CO 80309 USA}
\affiliation{Department of Physics, Stanford University, Stanford, CA 94305, USA}
\date{March, 2022}

\begin{abstract}
    \setstretch{1.1}
    We discuss quantum dynamics in the transverse field Ising model in two spatial dimensions. We show that, up to a prethermal timescale, which we quantify, the Hilbert space `shatters' into dynamically disconnected subsectors. We identify this shattering as originating from the interplay of a $\U1$ conservation law and a one-form $\Z2$ constraint. We show that the number of dynamically disconnected sectors is exponential in system \emph{volume}, and includes a subspace exponential in system volume within which the dynamics is exactly localized, even in the absence of quenched disorder. Depending on the emergent sector in which we work, the shattering can be weak (such that typical initial conditions thermalize with respect to their emergent symmetry sector), or strong (such that typical initial conditions exhibit localized dynamics). We present analytical and numerical evidence that a first order-like `freezing transition' between weak and strong shattering occurs as a function of the symmetry sector, in a non-standard thermodynamic limit. We further numerically show that on the `weak' (melted) side of the transition domain wall dynamics follows ordinary diffusion.
\end{abstract}

\maketitle


\section{Introduction}

Motivated by recent advances in experimental capability~\cite{Kinoshita2006NewtonsCradle,trotzky2012probing,Gring2012Relaxation,Schreiber2015Observation,Smith2016MBL,Kaufman2016Thermalization,Kucsko2018Critical}, the far-from-equilibrium dynamics of quantum many-body systems has emerged as one of the central problems for contemporary physics.
A key question relates to whether an isolated quantum system will thermalize under its own dynamics, and continues to produce surprising answers.
Starting from generic initial conditions, an isolated system may robustly fail to reach local equilibrium through the mechanism of ``many-body localization'' (MBL) in the presence of sufficiently strong quenched disorder~\cite{GMP, Basko2006Metal,Nandkishore2015MBL,Abanin2019RevModPhys,Gopalakrishnan2020Dynamics}.
The origin of this ergodicity breaking in MBL systems is the existence of an extensive set of emergent, \emph{local} integrals of motion~\cite{Imbrie2016many,Huse2014Phenomenology,Serbyn2013Local}.
While the MBL phase is predicated on the existence of quenched disorder, a number of new avenues for avoiding thermalization in systems possessing translational invariance have recently come to light.
These include quantum many-body ``scarring''~\cite{ShiraishiMori,Moudgalya2018,turner2018weak,serbyn2021quantum,moudgalya2021quantum}, where a small number of area-law entangled, athermal states are embedded in an otherwise thermalizing spectrum, disorder-free localization~\cite{knolle1, knolle2, knolle3, Brenes2018, nonfermiglasses, Smith2019, Russomanno2020, karpov2020disorderfree,Hart2021Logarithmic}, where local symmetries emulate the effects of disorder in typical symmetry sectors, and \emph{Hilbert space shattering}~\cite{Khemani2020,SalaFragmentation2020}.
In models that exhibit Hilbert space shattering, a finite list of additional constraints on the mobility of excitations -- typically arising in `fractonic' systems~\cite{ChamonQuantumGlassiness, Haah2011, CastelnovoGlassiness2012, VijayTopoOrder2015, VijayFractonTopo2016, PretkoSubdimensional2017, GromovMultipole2019, NandkishoreFractons2019, PPN} -- lead to an exponential number of dynamically disconnected Krylov sectors~\cite{Khemani2020,SalaFragmentation2020,RakovszkySLIOM2020,MorningstarFreezing2020,DeTomasiDynamics2019,Moudgalya2021,moudgalya2021quantum,Moudgalya2021Commutant,tiwari,Sen}.

The typical size of these dynamically disconnected sectors relative to their corresponding global symmetry sectors gives rise to
two distinct flavors of shattering: `weak' and `strong'~\cite{Khemani2020,SalaFragmentation2020}.
For weak shattering, a state selected at random from a typical symmetry sector will, with probability one, belong to the largest Krylov sector therein.
Atypical, area-law entangled states are therefore measure zero in the thermodynamic limit, and the system will almost surely reach a local equilibrium state. 
In contrast, in systems exhibiting strong shattering, the largest Krylov sector does not include almost all of the corresponding global symmetry sector, and the system can then exhibit localized dynamics from typical low-entanglement initial conditions, failing to explore an appreciable fraction of states with the same quantum numbers, thereby evading thermalization.
It is also possible for different symmetry sectors to exhibit disparate shattering properties.
For instance, one-dimensional spin-1 lattice models that conserve both charge and dipole moment with strictly local dynamics exhibit weak shattering at half-filling, at least for $k$-local gates (or Hamiltonians) with $k \geq 4$~\cite{Khemani2020,SalaFragmentation2020,MorningstarFreezing2020}.
However, as the charge density is altered from its infinite temperature value, a critical density is reached, beyond which the system freezes: For sufficiently high and sufficiently low charge densities, the system suffers a breakdown of connectivity between its various states, and the corresponding symmetry sectors exhibit strong shattering~\cite{MorningstarFreezing2020}.

In this manuscript we analyze the shattering properties and the putative freezing transition in the two-dimensional transverse field Ising model (TFIM) deep within its ferromagnetic phase.
In a similar manner to the tilted Fermi-Hubbard model, where a large tilt imposes strong kinetic constraints that give rise to emergent charge and dipole conservation~\cite{Khemani2020,SalaFragmentation2020,Moudgalya2021,vanNieuwenburg2019,Taylor2020Experimental}, Ref.~\cite{yoshinaga2021emergence} showed that a strong ferromagnetic Ising coupling imposes restrictions on domain wall motion in the TFIM in spatial dimensions $d \geq 2$ (illustrated for $d=2$ in Fig.~\ref{fig:allowed-moves-square}).
We argue that it is the combination of (i) domain wall number conservation, (ii) a local $\Z2$ constraint on domain wall configurations (which may alternatively be phrased as a $\Z2$ one-form constraint), and (iii) strict locality, which are ultimately responsible for the shattering of Hilbert space.
This is notably in contrast to other instances of shattering, which typically rely on the presence of \emph{two} mutually commuting global $\U1$ symmetries (e.g., charge and dipole moment~\cite{Khemani2020,SalaFragmentation2020}, two species of fermion~\cite{RakovszkySLIOM2020}, or domain wall number and the commuting component of total magnetization~\cite{YangIadecola2020,Chen2021}).
At half-filling for domain walls, we show using exact enumeration of states that the system exhibits \emph{weak} shattering, in spite of an exponential (in volume) number of fully frozen states.
Since shattering occurs in a tensor product basis, the restrictions on domain wall motion are essentially classical in nature~\cite{Moudgalya2021Commutant}, and can be phrased as kinetic constraints~\cite{Ritort2003Glassy,berthier2011dynamical}. This allows us to efficiently simulate the system numerically using stochastic cellular automaton circuits~\cite{IaconisSubsystem,FeldmeierAnomalousDiffusion},
using which we show that transport of domain walls is diffusive in the weak shattering regime.
It also allows us to view the `freezing' transition as an \emph{irreducibility} transition of the corresponding classical Markov process~\cite{Ritort2003Glassy}.

In contrast to Ref.~\cite{MorningstarFreezing2020} we show that the 2D TFIM exhibits no freezing transition for any nonzero \emph{density} of domain walls in the strict thermodynamic limit, i.e., where the number of domain walls scales with system volume, $N_\text{DW}\propto L^2$.
However, we provide analytical and numerical evidence in favor of a genuine transition between weak and strong shattering that occurs in a nonstandard thermodynamic limit.
Specifically, there exists a sharp freezing transition when domain wall number scales in a subextensive manner with system size as $N_\text{DW} \propto L^2/\ln L$.
We argue that the origin of the slow, logarithmic decay of the critical domain wall density is a consequence of so-called `large void instabilities', known to occur, for instance, in bootstrap percolation~\cite{adler1991bootstrap,DeGregorio2016} and other models with kinetic constraints~\cite{Ritort2003Glassy}.
This propensity for weak shattering is further confirmed by a strong even-odd effect in the presence of periodic boundary conditions. For antiferromagnetic coupling, the system exhibits ring frustration on the square lattice, which guarantees a subextensive number of defects in the classical ground state. This is another context in which a subextensive number of defects, here $N_\text{DW} \propto L$, is sufficient to prevent the system from being frozen.
By the same token, we additionally show that no freezing transition occurs for geometrically frustrated lattices (e.g., the triangular lattice) with antiferromagnetic coupling, since the system always possesses a nonzero \emph{density} of defects in its classical ground states. We work throughout on the lattice, avoiding the complications inherent with analyses of quantum dynamics in the continuum~\cite{mblcontinuum1, mblcontinuum2, mblcontinuum3}. 

The manuscript is structured as follows.
We begin by introducing the model in Sec.~\ref{sec:model}.
We discuss the limit of strong Ising coupling and the corresponding effective Hamiltonian that can be obtained in this limit by means of a Schrieffer-Wolff transformation. 
When the Schrieffer-Wolff transformation is truncated at first order in the transverse field, we discuss the kinetic constraints on domain wall motion, the resulting frozen states, and the relevant timescales for melting and for thermalization.
In Sec.~\ref{sec:enumeration} we perform an exact enumeration of the system's Krylov sectors.
When the enumeration is resolved by symmetry sector, we show that a finite size freezing transition occurs for a sufficiently low density of domain walls.
Section~\ref{sec:automaton} is concerned with quantifying ``sufficiently low''.
First, we benchmark the automaton circuits by showing that domain wall density diffuses when the system is weakly shattered, and then move on to characterizing the freezing transition numerically.
In Sec.~\ref{sec:void-instability}, we provide analytical evidence that the (un)freezing transition observed numerically in Sec.~\ref{sec:automaton} is a consequence of `large void instabilities'.
Finally, we discuss our results and their experimental implications in Sec.~\ref{sec:discussion}.


\section{Model}
\label{sec:model}

We consider the transverse field Ising model (TFIM), deep in the ferromagnetic phase
\begin{equation}
    \hat{H} = -J\sum_{\langle i j \rangle} \hat{\sigma}_i^z \hat{\sigma}_j^z - h \sum_i \hat{\sigma}_i^x
    \, .
    \label{eqn:TFIM}
\end{equation}
In this manuscript, the spin-1/2 degrees of freedom $\hat{\sigma}_i$ live on the sites of either a square or a triangular lattice in $d=2$ spatial dimensions, as canonical examples of bipartite and nonbipartite lattices, respectively.
In contrast to $d=1$, where the Hamiltonian can be mapped to free fermions~\cite{Lieb1961Soluble, Pfeuty1970}, the model is interacting in $d\geq 2$.
Unless otherwise specified, we will impose periodic boundary conditions on the spins throughout.
Absent the transverse field, i.e., $h=0$, the Hamiltonian~\eqref{eqn:TFIM} is diagonalized by $\hat{\sigma}^z_i$ product states $\ket{\{\sigma_i^z\}}$ (colloquially, the `computational basis').
The eigenstates $\ket{\{\sigma_i^z\}}$ have a definite number of domain walls.
It is therefore convenient to define charges $\hat{Q}_{ij}$ living on the links $\langle ij \rangle$, defined by $\hat{Q}_{ij} = \frac12(\mathds{1}-\hat{\sigma}_i^z \hat{\sigma}_j^z)$, such that a link hosting a ferromagnetic (antiferromagnetic) arrangement of neighboring spins satisfies $Q_{ij}=0$ ($Q_{ij}=1$) [operators (eigenvalues) are distinguished by the presence (absence) of a `hat']. Equivalently, the idempotent operator $\hat{Q}_{ij}$ counts the number of (bare) domain walls living on the link $\langle ij \rangle$.
For sufficiently weak magnetic fields, $h \ll J$, we may apply a Schrieffer-Wolff transformation that is chosen in such a way that the number of (dressed) domain walls is conserved.

\begin{figure}[t]
    \centering
    \includegraphics[width=0.7\linewidth]{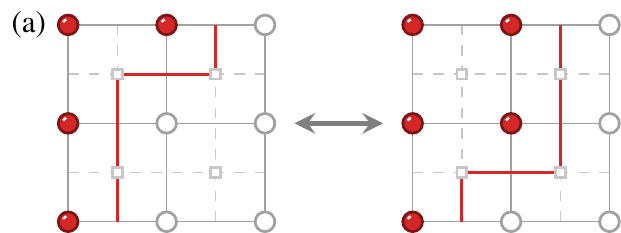}\\[0.2cm]
    \includegraphics[width=0.7\linewidth]{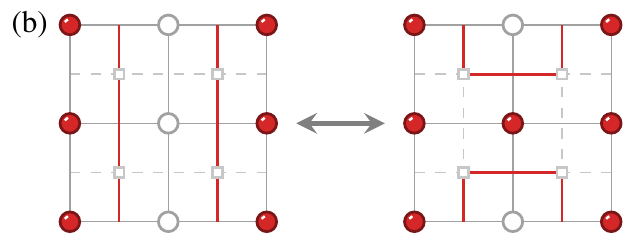}
    \caption{\textbf{Domain wall conserving moves}. (a) `Diagonal' domain wall motion (kink propagation) preserves the total number of domain walls and the number of domain walls intersecting each line ($x=\text{const.}$ or $y=\text{const.}$) on the direct lattice. (b) Plaquette flipping motion preserves the number of domain walls emanating from each dual lattice site, and the \emph{parity} of domain walls intersecting each line on the direct lattice. The empty (red) circles denote positive (negative) spins, while the solid red lines denote domain walls, which live on the dual lattice.}
    \label{fig:allowed-moves-square}
\end{figure}

The transverse field contribution can be written in terms of operators $\hat{T}_n$ that increment the number of domain walls by $n$~\cite{stein1997flow,VidalParallel2009,VidalTransverse2009}. Up to a trivial energy shift,
\begin{equation}
    \hat{H} = 2J\hat{Q} - h \sum_{n\in \mc{N}} \hat{T}_n
    \, ,
\end{equation}
where $\mc{N}$ is the set of permitted changes in the domain wall number (e.g., for the square lattice, $\mc{N} = \{0, \pm 2, \pm 4\}$), the total number of domain walls is denoted $\hat{Q} = \sum_{\langle ij \rangle} \hat{Q}_{ij}$, and the operators $\hat{T}_n$ satisfy $[\hat{Q}, \hat{T}_n] = n \hat{T}_n$.
As described in Appendix~\ref{sec:Schrieffer-Wolff}, the effective Hamiltonian that conserves the number of dressed domain walls up to second order in the magnetic field is given by
\begin{equation}
    \hat{H}' = 2J\hat{Q} - h\hat{T}_0 +
               \frac{h^2}{4J} [\hat{T}_2, \hat{T}_{-2}] +
               \frac{h^2}{8J} [\hat{T}_4, \hat{T}_{-4}] + \ldots
               \, .
    \label{eqn:SW-effective-H}
\end{equation}
This procedure can, in principle, be carried out to very high orders in the field using the method of perturbative continuous unitary transformations~\cite{stein1997flow,knetter2000perturbation,Knetter2003Structure}.
If the higher order terms denoted by the ellipsis are dropped, the effective Hamiltonian by construction commutes with total domain wall number: $[\hat{H}', \hat{Q}]=0$.
The operators $\hat{T}_n$ can be written in terms of the spins as
\begin{equation}
    \hat{T}_n = \sum_i \hat{\sigma}_i^x \hat{\Pi}_n(i)
    \, ,
    \label{eqn:Tn-definition}
\end{equation}
where the local projector $\hat{\Pi}_n(i)$ projects out spin configurations that violate the constraint $\sigma_i^z \sum_{j : \langle ij \rangle} \sigma_j^z = n $, where $j : \langle ij \rangle$ denotes the nearest neighbors of the $i$th spin.
For a $d$-dimensional hypercubic lattice, the projector $\hat{\Pi}_0(i)$ can be written explicitly as~\cite{yoshinaga2021emergence}
\begin{equation}
    \hat{\Pi}_0(i) = \prod_{m=1}^d \frac{1}{(2m)^2}
    \left[ (2m)^2 - \left(\sum_{j : \langle ij \rangle} \hat{\sigma}_j^z\right)^2\right]
    \, .
    \label{eqn:explicit-projector}
\end{equation}
The generalization of~\eqref{eqn:explicit-projector} to the triangular lattice is straightforward: Spin configurations satisfying $\sum_{j : \langle ij \rangle} \sigma_j^z = n$ with $n \in \mc{N} \setminus \{ 0 \}$ must be projected out. 
First, we keep only the leading order term $\propto \hat{T}_0$ [note that since the effective Hamiltonian $\hat{H}'$ conserves $\hat{Q}$, the first term in~\eqref{eqn:SW-effective-H} is trivial and can be dropped without consequence, at least for states with a definite number of quasi-domain walls]. We are therefore left with the effective Hamiltonian
\begin{equation}
    \hat{H}_1 = 2J\hat{Q} - h\hat{T}_0 
    \, .
    \label{eqn:SW-first-order}
\end{equation}
We will work throughout the manuscript with the effective Hamiltonian $\hat{H}_1$, truncated at first order in the magnetic field, which may
be considered as a two-dimensional generalization of the domain wall conserving model in Ref.~\cite{Iadecola2020}.
Much of the phenomenology that we will discuss hinges on the strictly finite support of the local operators in~\eqref{eqn:Tn-definition}.

We now comment on timescales. The Schrieffer-Wolff transformation procedure is valid up to an order $n_*$ set by $n_* \sim  J/h$, up to logarithmic corrections, which determines a prethermal timescale that is exponentially large in $n_*$, i.e., $\tau \sim e^{\Upsilon n_*}$ for some $\Upsilon > 0$, as shown in Ref.~\cite{abanin2017rigorous}. 
For times beyond $\tau$, the conservation of quasi-domain wall number breaks down and the system is able to thermalize. However, the truncation of the Schrieffer-Wolff transformation to leading order in magnetic fields is only valid up to a timescale $\tau' \sim J/h^2$. Nevertheless, we expect some of our results to remain applicable up to the true (exponentially long) prethermal timescale $\tau$, and we will discuss the relevant timescales as we present our results.

The dynamics implied by an application of the term $ \hat{T}_0$ on a computational basis state $\ket{\{\sigma_i^z\}}$ on the square lattice is depicted in Fig.~\ref{fig:allowed-moves-square}.
A given spin is flippable if and only if its neighboring spins sum to zero. If this constraint is satisfied, there are two possibilities: (i) spins of opposite sign are diametrically opposite one another with respect to the central spin, or (ii) the spins of opposite sign neighbour one another.
\begin{figure}
    \centering
    \includegraphics[width=\linewidth]{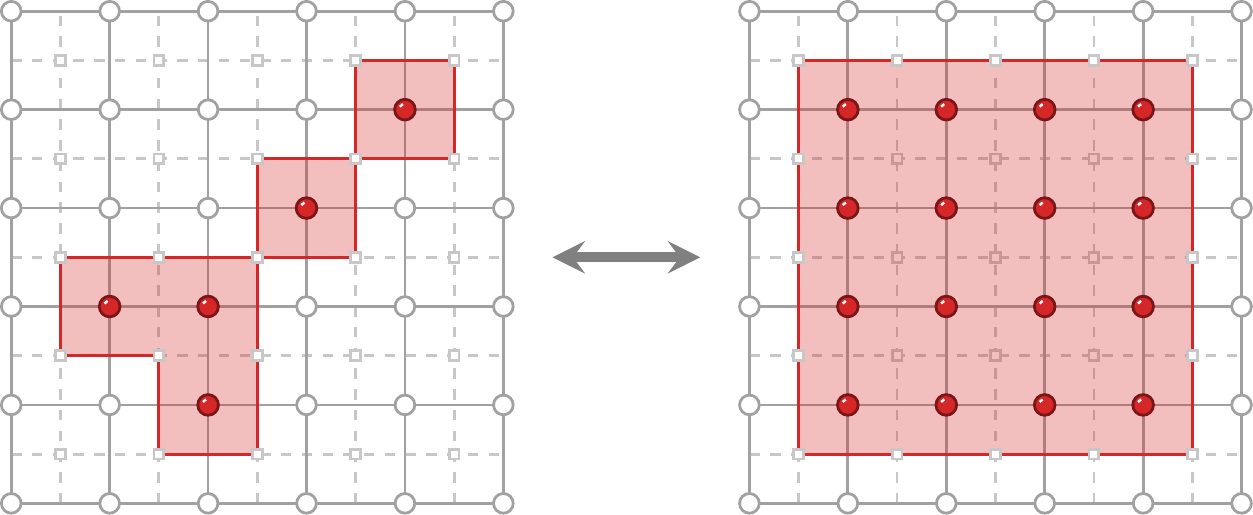}
    \caption{\textbf{Growth of a minority cluster}. The domain wall conserving dynamics permits a minority spin cluster to grow in area, but the maximum extent of the domain is set by the perimeter of the cluster, which is fixed by energy conservation.}
    \label{fig:domain-growth}
\end{figure}
For case (i) [Fig.~\ref{fig:allowed-moves-square}(a)] the domain walls move `diagonally' across the central spin (giving rise to propagation of domain wall `kinks'),
and the update to the domain wall configuration therefore preserves 
\begin{equation}
    \sum_{\substack{\langle ij \rangle_x : \\ y=\text{const}. }} Q_{ij} \qquad \text{and} \qquad 
    \sum_{\substack{\langle ij \rangle_y : \\ x=\text{const}. }} Q_{ij}
    \, ,
    \label{eqn:kink-propagation-symmetry}
\end{equation}
where, e.g., $\langle ij \rangle_x$ denotes an $x$-oriented bond.
The conservation laws~\eqref{eqn:kink-propagation-symmetry} therefore correspond to a subsystem symmetry, which commonly appear in the context of fractonic systems~\cite{ChamonQuantumGlassiness, Haah2011, CastelnovoGlassiness2012, VijayTopoOrder2015, VijayFractonTopo2016, PretkoSubdimensional2017, GromovMultipole2019, NandkishoreFractons2019}.
For case (ii) [Fig.~\ref{fig:allowed-moves-square}(b)] the domain walls exhibit a dimerlike~\cite{RokhsarKivelson} flipping motion across a plaquette, and the update preserves the number of domain walls emanating from each dual lattice site.
When (i) and (ii) are combined, however, they conserve only the parity of domain walls intersecting any closed loop $\gamma$ on the direct lattice, although this follows directly from the definition of domain walls in terms of the underlying spins: $\prod_{\langle ij \rangle \in \gamma} \hat{\sigma}_i^z \hat{\sigma}_j^z = \prod_{i \in \gamma} \hat{\sigma}_i^2 = \mathds{1}$.
Since we are able to write $\prod_{\langle ij \rangle \in \gamma}  {\sigma}_i^z {\sigma}_j^z  = \exp(i \pi \sum_{\langle ij \rangle \in \gamma} q_{ij})$, the constraint can be rephrased as $\sum_{\langle ij \rangle \in \gamma} q_{ij} = 0 \, \mod \, 2$,
and can therefore be thought of as a one-form $\Z2$ constraint on the Hilbert space of domain wall configurations.
Additional discrete global symmetries possessed by the Ising model, such as the $\Z2$ Ising symmetry $\prod_i \hat{\sigma}_i^x$, and various mirror symmetries (see, e.g., Ref.~\cite{MondainiRigolTFIM_II}) do not affect the shattering properties, and as a result we will not discuss them further.

An alternative way to view this constraint makes use of the duality between the two-dimensional transverse field Ising model and $\Z2$ lattice gauge theory~\cite{WegnerDuality1971,Kogut_RevModPhys1979}.
If we introduce gauge spins $\hat{\tau}_{ij}$ living on the links $\langle ij \rangle$ of the lattice, satisfying $\hat{\sigma}_i^x = \prod_{j : \langle ij \rangle} \hat{\tau}_{ij}^x$, and $\hat{\tau}_{ij}^z = \hat{\sigma}_i^z \hat{\sigma}_j^z$, then the Hamiltonian~\eqref{eqn:TFIM} becomes
\begin{equation}
    \hat{H} = -h \sum_i  \prod_{j : \langle ij \rangle} \hat{\tau}_{ij}^x - J \sum_{\ell} \hat{\tau}_\ell^{z}
    \, .
    \label{eqn:Z2-LGT}
\end{equation}
The Hamiltonian exhibits a $\Z2$ gauge symmetry, $[\hat{H}, \hat{B}_p]=0$, where $\hat{B}_p = \prod_{\langle ij \rangle \in p} \hat{\tau}_{ij}^z$, for each plaquette $p$ of the lattice.
The Hamiltonian is therefore supplemented by the Gauss law constraint $\hat{B}_p \ket{\Psi} = \ket{\Psi}$, which restricts the Hilbert space to gauge-invariant states (equivalently, states of the gauge spins that correspond to configurations of the local degrees of freedom, $\hat{\sigma}_i^z$).
Since we consider the ferromagnetic phase of the Ising model, $J \gg h$, this maps to the confining phase of the $\Z2$ lattice gauge theory~\cite{fradkin_2013}.
Indeed, we will show that it is the combination of nonzero line tension (the defining feature of the confining phase) and strict locality that are responsible for the `shattering' of Hilbert space.

The dynamics depicted in Fig.~\ref{fig:allowed-moves-square} differs fundamentally from the behaviour of pointlike quasiparticles.
This may be illustrated by the behaviour of \eqref{eqn:TFIM} in the paramagnetic phase, $h \gg J$, or, alternatively, by~\eqref{eqn:Z2-LGT} in its deconfined phase, where the toric code~\cite{Kitaev2003} emerges perturbatively.
In this opposite limit, the quasiparticles are not domain walls but isolated flipped spins relative to the state $\ket{\{\sigma_i^x\}}$ aligned with the magnetic field.
Repeating the Schrieffer-Wolff transformation in this limit, we obtain a simple two-dimensional tight-binding model, which by construction conserves the number of quasiparticles (the Ising interaction $\propto \hat{\sigma}_i^z \hat{\sigma}_j^z$ hops the flipped spin to a neighboring site along the bond $\langle ij \rangle$).
An isolated flipped spin can therefore propagate freely throughout the lattice, since particle number conservation alone imposes no restrictions on the mobility of quasiparticles.
In contrast, in the ferromagnetic regime, an isolated domain of minority spins can only grow to be as large as its perimeter allows, under dynamics generated by the effective Hamiltonian $\hat{H}_1$, since the number of domain walls must remain fixed.
This phenomenon is illustrated in Fig.~\ref{fig:domain-growth}.
Infinite line tension, which fixes univocally the length of domain walls, leads to `frozen states', which exhibit no dynamics under $\hat{H}_1$, as we will shortly show.


\subsection{Frozen states}
\label{sec:frozen-states}

\begin{figure}[t]
    \centering
    \includegraphics[height=0.3\linewidth]{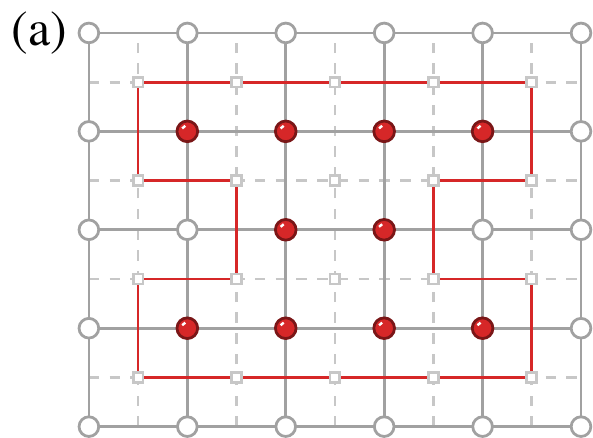}%
    \hspace{20pt}%
    \includegraphics[height=0.3\linewidth]{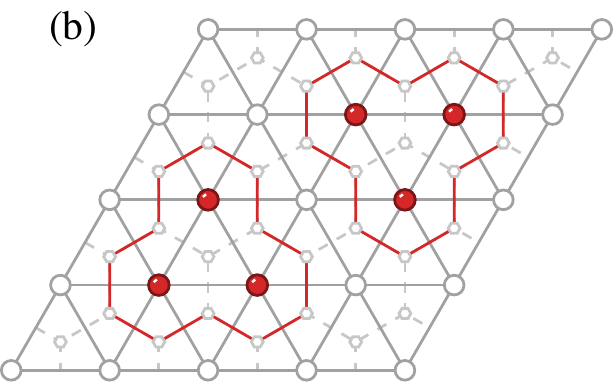}\\[10pt]
    \includegraphics[height=0.3\linewidth]{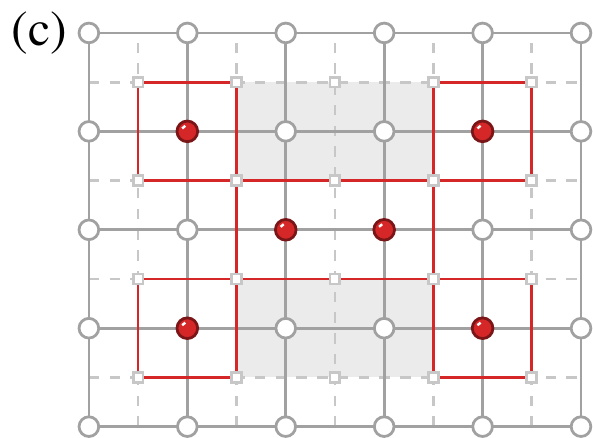}%
    \hspace{20pt}%
    \includegraphics[height=0.3\linewidth]{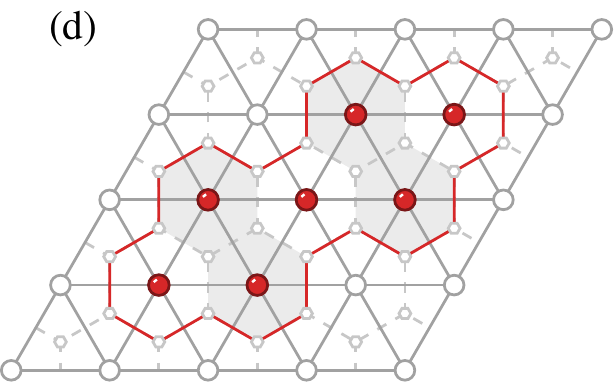}
    \caption{\textbf{Frozen and active spin configurations}. (a) A region of spins on the square lattice surrounded by `crenellations' is fully frozen.
    (b) For the triangular lattice, one can similarly construct frozen states from enclosed regions of flipped spins. Again, the region can be compact, as long as the domain walls make coarse-grained 120\degree turns only. The number of such configurations scales with the volume of the system.
    On the other hand, if any spins are surrounded by exactly $z/2$ domain walls, then they are active. In (c) and (d) we show examples of spin configurations that contain active spins for the square and triangular lattices, respectively. Spins that can be flipped are shaded in light gray.}
    \label{fig:frozen-states}
\end{figure}

A state $\ket{\boldsymbol{\sigma}_\text{f}}$ is fully frozen if it is not dynamically connected to any other states, i.e., $\langle \boldsymbol{\sigma}' |\hat{H}_1|\boldsymbol{\sigma}_\text{f} \rangle = 0$, $\forall \, \boldsymbol{\sigma}' \neq \boldsymbol{\sigma}_\text{f}$~\cite{Khemani2020, SalaFragmentation2020}.
If the state $\ket{\boldsymbol{\sigma}_\text{f}}$ satisfies this condition, it is an eigenstate of $\hat{H}_1$, and it belongs to its own one-dimensional \emph{Krylov sector}~\cite{Moudgalya2021}.
If the system is initialized in such a state, it will retain the same spin configuration for all times thereafter, under time dynamics generated by $\hat{H}_1$.
One may immediately observe that ferromagnetic spin configurations, with all spins pointing along $\pm \hat{\vec{z}}$, correspond to completely frozen states of the Hamiltonian~\eqref{eqn:SW-first-order}, since $ \sigma_i^z \sum_{j : \langle ij \rangle} \sigma_j^z = z > 0$ (the coordination number of the lattice).
For the square lattice, fully antiferromagnetic configurations of spins are also fully frozen, since $ \sum_{j : \langle ij \rangle} \sigma_j^z = \pm 4$, where the sign depends on the sublattice to which the central spin belongs. No such antiferromagnetic configuration is possible for the triangular lattice, a consequence of geometric frustration~\cite{lacroix2011introduction}.
Such translationally invariant spin configurations do not however exhaust the list of fully frozen states. Alternative stripes of `all up' and `all down' ferromagnetic domains, with perfectly straight domain boundaries, will also be frozen, yielding a number of frozen states exponential in linear system size, as anticipated in Ref.~\cite{yoshinaga2021emergence}. Other possibilities are shown in the top row of Fig.~\ref{fig:frozen-states}: a `snaking' pattern of domain walls along the boundary of a compact closed region of the lattice also leads to fully frozen states of $\hat{H}_1$ in~\eqref{eqn:SW-first-order}.
Since such compact regions can tile the lattice, e.g., Fig.~\ref{fig:frozen-states} can be regarded as a unit cell that tiles the lattice periodically, the number of frozen states that such a configuration can give rise to is $\sim 2^{N / N_a}$, where $N_a$ is the number of spins per frozen lattice animal, since each animal can be either present or absent in every location.
Hence, the number of frozen states, $N_\text{f}$, scales exponentially with the \emph{volume} of the system, although at a slower rate than the total dimension of the Hilbert space, i.e., $N_\text{f} \propto \exp(\alpha N)$, up to polynomial corrections, with $0 < \alpha < \ln 2$. This phenomenology is quite similar to the `shattering by charge and dipole conservation' discussed in Refs.~\cite{Khemani2020, SalaFragmentation2020}, but generated by the combination of a global $\U1$ conservation law (on domain wall number) and a $\Z2$ one-form constraint, instead of two global $\U1$ conservation laws as in the dipole conserving case.
Since the shattering occurs in a product state basis, it is ``classical'' in nature in the sense of Ref.~\cite{Moudgalya2021Commutant}, a feature that we will exploit to efficiently simulate the model in Sec.~\ref{sec:automaton}.

As in other models that exhibit Hilbert space shattering~\cite{YangIadecola2020}, the number of frozen states depends on the order at which the Schrieffer-Wolff transformation is truncated.
Spin configurations that are frozen at a particular order in the Schrieffer-Wolff transformation may become active in the presence of higher order terms that act on larger regions of the lattice, e.g., an isolated flipped spin surrounded by four domain walls becomes mobile at second order in the magnetic field. In contrast, a `stripelike' pattern, with perfectly straight domain walls at least a distance $\ell \gg 1$ apart, will only become mobile at order $\sim \ell$. Generically, if a spin configuration becomes mobile at order $n$ in the Schrieffer-Wolff transformation, then this movement will manifest at times $ht \sim (J/h)^{n-1}$. If $n>n_* \sim J/h$ then the timescale will be set by the prethermal timescale associated with the breakdown of the Schrieffer-Wolff procedure, $\tau \sim \exp(\Upsilon n_*)$.

Even if the state of the system is not fully frozen, there may be distinct frozen and active \emph{regions}. A simple example is a minority spin cluster embedded in an otherwise ferromagnetically ordered system, as shown in Fig.~\ref{fig:domain-growth}: All flipped spins in the right panel are `active' (flippable), and the remainder are frozen (unflippable).
Since each symmetry sector (i.e., the set of states with fixed domain wall number) contains many disconnected Krylov sectors, the Hamiltonian~\eqref{eqn:SW-first-order} exhibits Hilbert space shattering~\cite{Khemani2020} (also known as fragmentation~\cite{SalaFragmentation2020}).
We may quantify the extent to which the system is `shattered' by inspecting the distribution of Krylov sector sizes within each symmetry sector. In all that follows, we will be working with the effective Hamiltonian in Eq.~\eqref{eqn:SW-first-order}.


\section{Exact enumeration}
\label{sec:enumeration}

\begin{figure}
    \centering
    \includegraphics[width=\linewidth]{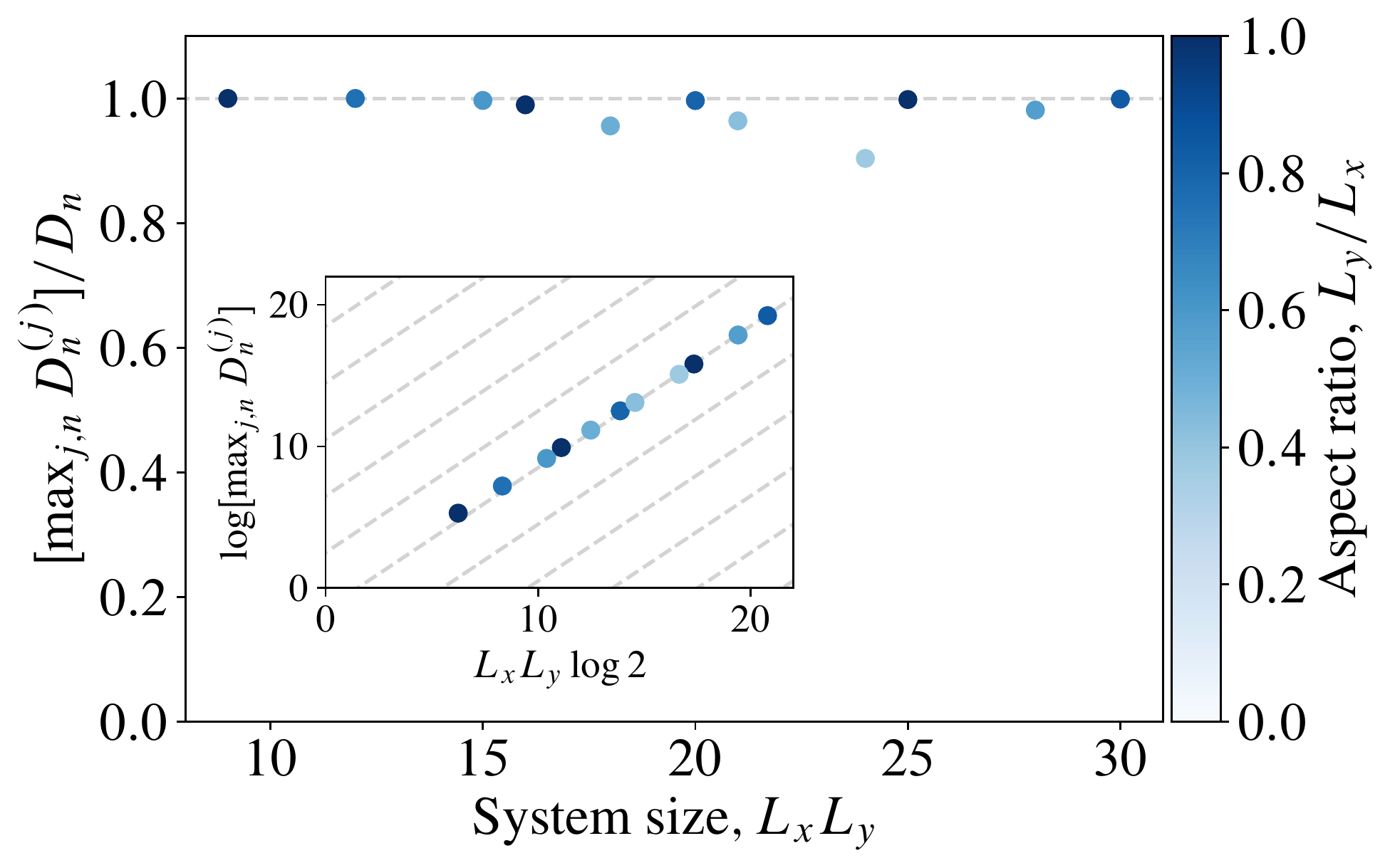}
    \caption{\textbf{Exact enumeration of Krylov sectors}. Ratio of largest Krylov sector size to the dimension of its corresponding symmetry sector for the Hamiltonian $\hat{H}_1$~\eqref{eqn:SW-first-order} on the square lattice. There is no appreciable variation of the ratio with system size for a fixed aspect ratio, consistent with \emph{weak} shattering of Hilbert space. The inset shows the absolute value of the maximum Krylov sector dimension, which is consistent with the scaling $\max_{j,n}D_n^{(j)} \propto 2^N$ (gray dashed lines).}
    \label{fig:largest-sector-square}
\end{figure}

To quantify the extent to which the Hilbert space shatters, we begin with exact enumeration: For sufficiently small systems, it is possible to explicitly construct all Krylov sectors within each symmetry sector. We denote the size of the symmetry sector defined by a domain wall density $n$ by $D_n$, and the individual Krylov sectors that live within this symmetry sector by $D_n^{(j)}$, labelled by the index $j$. When defined in this way, the dimensions satisfy $\sum_j D_n^{(j)} = D_n$.
Technically, this is carried out by performing a breadth first search of the system's adjacency matrix graph (i.e., the system's Hamiltonian represented in the basis of $\ket{\{\sigma_i^z\}}$ tensor product states) to find all connected sub-graphs of $\hat{H}_1$.

\begin{figure}
    \centering
    \includegraphics[width=0.85\linewidth]{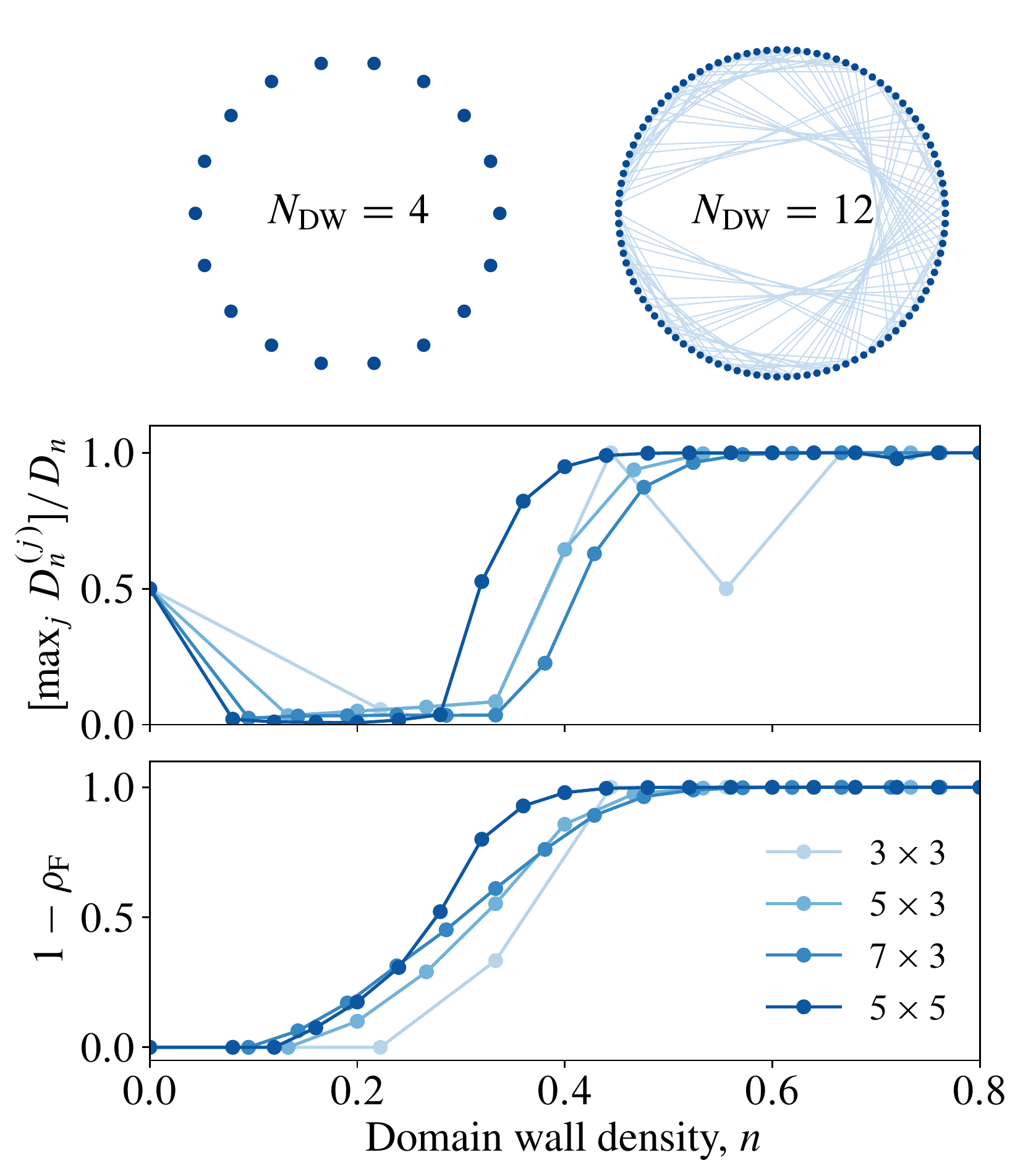}
    \caption{\textbf{Finite size freezing transition}. Top: adjacency graph of the $N_\text{DW}=4, 12$ symmetry sectors for a system of size $3\times 3$. Middle: ratio of maximum Krylov sector dimension to corresponding symmetry sector dimension, resolved by domain wall density (symmetry sector). Bottom: active site density $\rho_\text{A} = 1- \rho_\text{F}$ resolved by symmetry sector. Domain wall density is normalised per bond such that an (anti)ferromagnetic spin configuration has density 0 (1). Odd linear system size gives rise to ring frustration, forbidding the frozen N\'{e}el spin configuration.}
    \label{fig:frozen-sites-exact}
\end{figure}

If the Hamiltonian exhibits \emph{strong} shattering, then $\max_{j} [D_n^{(j)}]/D_n$ vanishes for typical values of $n$ as system size is increased, $L \to \infty$~\cite{Khemani2020, SalaFragmentation2020}.
That is, the largest Krylov sector comprises a vanishingly small fraction of its corresponding symmetry sector, and a typical initial state
will be unable to efficiently explore an appreciable fraction of the symmetry sector (correspondingly, the time-evolved state will exhibit an anomalously high overlap with the initial state).
This scenario is illustrated in the top left panel of Fig.~\ref{fig:frozen-sites-exact}, where all states in the symmetry sector are dynamically disconnected.
If, conversely, the system exhibits \emph{weak} shattering then $\max_{j} [D_n^{(j)}]/D_n \to 1$ as $L \to \infty$~\cite{Khemani2020, SalaFragmentation2020} for typical $n$.
Now, a state selected at random from a typical symmetry sector will, with probability one, belong to the largest Krylov sector, and -- given sufficient time -- will explore the states belonging to the symmetry sector densely.
This scenario is demonstrated in the top right panel of Fig.~\ref{fig:frozen-sites-exact}.
Note that the number of states that do not belong to the largest Krylov sector can still be exponentially large in the volume of the system, as long as they still correspond to a vanishingly small \emph{fraction} of total states in the thermodynamic limit: $\sum_{j \neq j_m} D_n^{(j)}/D_n^{(j_m)} \to 0$ as $L \to \infty$, where $j_m$ denotes the index of the largest Krylov sector.

The results obtained by performing an exact enumeration of sectors for systems of size up to and including $N=30$ are shown in Fig.~\ref{fig:largest-sector-square}.
Since we work directly with the system's Hamiltonian, which is sparse, rather than with its eigenstates, we are able to reach significantly larger system sizes than those accessible to exact diagonalisation (as in, e.g., Ref.~\cite{yoshinaga2021emergence}). Further details pertaining to the numerical simulations are presented in Appendix~\ref{sec:numerical-details}. We note in passing that Ref.~\cite{MorningstarFreezing2020} presented analogous results on a spin-1 chain with $N=18$, which corresponds to a similar Hilbert space size to $N=30$ with spin-$1/2$ degrees of freedom.

We observe that there is no appreciable variation in the ratio $\max_{j,n} [D_n^{(j)}]/D_n$ with system size, at least for a fixed aspect ratio. Note that while we plot the results for the largest Krylov sector, where the maximum is taken over \emph{all} symmetry sectors, the obtained ratio is not atypical; analogous results are found by averaging over symmetry sectors at infinite temperature\footnote{Here, by infinite temperature average we mean that the probability of picking a given symmetry sector is proportional to the number of states that it contains.} (see the middle panel of Fig.~\ref{fig:frozen-sites-exact} in the vicinity of $n=1/2$, the infinite temperature value).

We can gain further insight into the behaviour of the system by looking at the ratio $\max_{j} [D_n^{(j)}]/D_n$ resolved by symmetry sector, parameterized by $n$.
As explained in Sec.~\ref{sec:frozen-states}, in the limit of low domain wall density, in which domain walls form small, isolated clusters, we expect that the states should be
predominantly frozen and therefore the corresponding symmetry sectors should exhibit strong shattering.
This intuition is borne out in the middle panel of Fig.~\ref{fig:frozen-sites-exact}.
Note however that the ratio assumes the value $1/2$ at zero domain wall density: The two ferromagnetic spin configurations both belong to the $n=0$ sector and are dynamically disconnected from one another.
We additionally plot the average density of \emph{active} sites, the complement of the frozen site density.
A site is frozen if it is not dynamical, i.e., a given spin is frozen in the context of a Krylov sector if \emph{all} states belonging to the Krylov sector share the same spin direction.
A site that is not frozen is active; once all states within the Krylov sector have been explored, the spin will have flipped.
The average is over all Krylov sectors, with each Krylov sector weighted by the number of states that it hosts.
In both plots we observe a crossover from strong shattering (where a vanishing density of spins are active) at low domain wall density to weak shattering (where a given spin is active with probability one) at high densities. However, the system sizes accessible to exact enumeration are rather limited, and the existence of a putative weak--strong transition versus a smooth crossover cannot be established from the data in Fig.~\ref{fig:frozen-sites-exact} alone.
That the crossover appears to drift towards smaller values of domain wall density $n$ with increasing system size is worthy of note, and will be explained in detail in Sec.~\ref{sec:void-instability}.

Observe that Fig.~\ref{fig:frozen-sites-exact} is not symmetric under $n \to 1-n$, as one might have expected.
The origin of this asymmetry is \emph{ring frustration}; since we plot systems with both $L_x$ and $L_y$ odd, and periodic boundary conditions are applied, it is not possible for the system to exhibit perfect N\'{e}el order.
Instead, each row and each column must have (at least) one ferromagnetic bond, which bridges two antiferromagnetic regions with opposite parity.
If even $L_x$, $L_y$ are used instead then Fig.~\ref{fig:frozen-sites-exact} becomes exactly symmetric under $n\to 1-n$.
Typically, this even-odd effect is inconsequential in the thermodynamic limit, since the minimal number of ferromagnetic bonds scales as $\sim L_x+L_y$, so that the density of `defective' ferromagnetic bonds is subextensive, $\sim L^{-1}$.
In the context of the existence of a finite size freezing transition, however, ring frustration plays an important role.
With both $L_x$, $L_y$ odd, the state that maximizes the number of domain walls has two straight winding loops of ferromagnetic bonds that intersect at a point.
It is possible to show that the linear number of defective bonds in this state are sufficient to make \emph{all} sites active, as is observed in Fig.~\ref{fig:frozen-sites-exact}.
Consequently, there is no freezing transition as $n \to 1$ for both $L_x$, $L_y$ odd.
If exactly one of $L_x$, $L_y$ is odd, then the state that maximizes the domain wall density now has a single winding ferromagnetic loop.
Since an intersection point is no longer present, the single winding loop remains frozen, and a finite size freezing transition can again occur in the vicinity of both $n \to 0$ and $n \to 1$.\footnote{While the classical ground states with antiferromagnetic coupling are frozen when exactly one of $L_x$ and $L_y$ is odd, the fraction of frozen sites is still not symmetric under $n \to 1-n$ due to the presence of a macroscopic number of ferromagnetic bonds in the ground states, equal to $\min(L_x, L_y)$.}

The triangular lattice does not support antiferromagnetic N\'{e}el order, irrespective of the parity of its linear dimensions;
geometric frustration gives rise to a nonzero density of ferromagnetic bonds in the classical antiferromagnetic ground states.
This results in behaviour analogous to that shown in Fig.~\ref{fig:frozen-sites-exact}, whereby no freezing transition occurs
for the largest attainable values of domain wall density ($n = 2/3$).
No such obstruction exists as $n \to 0$, however, and a finite size freezing transition can occur, which will be explored in further detail
in the next section.


\section{Automaton numerics}
\label{sec:automaton}

To analyze the putative weak--strong shattering transition that occurs at low domain wall density, we make use of classical cellular automaton circuits, following Ref.~\cite{IaconisSubsystem}.
This allows us to access significantly larger system sizes than those accessible to the exact enumeration performed in Sec.~\ref{sec:enumeration}.
While finite size no longer represents an insurmountable barrier, the automaton numerics are instead limited principally by finite \emph{time} -- while the automaton circuits will sample a representative selection of states, the Hilbert space dimensions ($2^N$) are so large that not \emph{all} states will be sampled.

Cellular automaton dynamics is a class of discrete, unitary time evolution in which entanglement does not grow in a particular, privileged basis.
In this manuscript, the privileged basis corresponds to the `computational' basis $\ket{\{\sigma_i^z\}}$.
An automaton gate $\hat{U}$ acts as a permutation operator when acting on states belonging to the privileged basis, returning another state belonging to the same basis, up to a phase: $\hat{U}\ket{\{\sigma_i^z\}} = e^{i\theta} \ket{\{\tilde{\sigma}_i^z\}}$.
In this manuscript we make use of \emph{stochastic} automata, in which the permutations are chosen stochastically.

The simplest dynamics that one can implement is `single spin flip' (SSF).
In SSF dynamics, a candidate spin is chosen at random, and it is flipped only if its neighboring spins sum to zero: This corresponds to a stochastic automaton with five-site gates (for the square lattice).
On the triangular lattice, the $z=6$ neighboring spins determine whether the central spin can be flipped.
To help ameliorate the finite time limitation, we work with gates of larger size.
On the square lattice, we can consider gates of size $G_n = n^2 + 4n$, i.e., $n^2$ flippable spins, and $4n$ boundary spins that determine which of the `bulk' spins are flippable (SSF dynamics then corresponds to the choice $n=1$).
For each state of the $4n$ boundary spins, we find \emph{all} spin states that can, in principle, be accessed by applying the fundamental SSF gates.
More precisely, we find the Krylov sectors of the subregion of size $G_n$ that can be accessed by flipping the $n^2$ `bulk' spins, subject to the domain-wall-conserving constraints in Fig.~\ref{fig:allowed-moves-square}.
The dynamics then proceeds by picking with uniform probability a random state from the configurations that are dynamically accessible.
This procedure is equivalent to performing infinite temperature Monte Carlo on the gate subregion for infinite time, and then selecting the output: each state -- if it is dynamically connected to the initial state -- is selected with equal probability, including the initial state.
In this way, the dynamics satisfies detailed balance.

\begin{figure}
    \centering
    \includegraphics[width=0.85\linewidth]{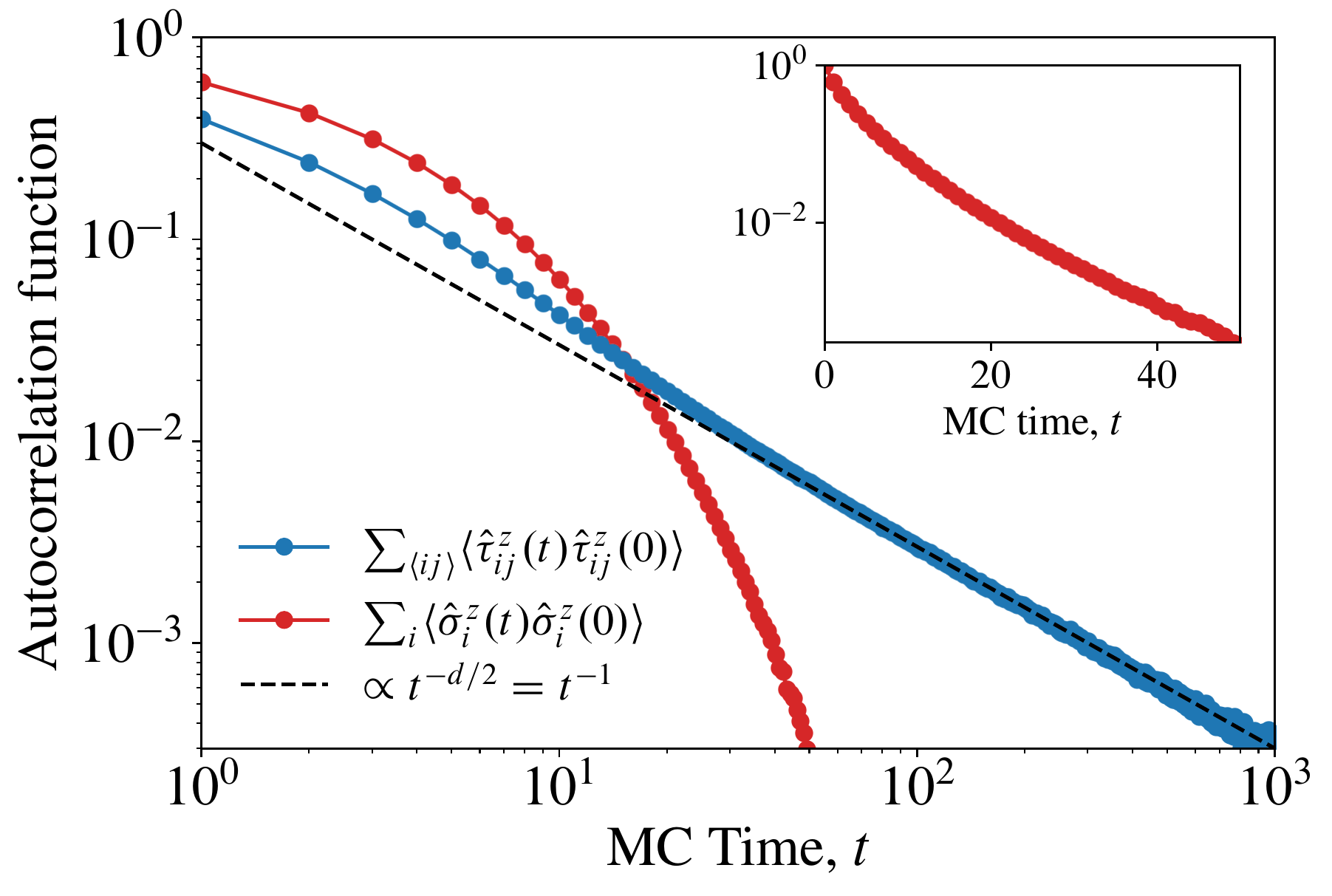}
    \caption{\textbf{Diffusion of conserved charge}. Autocorrelation function for the bond variables $\hat{\tau}^z_{ij} \equiv \hat{\sigma}_i^z \hat{\sigma}_j^z$ (blue), which are locally conserved by the dynamics, and for the spins $\hat{\sigma}_i^z$ (red). The bond variables exhibit a diffusive $\propto 1/t$ decay (dashed line), while the nonconserved $\hat{\sigma}_i^z$ correlator decays exponentially in time, as shown in the inset. The dynamics makes use of the gate $G_3$ in a system of size $L_x=L_y=256$.}
    \label{fig:diffusion}
\end{figure}


\subsection{Autocorrelation functions}

In Fig.~\ref{fig:diffusion} we plot the autocorrelation functions for the spins $\hat{\sigma}_i^z$, and for the locally conserved operators $\hat{\tau}_{ij}^z \equiv \hat{\sigma}_i^z \hat{\sigma}_j^z = 2\hat{Q}_{ij}-\mathds{1}$, related to the local density of the conserved charge $\hat{Q}$. Before presenting our results, we place some bounds on what could happen. In the absence of a one-form constraint, we would expect the locally conserved operators to relax via diffusion. With a subsystem $\U1$ conservation law in addition to the global $\U1$ conservation law, we would expect $k^4$ subdiffusion {\it\`{a} la} `fracton hydrodynamics'~\cite{IaconisSubsystem, GLN}. A global $\U1$ conservation law plus a subsystem $\Z2$ constraint should presumably produce relaxation no faster than diffusion and no slower than $k^4$ subdiffusion.

We now determine numerically what obtains. We use the operators $\hat{\tau}_{ij}^z$ in place of the charges $\hat{Q}_{ij}$ since the average $\langle \tau_{ij}^z \rangle$ vanishes at infinite temperature.
Since the autocorrelation function can be interpreted as a return probability, normalization of the domain wall density distribution gives rise to the asymptotic behaviour $\langle \hat{\tau}_{ij}^z(t)\hat{\tau}_{ij}^z(0) \rangle \sim 1/t^{d/2} = 1/t$ if domain wall density spreads diffusively, as is observed in Fig.~\ref{fig:diffusion}.
The numerics therefore shows that the weak one-form $\Z2$ constraint on the Hilbert space of domain wall configurations, $\prod_{\langle ij \rangle \in \gamma} \hat{\tau}_{ij} = \mathds{1}$ for all closed loops $\gamma$, has no measurable impact on the decay of the autocorrelation function, which follows ordinary diffusion. 
This result is to be contrasted with the behaviour of the autocorrelation function of the local magnetization, $\langle \hat{\sigma}_i^z(t) \hat{\sigma}_i^z(0) \rangle$: Since $\hat{\sigma}_i^z$ is not locally conserved, its associated autocorrelation function instead (asymptotically) decays exponentially in time, as shown in the inset of Fig.~\ref{fig:diffusion}.


\subsection{Frozen sites}


\subsubsection{Definition of frozen site density}

To study in further detail the crossover from weak to strong shattering as domain wall density is reduced from $n=1/2$, as observed in Fig.~\ref{fig:frozen-sites-exact}, we must introduce a finite time generalization of the frozen site density. We make use of the definition proposed in Ref.~\cite{MorningstarFreezing2020}.
At time $t=0$, all sites are classified as frozen. At a later time, a site is classified as frozen if it has flipped in \emph{any} of the prior configurations that the system has passed though:
\begin{equation}
    \rho_\text{F}(t) = \frac{1}{N}  \big\lvert \{ \sigma_i^z \: | \: \sigma_i^z(t) = \sigma_i^z(t-1) = \cdots = \sigma_i^z(0) \} \big\rvert
    \, .
    \label{eqn:frozen-site-automaton}
\end{equation}
Correspondingly, the number of active sites is given by the complement: $\rho_\text{A}(t) \equiv 1 - \rho_\text{F}(t)$.
When defined in this way, a frozen site can become active, but the converse is not true: an active site can never become frozen.
If it were possible to run the automaton circuits for infinite time, the the definition~\eqref{eqn:frozen-site-automaton} would agree exactly with the definition used previously in Sec.~\ref{sec:enumeration}. Note also that $\lim_{t\to \infty} \rho_\text{F}(t)$ does not depend on the size of the gate $G_n$ chosen to evolve the system.
In practice, we run the automaton circuits until $\rho_\text{F}(t)$ has plateaued. This does not, however, rule out the existence of states with a significantly higher density of active sites, although such states would need to be atypical. To mitigate the possibility of a diverging timescale -- invisible to large automaton simulations -- after which the behaviour of $\rho_\text{F}(t)$ changes drastically, we start with system sizes that can be accessed using exact enumeration, and show that there exists a unique
plateau (i.e, no metastable behaviour) in $\rho_\text{F}(t)$, whose value coincides with the exact result obtained using the enumeration results from Sec.~\ref{sec:enumeration}.
While this procedure does not rigorously rule out the existence of a diverging timescale that only manifests in larger system sizes, it makes it plausible that the plateaux observed in our simulations do indeed coincide with the true asymptotic behaviour of the frozen site density $\rho_\text{F}(t)$.

\begin{figure}
    \centering
    \includegraphics[width=\linewidth]{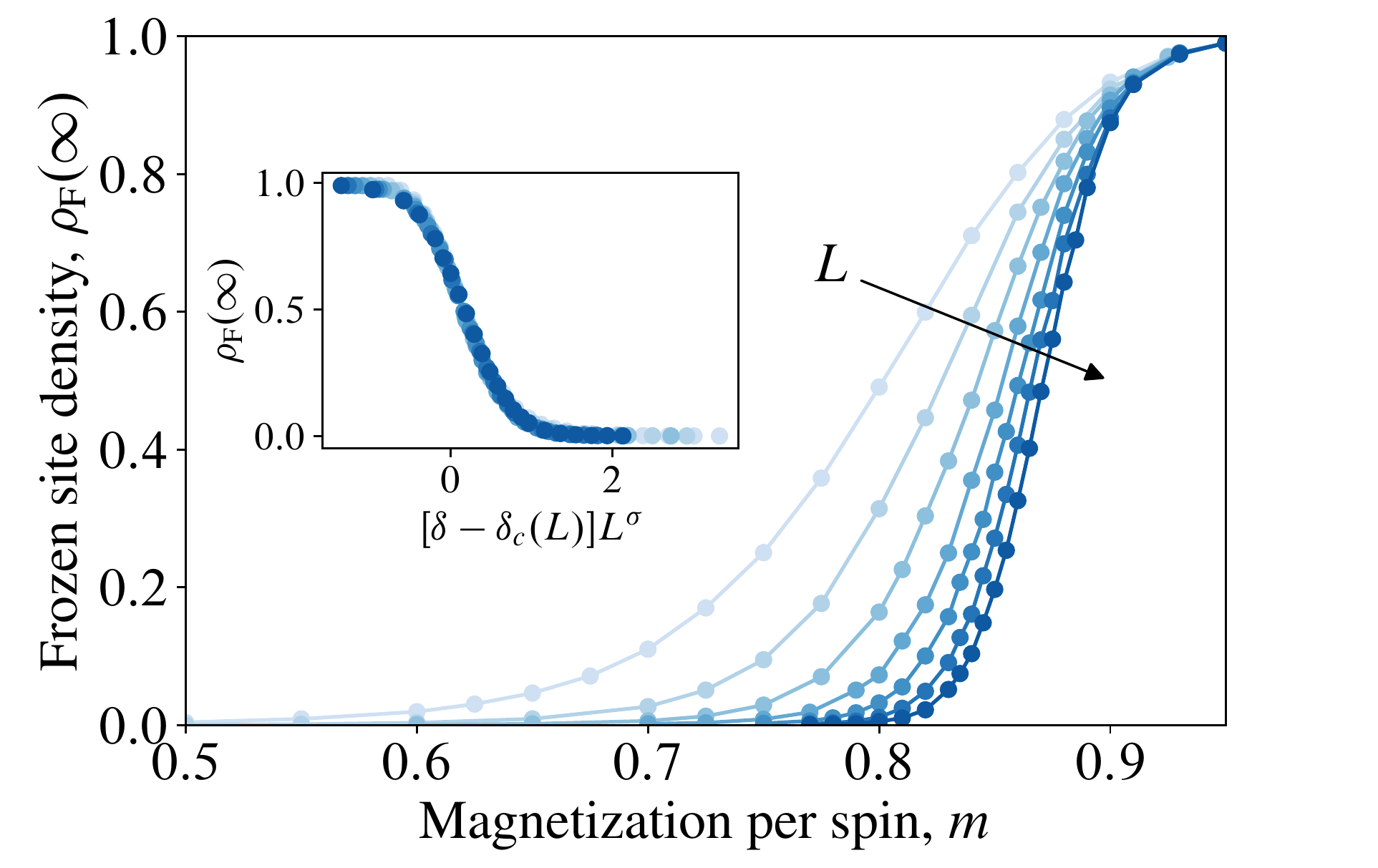}
    \caption{\textbf{Finite size scaling of frozen site density}. Asymptotic value of the frozen site density obtained from the square lattice automaton circuit as a function of the initial magnetization density $m$. The data are for systems of size $L_x=L_y = 12, 16, \ldots, 36$ in equally spaced intervals. The data are collapsed as a function of $\delta = (1-m)/2$ in the inset using a system-size-dependent critical magnetization density, $m_c(L) = 1 - 2\lambda/\ln L$. Each data point is taken once $\rho_\text{F}$ has equilibrated, and is averaged over at least $2^{13}$ independent histories.}
    \label{fig:frozen-collapse}
\end{figure}


\subsubsection{Initial state distribution}

To access states with domain wall density $n<1/2$, i.e., away from infinite temperature, we must specify a distribution of initial states with unequal weighting.
We choose to work with uncorrelated states in the computational basis with nonvanishing magnetization density.
That is, each spin is drawn from a biased, bimodal probability distribution $P(\sigma_i^z) = \frac12(1 + \sigma_i^z m)$, such that each spin individually satisfies
$\langle \sigma_i^z \rangle = m$ in the initial state.
Since the distribution of domain walls is symmetric under $m\to -m$, we work without loss of generality with initial states satisfying $m>0$.
An alternative choice would be to work with eigenstates of $\hat{H}_{0} = 2J\hat{Q}$, weighted according to the Boltzmann distribution $\propto \exp(-\beta \hat{H}_0)$, where the temperature $T=\beta^{-1}$ controls the density of domain walls.
However, at low temperatures, $T \ll J$, where the correlation length is $O(1)$, the two distributions should agree quantitatively with one another.


\subsubsection{Automaton results}

The asymptotic frozen site density obtained using the automaton circuit is shown in Fig.~\ref{fig:frozen-collapse}.
It is clear from Fig.~\ref{fig:frozen-collapse} that the crossover from strong to weak shattering
(i) becomes sharper with increasing system size, and (ii) drifts towards larger values of the magnetization density $m$ (smaller values of the domain wall density). 
These two features are verified quantitatively in the scaling collapse shown in the inset of Fig.~\ref{fig:frozen-collapse},
which establishes that $\rho_\text{F}(m, L) \simeq \mc{F}[(\delta - \delta_c(L)) L^{\sigma}]$, where $\delta$ parametrizes the number of domain walls via $m \equiv 1-2\delta$, $\sigma = 1.0(1)$, and $\delta_c(L) = \lambda/\ln L$, with $\lambda = 0.22(2)$. This scaling of the critical domain wall density will be justified analytically in Sec.~\ref{sec:void-instability}.
We therefore conclude that there is no phase transition at nonzero excitation density in the standard thermodynamic limit, i.e., $\lim_{L \to \infty} \rho_\text{F}(m, L) = 0$ for all $m<1$ (i.e., $\delta > 0$).
While there exists no transition in the standard thermodynamic limit, two comments are in order.
First, the slow decay of $\delta_c(L)$ with system size $L$ implies that finite size effects are extremely important; even macroscopically large systems will exhibit a finite size transition due to the slow (logarithmic) dependence of $\delta_c$ on $L$.
Second, the transition becomes sharp as a function of the rescaled density $\delta \ln L$. Although the critical \emph{density} of domain walls vanishes in the strict thermodynamic limit,
the system requires a macroscopic \emph{number} of defects relative to the ferromagnetic spin configurations to lead to melting.
One may therefore view the behaviour in Fig.~\ref{fig:frozen-collapse} as a freezing transition in a nonstandard thermodynamic limit.

\begin{figure}
    \centering
    \includegraphics[width=\linewidth]{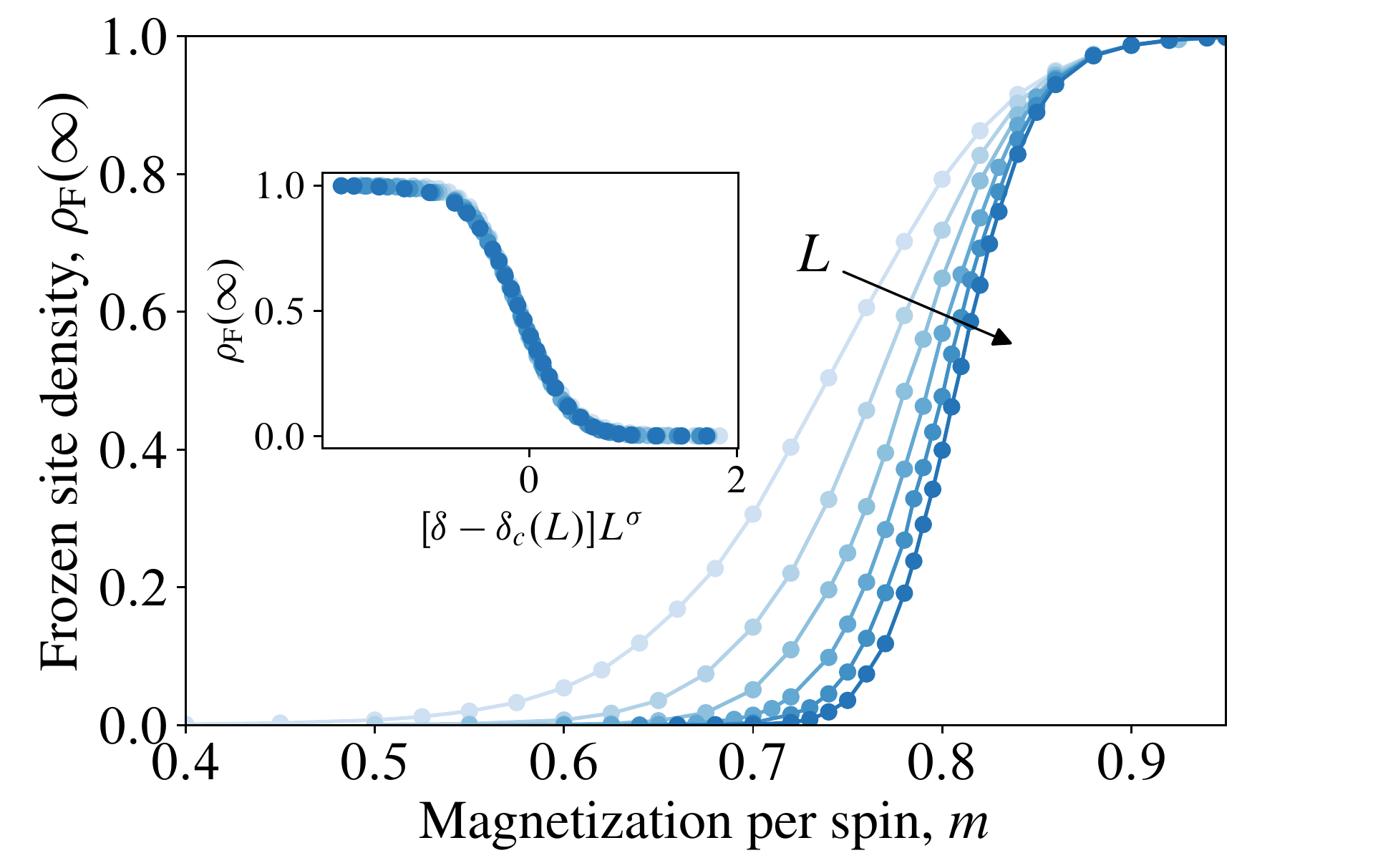}
    \caption{\textbf{Finite size scaling of triangular lattice frozen site density}. Asymptotic value of the frozen site density obtained from the triangular lattice automaton circuit as a function of the initial magnetization density $m$. The data are for systems of size $L_x=L_y = 12, 16, \ldots, 32$ in equally spaced intervals. The data are collapsed in the inset using a system-size-dependent critical magnetization density, $m_c(L) = 1 - 2\lambda/\ln L$. Each data point is taken once $\rho_\text{F}$ has equilibrated, and is averaged over at least $2^{14}$ independent histories.}
    \label{fig:frozen-collapse-tri}
\end{figure}

Equivalent results for the triangular lattice are shown in Fig.~\ref{fig:frozen-collapse-tri}, which also exhibits a finite size freezing transition as $m \to 1$. We observe that the data again exhibit a high quality collapse upon rescaling the magnetization density $m \to [m - m_c(L)]L^{\sigma}$, with $\sigma \approx 0.9(1)$ and a system-size-dependent $\delta_c(L) = (1-m_c)/2 =\lambda/\ln L$, with $\lambda=0.34(2)$. While $\lambda$ differs substantially between the square and triangular lattices, the values for $\sigma$ are consistent with one another.


\subsubsection{Analyzing the transition}

To gain further information about the weak--strong transition, we perform an analysis of the system's spin clusters once $\rho_\text{F}$ has reached equilibrium.
We will show that the freezing transition coincides with a percolation threshold for minority spin clusters.
A cluster is defined as a contiguous region of spins with the same sign, separated from spins of the opposite sign by domain walls. The system is percolating if it contains a so-called wrapping cluster of minority spins (the analogue of spanning clusters in the context of open boundary conditions).
A wrapping cluster is defined as a spin cluster that contacts itself around the periodic boundary conditions in either direction, i.e., it has a nontrivial winding number.
The number of non-wrapping clusters of size $s$ \emph{per site} is given by $n_s(m)$,
which implies that the probability that a given spin belongs to a non-wrapping cluster of size $s$ is given by
\begin{equation}
    w_s(m) = 
    \frac{ sn_s(m)}{\sum_{s} sn_s(m)}
    \, .
\end{equation}
This probability distribution can then be used to define the mean cluster size via $S(m) = \sum_s s w_s(m)$.
Note that while we parametrize the cluster properties by the initial state magnetization $m$, the measurements are made in the plateaux of $\rho_\text{F}$ where the magnetization of the system is related nontrivially to $m$.
Finally, we introduce the percolation probability $\Pi(m)$ defined by the fraction of histories that contain (at least) one wrapping cluster.
Since the mapping between spins and domain walls is not one-to-one (i.e., there are two spin configurations $\ket{\{\sigma_i^z\}}$ and $\ket{\{ \varsigma_i^z \}}$ for each configuration of domain walls, $\{\tau_{ij}^z\}$, related by the Ising symmetry, $\ket{\{\sigma_i^z\}} = \prod_j \hat{\sigma}_j^x \ket{\{ \varsigma_i^z \}}$), the spin species that is initially in the minority \emph{may}, if the system is only weakly fragmented, become the majority spin species. If the minority species is able to percolate, we therefore expect the percolation probability $\Pi \simeq 1/2$. That is, if $\rho_\text{F} \ll 1$, we expect that the system will be able to explore states with magnetization $\simeq m$ and $\simeq 1-m$. If the system spends roughly equal amounts of time in each configuration, then the
probability that the spin species that was initially in the minority has become the majority species -- at a particular point in time -- is roughly $1/2$.

\begin{figure}
    \centering
    \includegraphics[width=\linewidth]{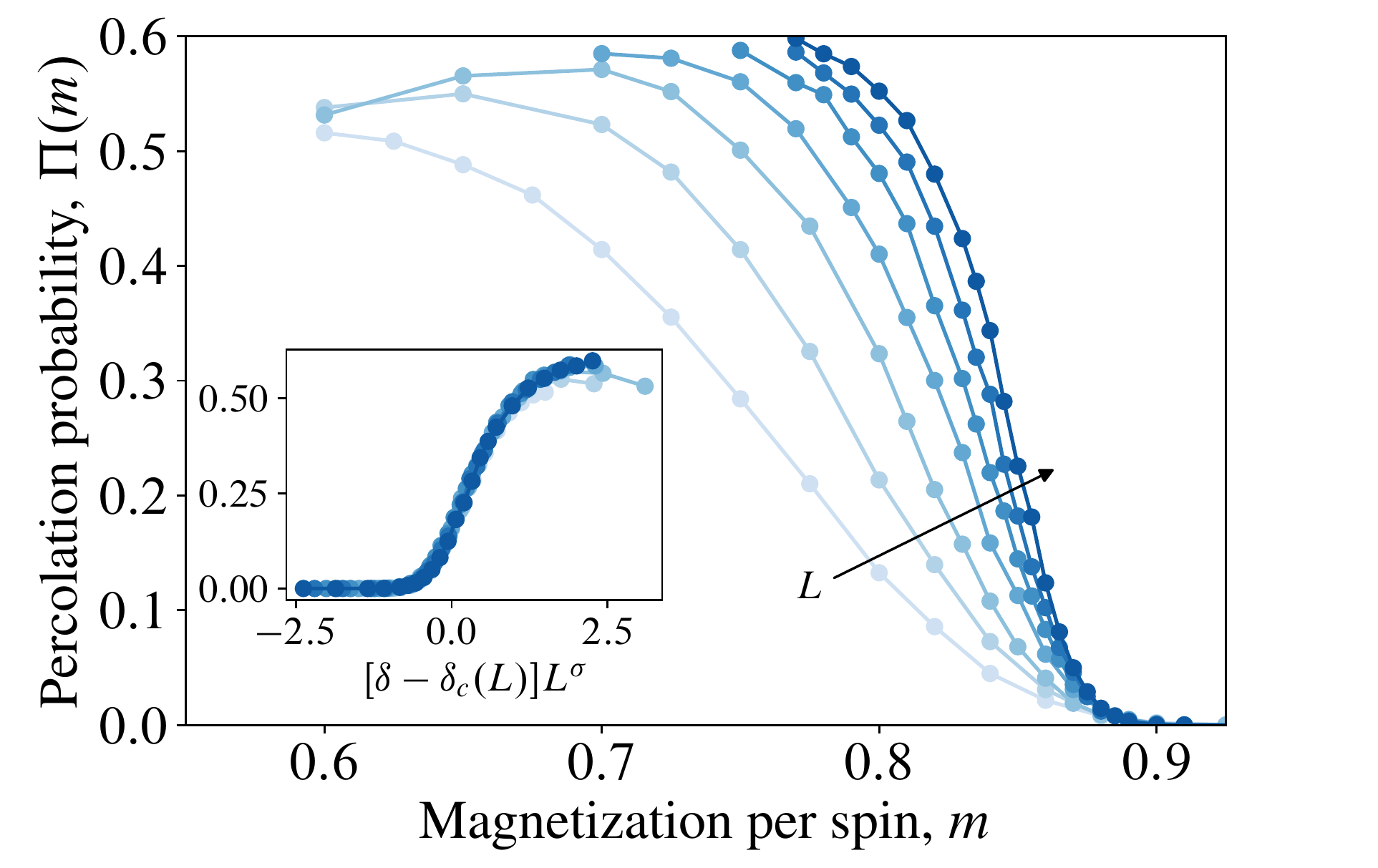}
    \caption{\textbf{Square lattice percolation probability}. The fraction of histories that contain a wrapping cluster, $\Pi(m)$, as a function of initial state magnetization on the square lattice. In the inset, we observe an excellent collapse of the data as a function of $\delta=(1-m)/2$ using parameters consistent with Fig.~\ref{fig:frozen-collapse}. Each data point is taken once $\rho_\text{F}$ has equilibrated, and is averaged over at least $2^{13}$ independent histories. The data are for systems of size $L_x=L_y = 12, 16, \ldots, 36$ in equally spaced intervals.}
    \label{fig:perc-probability}
\end{figure}

The percolation probability $\Pi$ as a function of initial state magnetization $m$ for various system sizes is shown in Fig.~\ref{fig:perc-probability}.
The data exhibit a high quality collapse using parameters that are consistent with those used to collapse the frozen site density in Fig.~\ref{fig:frozen-collapse}.
Figure~\ref{fig:perc-probability} therefore suggests that the unfreezing transition as domain wall number is increased can alternatively be viewed as a percolation transition for the minority spin clusters.
For a density of excitations less than the critical value, $n_c \sim \lambda /\ln L$, isolated minority clusters remain disconnected, with a finite mean size $S(m)$, and are unable to make the system active.
While the disconnected clusters may exhibit dynamics, any active spins represent a vanishing fraction of the total system volume.
Conversely, above $n_c$, the initially disconnected clusters are able to grow under the domain-wall-conserving dynamics, maintaining a constant perimeter, and are able to coalesce to form a percolating cluster of spins that covers a macroscopic fraction of the system, making it active in the process. In the next section we will propose a mechanism by which this growth process is able to occur.


\section{Large void instabilities}
\label{sec:void-instability}

In the limit of low domain wall density, $\delta \equiv (1-m)/2 \ll 1$, the average separation between flipped spins is large, $\sim \delta^{-1/d} \gg 1$.
However, in sufficiently large systems, there will exist rare regions where the local density of flipped spins significantly exceeds $\delta$.
We will argue that such rare regions, which must be present in an infinite system, provide a mechanism through which the entire system is able to melt (i.e., become active).
Equivalently, for a fixed system size $L$, there will exist a domain wall density $\delta_c(L)$ above which there exists an appreciable probability of such a rare region, allowing the system to melt.

Suppose that there exists a square of minority spins of size $\ell \times \ell$ somewhere in the system.
If the square is surrounded by boundary layer of the majority spin species of width two, then the square is unable to grow.
Conversely, if there exists a single flipped spin on one of its edges, creating a neighboring kink and antikink,
\begin{equation}
    \includegraphics[width=0.5\linewidth, valign=c]{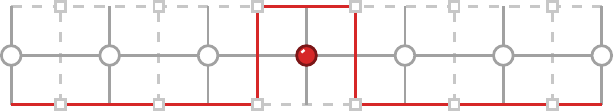}
    \, ,
    \label{eqn:bump}
\end{equation}
both the kink and the antikink can be propagated outwards, away from the arbitrary initial position of the flipped spin, increasing the width or height of the $\ell\times\ell$ square by one unit:
\begin{equation}
    \includegraphics[width=0.5\linewidth,valign=c]{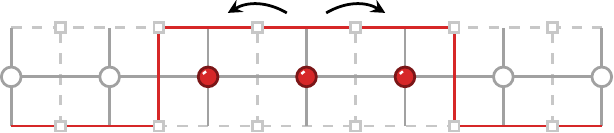}
    \, .
    \label{eqn:bump_grow}
\end{equation}
If there is one flipped spin on each of the four edges [and there are no spins in outer rows to obstruct the growth process shown in~\eqref{eqn:bump_grow}], then the square can be grown from size $\ell\times\ell \to (\ell+2)\times(\ell+2)$ using the moves in~\eqref{eqn:bump_grow}.

\subsection{One spin per edge}

Having shown that a single flipped spin on each edge of the square is able to enlarge the width and height of a square region by two units, we now show that flipped spins in outer rows -- required for further growth -- do not significantly impede this growth process.
We will begin by showing constructively that \emph{exactly} one spin per row allows the system to melt.
The pathways that we provide are certainly not unique; there are other ways in which the minority spin cluster can grow for a given initial condition of i.i.d.~spins.
Also note that we are not concerned with the timescale over which this growth occurs, i.e., we characterize whether the system's dynamics is irreducible (two states selected at random will be connected by the dynamics) as opposed to ergodic (two states selected at random will be connected in \emph{finite time} by the dynamics)\footnote{A nonzero frozen fraction is a sufficient condition for reducibility and the breaking of ergodicity, but it is not necessary. E.g., the one magnon sector of the \textit{XY} model is reducible, but $\rho_\text{F}=0$.}.
First, we consider the interaction of the kink structure in~\eqref{eqn:bump} with spins in outer rows.
Making use of reflection symmetry of an edge and the ability to translate the isolated flipped spin, there are just two situations that we need to consider (up to corner cases considered in the supplemental material (SM)~\cite{SM}).
First, consider the special case that the flipped spins in adjacent rows are nearest neighbors.
In this case, the baseline can immediately be increased by one unit using only the permitted local rearrangements of domain walls:
\begin{equation}
    \includegraphics[width=0.85\linewidth,valign=c]{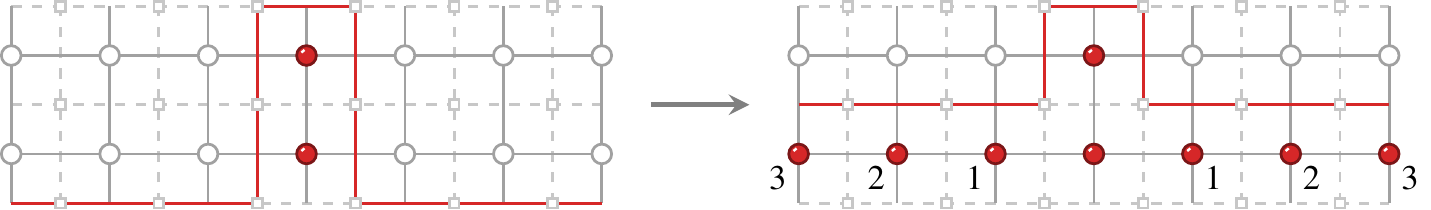}
    \, .
    \label{eqn:vertically}
\end{equation}
The spins have been flipped in the order 1, 2, 3, denoted by the integers next to the spins on the right-hand side.
The state of the system has therefore been reduced that of state~\eqref{eqn:bump}, and requires no further classification.
If the spins are not nearest neighbors, then the bump can (almost) always be translated such that the red spins are arranged with the following relative positions (special cases, in which the spin in the top row hangs over the corner of the square, are dealt with separately in the SM~\cite{SM}):
\begin{equation}
    \includegraphics[width=0.85\linewidth,valign=c]{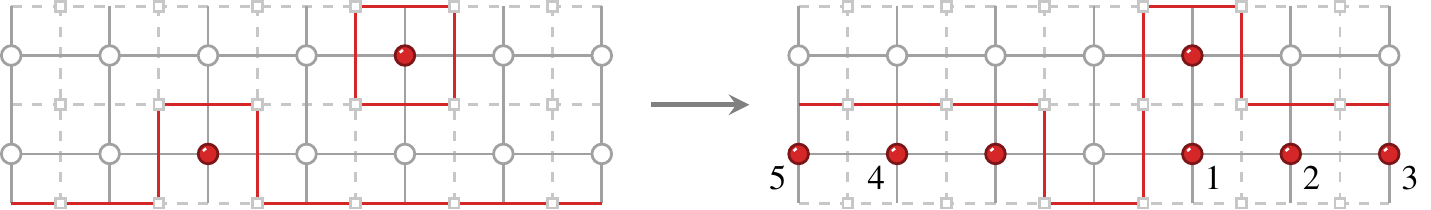}
    \, .
    \label{eqn:diagonally}
\end{equation}
In the above, the isolated flipped spin that is separated from the cluster is attached by flipping the intervening spin. The baseline is subsequently increased by one unit to produce a new type of kink structure.
We must therefore classify how the new structure in~\eqref{eqn:diagonally} interacts with flipped spins in adjacent rows.
Note that, in the absence of spins in outer rows, the structure on the right-hand side of~\eqref{eqn:diagonally} can be freely translated perpendicular to the surface normal.
First, we consider the case where the flipped spin appears to the left of the kink structure
\begin{equation}
    \includegraphics[width=0.85\linewidth,valign=c]{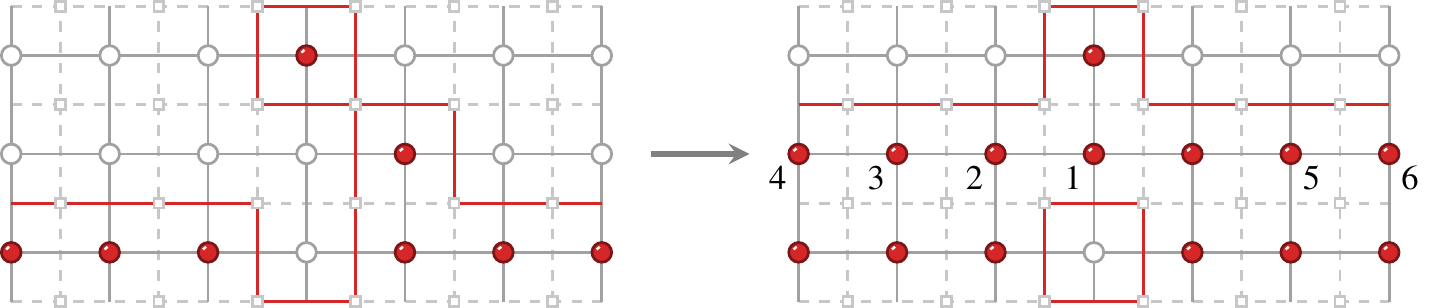}
    \, .
    \label{eqn:double-kink-left}
\end{equation}
The isolated spin is incorporated into the cluster by first flipping the spin underneath, and the baseline can then be increased by propagating kinks outwards, as in~\eqref{eqn:bump_grow}.
The configuration is therefore reduced to that found in~\eqref{eqn:bump}.
While the void that is left behind in the spin cluster may appear frozen in~\eqref{eqn:double-kink-left}, note that, for instance, the kink structure could have been translated to the right prior to incorporating the isolated spin.
If the isolated spin instead appears to the right of the kink structure, a more intricate rearrangement of domain walls is required
\begin{equation}
    \includegraphics[width=0.85\linewidth,valign=c]{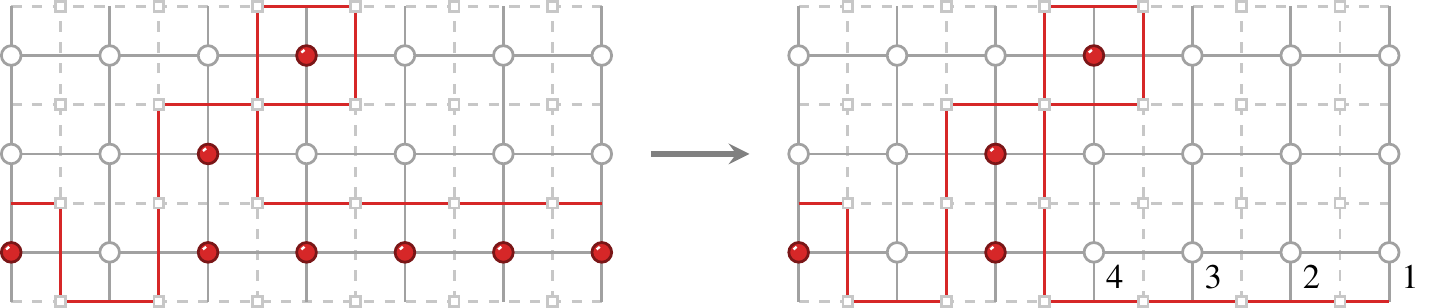}
    \, .
    \label{eqn:double-kink-right-pre}
\end{equation}
The first step corresponds to bringing in a kink from the right corner\footnote{Bringing in a kink from the right corner can alternatively be thought of as undoing the moves 2, 3 in~\eqref{eqn:diagonally} that were used to propagate the kink away.}.
The `tower' of minority spins can then be disconnected from the main cluster and subsequently reconnected in a different location to allow the baseline to be increased by one unit:
\begin{equation}
    \includegraphics[width=0.85\linewidth,valign=c]{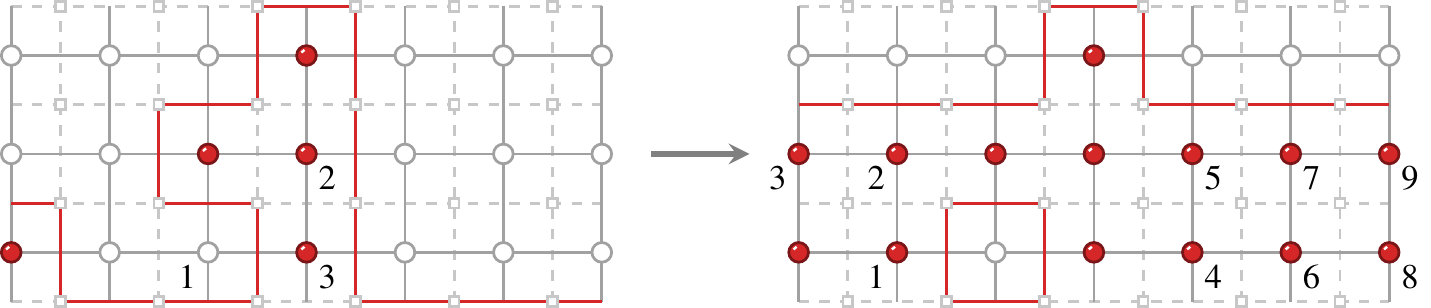}
    \, ,
    \label{eqn:double-kink-right-post}
\end{equation}
reducing the configuration to~\eqref{eqn:bump} with a void. As above, the spin in the void is not frozen, since the kink structure can be translated to the left prior to incorporating the spin.
The final situation to consider corresponds to the case where the the spin in the top row of either~\eqref{eqn:double-kink-left} or \eqref{eqn:double-kink-right-pre} is a nearest neighbor of the spin in the row below. In this case, spins in rows above must facilitate the inclusion of the `tower' of spins into the cluster. It is possible to use an analogous sequence of moves to those presented in \eqref{eqn:double-kink-left}--\eqref{eqn:double-kink-right-post} to absorb a tower of arbitrary height into the cluster.
This sequence of moves is presented explicitly in the SM~\cite{SM}, along with a number of special cases that occur at the corners of the square.
While it may appear that the condition of exactly one spin per row is overly restrictive, and could instead be relaxed to the condition that no two adjacent rows are empty, there exist edge cases that do not permit growth in the manner described in this section. If there exist pathways that allow such edge cases to be melted, then the prefactor in~\eqref{eqn:critical-delta-asymptotic} below will be modified, but the scaling with system size will remain unchanged. 


\subsection{Scaling of the critical magnetization}

Given that exactly one flipped spin per edge is sufficient to melt the system, we now bound the
probability of melting by assuming that additional flipped spins will help to facilitate growth of the cluster, rather than impede it. This statement is certainly true on average, since the frozen fraction $\rho_\text{F}(\delta)$ is a monotonic decreasing function of domain wall density, parametrized by $\delta$, for all system sizes.
It is therefore plausible that \emph{at least} one defect per side is required in order for the square of size $\ell\times \ell$ to be able to grow to become infinitely large in the thermodynamic limit $L \to \infty$.
If this assumption is satisfied, then the probability that a region of size $\ell \times \ell$ is able to grow to size $(\ell + 2)\times (\ell + 2)$ is therefore
\begin{equation}
    P_{\ell\to\ell+2} = \left[1 - (1-\delta)^\ell \right]^4
    \, .
    \label{eqn:l->l+2}
\end{equation}
Analogous expressions for the growth probability arise in the context of, e.g., self-diffusion~\cite{jackie1991size}, bootstrap percolation~\cite{DeGregorio2016}, and other kinetically constrained models~\cite{Ritort2003Glassy}.
The probability for the this process to be able to continue indefinitely is given by iterating the above expression, implying that $P_{\ell \to \infty} = \prod_{k=0}^\infty P_{\ell+2k\to \ell + 2k + 2}$, or, using~\eqref{eqn:l->l+2},
\begin{equation}
    P_{\ell\to\infty} = \exp\left\{ 4 \sum_{k=0}^{\infty} \log\left[1 - (1-\delta)^{\ell+2k} \right] \right\}
    \, .
    \label{eqn:P_infty_exact}
\end{equation}
This expression should represent a lower bound on the probability for the entire system to be become active, since growth can proceed via alternate, e.g., rectangular, pathways.
For sufficiently large $\ell$, $(1-\delta)^\ell \ll 1$, allowing the logarithm to be expanded. This allows the summation to be performed exactly,
and leads to the approximate expression
\begin{equation}
    P_{\ell \to \infty} \simeq \exp\left( -4 \frac{ (1-\delta)^{\ell} }{2\delta - \delta^2} \right)
    \, .
    \label{eqn:P_to_infty}
\end{equation}
As explained in, e.g., Ref.~\cite{Ritort2003Glassy}, for $\ell \gtrsim \ell_*$ the probability that the $\ell \times \ell$ minority cluster grows to envelop the whole system saturates to unity; an ``unstable void''.
From Eq.~\eqref{eqn:P_to_infty}, we identify $\ell_* = \ln[(2\delta - \delta^2)/4] / \ln(1-\delta) \simeq -\log(\delta/2)/\delta$. The probability that such an unstable cluster is present in the system's initial condition is exponentially small in $\ell_*^2 \gg 1$. 
Instead, it is more likely that a small cluster grows to become unstable by reaching a size $\ell_* \times \ell_*$, after which its ability to grow is guaranteed.
The summation appearing in~\eqref{eqn:P_infty_exact}, for growth beginning from a single site, $P_{1\to\infty}$, can be bounded from below by turning the summation into an integral~\cite{AizenmanLebowitz1988,Ritort2003Glassy}
\begin{equation}
    \sum_{k=0}^{\infty} \ln[1-(1-\delta)^{1 + 2k}] \geq \frac{-1}{2\ln(1-\delta)} \int_{0}^{\infty} dx \, g(x) \\
\end{equation}
where $g(x) = \ln[1-e^{-x}]$. The integral can be evaluated exactly to give $-\int_0^\infty dx \, g(x) = \Li_2(1) = \pi^2/6$, with $\Li_n(z)$ the polylogarithm function.
Combining the above results, the probability that a single minority spin can grow to cover the entire system is bounded by
\begin{equation}
    P_{1\to\infty} \geq \exp\left\{  \frac{\pi^2}{3} \frac{1}{\ln(1-\delta)} \right\} > 0
    \, .
    \label{eqn:P1}
\end{equation}
The probability $P_{1\to\infty}$ vanishes as $P_{1\to\infty} \sim \exp\{ -\text{const.}/\delta \}$ as $\delta \to 0^+$. Nevertheless, it is still nonzero, and for sufficiently large systems, there will exist such a nucleation site.
The probability that the system contains a minority spin that is able to grow and melt the system is set by $\delta L^2 P_{1\to\infty}$~\cite{Ritort2003Glassy} (alternatively, the probability can be bounded more rigorously by subdividing the system into $L$ independent $\sqrt{L}\times \sqrt{L}$ regions~\cite{Ertel1988,AizenmanLebowitz1988}), and the critical value of $\delta$ is found by solving $\delta L^2 \exp( -\pi^2/(3\delta) ) = c = O(1)$ (valid for $\delta \ll 1$), leading to a critical value of $\delta$ that vanishes with increasing system size logarithmically
\begin{equation}
    \delta_c(L) = \dfrac{\pi^2}{3 W(\frac{L^2 \pi^2}{3c})} \sim \dfrac{\pi^2}{6 \ln L}
    \, ,
    \label{eqn:critical-delta-asymptotic}
\end{equation}
where $W(x)$ is the product-log function.
In the context of bootstrap percolation, the prefactor $\pi^2/6$ appearing in Eq.~\eqref{eqn:critical-delta-asymptotic} is in fact provably asymptotically \emph{exact} for $d=2$~\cite{holroyd2003sharp}.
However, the subleading corrections as $\delta \to 0^+$ can lead to sizable corrections to the asymptotic behavior in system sizes that are accessible numerically~\cite{DeGregorio2005exact,DeGregorio2016}. 

\begin{figure}
    \centering
    \includegraphics[width=\linewidth]{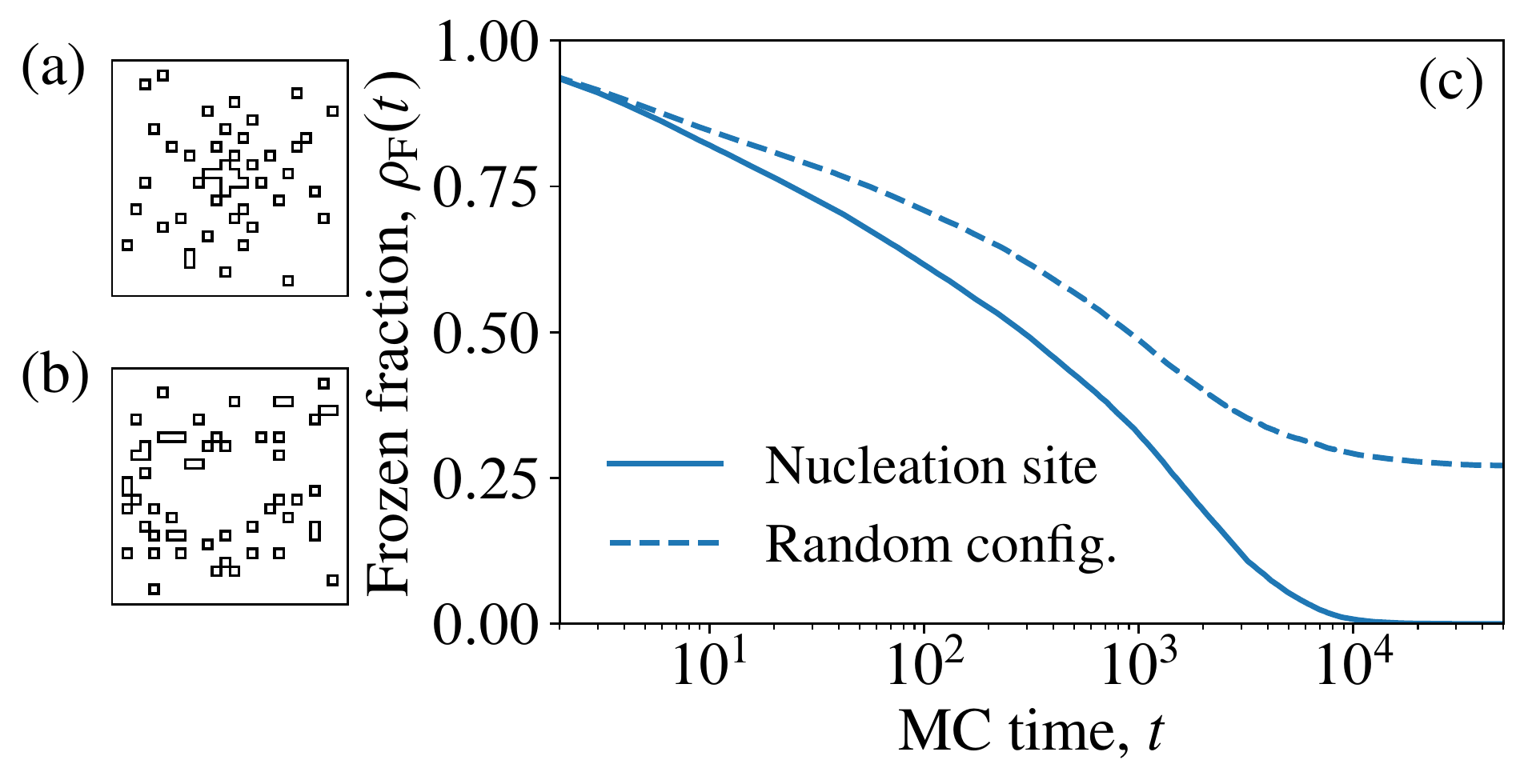}
    \caption{\textbf{Melting from nucleation site}. The configuration in (a) has a $2\times 2$ square surrounded by one spin per row, while (b) is a typical uncorrelated random state with the same magnetization (on average). In (c) we plot the frozen fraction $\rho_\text{F}(t)$ as a function of time for the configuration (a) (solid line) and averaged over typical uncorrelated states with $m\approx 0.83$ (dashed line). The initial state (a) saturates to $\rho_\text{F}=0$, while the average over random configurations has a nonzero asymptotic frozen fraction. Both curves are computed for a system of linear size $L=24$, and are averaged over $500$ circuit realizations.}
    \label{fig:nucleation}
\end{figure}

The above provides a justification of the slow $(\ln L)^{-1}$ scaling of $\delta_c(L)$ observed in Figs.~\ref{fig:frozen-collapse} and~\ref{fig:perc-probability}.
Intriguingly, the more constrained triangular lattice appears to exhibit similar logarithmic decay. To provide further numerical evidence that the above mechanism is responsible for the melting transition, we plot in Fig.~\ref{fig:nucleation} the frozen fraction as a function of time for a system that (a) contains a nucleation site, and (b) has uncorrelated initial states with the same magnetization as (a).
We observe that the nucleation site permits \emph{all} sites to become active for times $t\gtrsim 10^4$ (importantly, this time is system size dependent), while the uncorrelated initial states lead to a substantial fraction of asymptotically frozen sites.
This figure confirms that nucleation sites, if present, have the ability to melt the entire system.

\section{Discussion}
\label{sec:discussion}

We have shown that the transverse field Ising model in two space dimensions, deep within its ferromagnetic phase, exhibits a rich and hitherto largely unappreciated structure to its quantum dynamics. In particular, up to a prethermal timescale, the Hilbert space shatters into a number of disconnected Krylov subsectors exponentially large in system volume. The precise timescale on which the Krylov subsectors reconnect depends on the particular patterns that are being tiled to make the subsectors, but could be as small as $J/h^2$, and large as $\sim \exp(\Upsilon J/h),$ where $\Upsilon > 0$ is some undetermined numerical constant. An easy to visualize example of a pattern for which the timescale saturates the prethermal upper bound is a stripelike pattern with locally ferromagnetic stripes of width greater than $ J/h$. 

We have explored in detail the dynamics with the Hamiltonian obtained by truncating at leading non-trivial order in Schrieffer-Wolff perturbation theory, and have provided numerical evidence that the resulting Hamiltonian exhibits at least one weak--strong shattering `freezing' transition as a function of symmetry sector in a non-standard thermodynamic limit.
We have also provided analytical and numerical evidence that this transition is linked to an instability of sufficiently large minority spin clusters. It is also important to emphasize that the freezing transition occurs for the effective Hamiltonian in Eq.~\eqref{eqn:SW-first-order}. Whether the transition survives the inclusion of corrections at higher order in the Schrieffer-Wolff procedure, or whether the `strongly shattered' phase disappears upon inclusion of higher order terms (and the transition with it), remains to be resolved.

This paper has concentrated on Ising models in two space dimensions. Our construction of exponentially many (in system volume) Krylov subsectors by tiling compact motifs should extend to arbitrary higher spatial dimensions as well, such that we do expect that the transverse field Ising model in higher spatial dimensions will also display an exponential in volume shattering of Hilbert space. Whether a strong--weak freezing transition exists in higher dimensions, and whether it can also be described by the growth of unstable minority spin clusters, remains to be explored. 

It is interesting to wonder if the observations contained herein could be probed in solid state experiments. Effective Ising models are, of course, ubiquitous in solid state systems. However, solid state systems also generically have phonons, which open up `thermal' relaxation pathways. As one lowers the temperature and moves deeper into the ferromagnetic regime, however, one would generically expect the `thermal' relaxation channels to become subleading to the intrinsically quantum relaxation channels that we have discussed. Experimental probes of the \emph{dynamics} deep in the ferromagnetic phase may therefore be able to see signatures of the Hilbert space shattering discussed herein. This would be an interesting problem for experimental exploration. 

Finally, it would be interesting to explore whether the `freezing transition' discussed herein could be obtained in realistic models in a \emph{standard} thermodynamic limit. Given that the combination of a $\U1$ conservation law and a one-form $\Z2$ constraint does not seem to be enough, it is tempting to attempt to upgrade to a $\U1$ conservation law and a one-form $\U1$ constraint, which lattice gauge theories could in principle provide~\cite{Kogut_RevModPhys1979}. The mapping of the Ising model to the Ising gauge theory suggests that such a transition, if it exists, would exist in the confining phase of the gauge theory. Whether $\U1$ lattice gauge theories in the confining phase do indeed support a freezing transition in a conventional thermodynamic limit would be another fruitful problem to explore in future work.


\begin{acknowledgements}
We thank David Huse for feedback on the manuscript. This material is based upon work supported by the Air Force Office of Scientific Research under award number FA9550-20-1-0222. RN also thanks the Simons Foundation for support through a Simons Fellowship in Theoretical Physics, and thanks the Stanford Department of Physics for its hospitality during his sabbatical. 
\end{acknowledgements}


\appendix


\section{Schrieffer-Wolff transformation}\label{sec:Schrieffer-Wolff}

 The operators $\hat{T}_n$ can be expressed explicitly, if desired, in terms of the orthogonal projectors $\hat{\Pi}_{ij}^1 = \hat{Q}_{ij}$ ($\hat{\Pi}_{ij}^0 = \mathds{1}-\hat{Q}_{ij}$), which project onto states whose bond hosts (doesn't host) a charge. In 2D, we will adopt the notation that $\hat{Q}_{i1}$, $\hat{Q}_{i2}$, $\hat{Q}_{i3}$, $\hat{Q}_{i4}$ label the charges on the bonds surrounding the site $i$ on the square lattice in a clockwise direction, starting from the top. Then
\begin{subequations}
\begin{align}
    \hat{T}_4 &= \sum_i \hat{\sigma}_i^x \hat{\Pi}^0_{i1} \hat{\Pi}^0_{i2} \hat{\Pi}^0_{i3} \hat{\Pi}^0_{i4} \label{eqn:T2}\\
    \hat{T}_2 &= \sum_i \hat{\sigma}_i^x \hat{\Pi}^1_{i1} \hat{\Pi}^0_{i2} \hat{\Pi}^0_{i3} \hat{\Pi}^0_{i4} + \text{3 permutations} \label{eqn:T4} \\
    \hat{T}_0 &=  \sum_i \hat{\sigma}_i^x \hat{\Pi}^1_{i1} \hat{\Pi}^1_{i2} \hat{\Pi}^0_{i3} \hat{\Pi}^0_{i4} + \text{5 permutations}
    \, ,
\end{align}
\label{eqn:Tn-ops}%
\end{subequations}
where the permutations are over the superscript indices.
The operators $\hat{T}_{-4}$ and $\hat{T}_{-2}$ are obtained from Eqs.~\eqref{eqn:T2} and \eqref{eqn:T4}, respectively, by interchanging the projectors $\hat{\Pi}^1_{ij} \leftrightarrow \hat{\Pi}^0_{ij}$.
The generalization to other spatial dimensions and other lattices is transparent.
The property $\hat{\sigma}_i^x \hat{Q}_{ij} = (\mathds{1}-\hat{Q}_{ij}) \hat{\sigma}_i^x$ implies that the operators in~\eqref{eqn:Tn-ops} satisfy $\hat{T}_n^\dagger = \hat{T}^{\phantom{\dagger}}_{-n}$.
To construct an effective Hamiltonian that conserves quasi-domain wall number, we perform a Schrieffer-Wolff transformation parametrized by the Hermitian operator $\hat{S}=\hat{S}^\dagger$. That is,
\begin{equation}
    \hat{H}' = e^{i\hat{S}}\hat{H}e^{-i\hat{S}} = \hat{H} + [i\hat{S}, \hat{H}]+ \frac{1}{2!} [i\hat{S}, [i\hat{S}, \hat{H}]]+ \ldots
\end{equation}
The eigenstates of the effective Hamiltonian $\hat{H}'$ are then dressed by the operator $e^{-i\hat{S}}$ to obtain the eigenstates of the original Hamiltonian $\hat{H}$.
The operator $\hat{S}$ is chosen such that $\hat{H}'$ conserves the number of domain walls up to a particular order in $h/J$. Specifically, we write $\hat{S} = \sum_k \hat{S}^{(k)}$ where $\hat{S}^{(k)}$ is chosen such that $\hat{H}'$ conserves domain walls up to order $(h/J)^k$. 
At leading order, we find that
\begin{equation}
    i\hat{S} \simeq i\hat{S}^{(1)} = -\frac{h}{4J} (\hat{T}_2 - \hat{T}_{-2}) - \frac{h}{8J}(\hat{T}_4 - \hat{T}_{-4})
    \, .
    \label{eqn:SW-first-order-S}
\end{equation}
Note that this expression does not derive from the specific form of the operators $\hat{T}_n$, only from their mutual commutation relations and $[\hat{Q}, \hat{T}_n] = n\hat{T}_n$.
The property $\hat{T}_n^\dagger = \hat{T}^{\phantom{\dagger}}_{-n}$ ensures that the operator $i\hat{S}$ is anti-Hermitian.
The first order term in $\hat{S}$ gives us access to the \emph{second} order effective Hamiltonian (since the second order term in $\hat{S}$ is chosen in such a way as to remove the terms generated at second order in $h$ that \emph{do not} conserve domain wall number, leaving only the number-conserving terms that are generated by the first order $\hat{S}$). We are therefore left with the result
\begin{equation}
    \hat{H}' = 2J\hat{Q} - h\hat{T}_0 +
               \frac{h^2}{4J} [\hat{T}_2, \hat{T}_{-2}] +
               \frac{h^2}{8J} [\hat{T}_4, \hat{T}_{-4}] + \ldots
               \, .
\end{equation}
An isolated flipped spin, which is frozen at first order in the field, becomes mobile at second order as a result of the term $[\hat{T}_2, \hat{T}_{-2}]$, which allows an adjacent spin to flip followed by the reversal of the original isolated spin.


\section{Numerical details}
\label{sec:numerical-details}

Here we provide some additional details relating to the exact enumeration presented in Sec.~\ref{sec:enumeration} of the main text.
To identify the disconnected sub-graphs of the Hamiltonian, we make use of a breadth first search of the system's adjacency matrix. Since the adjacency matrix is sparse, we only need to store $O(L 2^L)$ connections. Furthermore, the kinetic constraints often forbid a substantial fraction of the $L$ possible states from being connected (e.g., for the $5\times 5$ square lattice, the average number of connections per state is $\approx 9$ of the possible $25$).

To perform the classification of sectors, we keep track of whether each state has been visited. In a loop over all states, if the state has not yet been visited, then it acts as the root node for a breadth first search. All neighbors of the root node are added to a queue. For all states in the queue, their neighbors are added to the queue if they have not yet been visited, and the state is subsequently dequeued. This procedure is repeated until the queue is empty, at which point all states that can be reached from the root node have been classified and added to a Krylov sector.
The loop over all states ensures that all states are classified as belonging to a unique Krylov sector.


\bibliography{library}

\begin{thebibliography}{84}%
\makeatletter
\providecommand \@ifxundefined [1]{%
 \@ifx{#1\undefined}
}%
\providecommand \@ifnum [1]{%
 \ifnum #1\expandafter \@firstoftwo
 \else \expandafter \@secondoftwo
 \fi
}%
\providecommand \@ifx [1]{%
 \ifx #1\expandafter \@firstoftwo
 \else \expandafter \@secondoftwo
 \fi
}%
\providecommand \natexlab [1]{#1}%
\providecommand \enquote  [1]{``#1''}%
\providecommand \bibnamefont  [1]{#1}%
\providecommand \bibfnamefont [1]{#1}%
\providecommand \citenamefont [1]{#1}%
\providecommand \href@noop [0]{\@secondoftwo}%
\providecommand \href [0]{\begingroup \@sanitize@url \@href}%
\providecommand \@href[1]{\@@startlink{#1}\@@href}%
\providecommand \@@href[1]{\endgroup#1\@@endlink}%
\providecommand \@sanitize@url [0]{\catcode `\\12\catcode `\$12\catcode
  `\&12\catcode `\#12\catcode `\^12\catcode `\_12\catcode `\%12\relax}%
\providecommand \@@startlink[1]{}%
\providecommand \@@endlink[0]{}%
\providecommand \url  [0]{\begingroup\@sanitize@url \@url }%
\providecommand \@url [1]{\endgroup\@href {#1}{\urlprefix }}%
\providecommand \urlprefix  [0]{URL }%
\providecommand \Eprint [0]{\href }%
\providecommand \doibase [0]{https://doi.org/}%
\providecommand \selectlanguage [0]{\@gobble}%
\providecommand \bibinfo  [0]{\@secondoftwo}%
\providecommand \bibfield  [0]{\@secondoftwo}%
\providecommand \translation [1]{[#1]}%
\providecommand \BibitemOpen [0]{}%
\providecommand \bibitemStop [0]{}%
\providecommand \bibitemNoStop [0]{.\EOS\space}%
\providecommand \EOS [0]{\spacefactor3000\relax}%
\providecommand \BibitemShut  [1]{\csname bibitem#1\endcsname}%
\let\auto@bib@innerbib\@empty
\bibitem [{\citenamefont {Kinoshita}\ \emph {et~al.}(2006)\citenamefont
  {Kinoshita}, \citenamefont {Wenger},\ and\ \citenamefont
  {Weiss}}]{Kinoshita2006NewtonsCradle}%
  \BibitemOpen
  \bibfield  {author} {\bibinfo {author} {\bibfnamefont {T.}~\bibnamefont
  {Kinoshita}}, \bibinfo {author} {\bibfnamefont {T.}~\bibnamefont {Wenger}},\
  and\ \bibinfo {author} {\bibfnamefont {D.~S.}\ \bibnamefont {Weiss}},\
  }\bibfield  {title} {\bibinfo {title} {A quantum {N}ewton's cradle},\ }\href
  {https://doi.org/10.1038/nature04693} {\bibfield  {journal} {\bibinfo
  {journal} {Nature}\ }\textbf {\bibinfo {volume} {440}},\ \bibinfo {pages}
  {900} (\bibinfo {year} {2006})}\BibitemShut {NoStop}%
\bibitem [{\citenamefont {Trotzky}\ \emph {et~al.}(2012)\citenamefont
  {Trotzky}, \citenamefont {Chen}, \citenamefont {Flesch}, \citenamefont
  {McCulloch}, \citenamefont {Schollw{\"o}ck}, \citenamefont {Eisert},\ and\
  \citenamefont {Bloch}}]{trotzky2012probing}%
  \BibitemOpen
  \bibfield  {author} {\bibinfo {author} {\bibfnamefont {S.}~\bibnamefont
  {Trotzky}}, \bibinfo {author} {\bibfnamefont {Y.-A.}\ \bibnamefont {Chen}},
  \bibinfo {author} {\bibfnamefont {A.}~\bibnamefont {Flesch}}, \bibinfo
  {author} {\bibfnamefont {I.~P.}\ \bibnamefont {McCulloch}}, \bibinfo {author}
  {\bibfnamefont {U.}~\bibnamefont {Schollw{\"o}ck}}, \bibinfo {author}
  {\bibfnamefont {J.}~\bibnamefont {Eisert}},\ and\ \bibinfo {author}
  {\bibfnamefont {I.}~\bibnamefont {Bloch}},\ }\bibfield  {title} {\bibinfo
  {title} {Probing the relaxation towards equilibrium in an isolated strongly
  correlated one-dimensional bose gas},\ }\href
  {https://doi.org/10.1038/nphys2232} {\bibfield  {journal} {\bibinfo
  {journal} {Nature physics}\ }\textbf {\bibinfo {volume} {8}},\ \bibinfo
  {pages} {325} (\bibinfo {year} {2012})}\BibitemShut {NoStop}%
\bibitem [{\citenamefont {Gring}\ \emph {et~al.}(2012)\citenamefont {Gring},
  \citenamefont {Kuhnert}, \citenamefont {Langen}, \citenamefont {Kitagawa},
  \citenamefont {Rauer}, \citenamefont {Schreitl}, \citenamefont {Mazets},
  \citenamefont {Smith}, \citenamefont {Demler},\ and\ \citenamefont
  {Schmiedmayer}}]{Gring2012Relaxation}%
  \BibitemOpen
  \bibfield  {author} {\bibinfo {author} {\bibfnamefont {M.}~\bibnamefont
  {Gring}}, \bibinfo {author} {\bibfnamefont {M.}~\bibnamefont {Kuhnert}},
  \bibinfo {author} {\bibfnamefont {T.}~\bibnamefont {Langen}}, \bibinfo
  {author} {\bibfnamefont {T.}~\bibnamefont {Kitagawa}}, \bibinfo {author}
  {\bibfnamefont {B.}~\bibnamefont {Rauer}}, \bibinfo {author} {\bibfnamefont
  {M.}~\bibnamefont {Schreitl}}, \bibinfo {author} {\bibfnamefont
  {I.}~\bibnamefont {Mazets}}, \bibinfo {author} {\bibfnamefont {D.~A.}\
  \bibnamefont {Smith}}, \bibinfo {author} {\bibfnamefont {E.}~\bibnamefont
  {Demler}},\ and\ \bibinfo {author} {\bibfnamefont {J.}~\bibnamefont
  {Schmiedmayer}},\ }\bibfield  {title} {\bibinfo {title} {{Relaxation and
  Prethermalization in an Isolated Quantum System}},\ }\href
  {https://doi.org/10.1126/science.1224953} {\bibfield  {journal} {\bibinfo
  {journal} {Science}\ }\textbf {\bibinfo {volume} {337}},\ \bibinfo {pages}
  {1318} (\bibinfo {year} {2012})}\BibitemShut {NoStop}%
\bibitem [{\citenamefont {Schreiber}\ \emph {et~al.}(2015)\citenamefont
  {Schreiber}, \citenamefont {Hodgman}, \citenamefont {Bordia}, \citenamefont
  {Lüschen}, \citenamefont {Fischer}, \citenamefont {Vosk}, \citenamefont
  {Altman}, \citenamefont {Schneider},\ and\ \citenamefont
  {Bloch}}]{Schreiber2015Observation}%
  \BibitemOpen
  \bibfield  {author} {\bibinfo {author} {\bibfnamefont {M.}~\bibnamefont
  {Schreiber}}, \bibinfo {author} {\bibfnamefont {S.~S.}\ \bibnamefont
  {Hodgman}}, \bibinfo {author} {\bibfnamefont {P.}~\bibnamefont {Bordia}},
  \bibinfo {author} {\bibfnamefont {H.~P.}\ \bibnamefont {Lüschen}}, \bibinfo
  {author} {\bibfnamefont {M.~H.}\ \bibnamefont {Fischer}}, \bibinfo {author}
  {\bibfnamefont {R.}~\bibnamefont {Vosk}}, \bibinfo {author} {\bibfnamefont
  {E.}~\bibnamefont {Altman}}, \bibinfo {author} {\bibfnamefont
  {U.}~\bibnamefont {Schneider}},\ and\ \bibinfo {author} {\bibfnamefont
  {I.}~\bibnamefont {Bloch}},\ }\bibfield  {title} {\bibinfo {title}
  {Observation of many-body localization of interacting fermions in a
  quasirandom optical lattice},\ }\href
  {https://doi.org/10.1126/science.aaa7432} {\bibfield  {journal} {\bibinfo
  {journal} {Science}\ }\textbf {\bibinfo {volume} {349}},\ \bibinfo {pages}
  {842} (\bibinfo {year} {2015})}\BibitemShut {NoStop}%
\bibitem [{\citenamefont {Smith}\ \emph {et~al.}(2016)\citenamefont {Smith},
  \citenamefont {Lee}, \citenamefont {Richerme}, \citenamefont {Neyenhuis},
  \citenamefont {Hess}, \citenamefont {Hauke}, \citenamefont {Heyl},
  \citenamefont {Huse},\ and\ \citenamefont {Monroe}}]{Smith2016MBL}%
  \BibitemOpen
  \bibfield  {author} {\bibinfo {author} {\bibfnamefont {J.}~\bibnamefont
  {Smith}}, \bibinfo {author} {\bibfnamefont {A.}~\bibnamefont {Lee}}, \bibinfo
  {author} {\bibfnamefont {P.}~\bibnamefont {Richerme}}, \bibinfo {author}
  {\bibfnamefont {B.}~\bibnamefont {Neyenhuis}}, \bibinfo {author}
  {\bibfnamefont {P.~W.}\ \bibnamefont {Hess}}, \bibinfo {author}
  {\bibfnamefont {P.}~\bibnamefont {Hauke}}, \bibinfo {author} {\bibfnamefont
  {M.}~\bibnamefont {Heyl}}, \bibinfo {author} {\bibfnamefont {D.~A.}\
  \bibnamefont {Huse}},\ and\ \bibinfo {author} {\bibfnamefont
  {C.}~\bibnamefont {Monroe}},\ }\bibfield  {title} {\bibinfo {title}
  {Many-body localization in a quantum simulator with programmable random
  disorder},\ }\href {https://doi.org/10.1038/nphys3783} {\bibfield  {journal}
  {\bibinfo  {journal} {Nature Physics}\ }\textbf {\bibinfo {volume} {12}},\
  \bibinfo {pages} {907} (\bibinfo {year} {2016})}\BibitemShut {NoStop}%
\bibitem [{\citenamefont {Kaufman}\ \emph {et~al.}(2016)\citenamefont
  {Kaufman}, \citenamefont {Tai}, \citenamefont {Lukin}, \citenamefont
  {Rispoli}, \citenamefont {Schittko}, \citenamefont {Preiss},\ and\
  \citenamefont {Greiner}}]{Kaufman2016Thermalization}%
  \BibitemOpen
  \bibfield  {author} {\bibinfo {author} {\bibfnamefont {A.~M.}\ \bibnamefont
  {Kaufman}}, \bibinfo {author} {\bibfnamefont {M.~E.}\ \bibnamefont {Tai}},
  \bibinfo {author} {\bibfnamefont {A.}~\bibnamefont {Lukin}}, \bibinfo
  {author} {\bibfnamefont {M.}~\bibnamefont {Rispoli}}, \bibinfo {author}
  {\bibfnamefont {R.}~\bibnamefont {Schittko}}, \bibinfo {author}
  {\bibfnamefont {P.~M.}\ \bibnamefont {Preiss}},\ and\ \bibinfo {author}
  {\bibfnamefont {M.}~\bibnamefont {Greiner}},\ }\bibfield  {title} {\bibinfo
  {title} {Quantum thermalization through entanglement in an isolated many-body
  system},\ }\href {https://doi.org/10.1126/science.aaf6725} {\bibfield
  {journal} {\bibinfo  {journal} {Science}\ }\textbf {\bibinfo {volume}
  {353}},\ \bibinfo {pages} {794} (\bibinfo {year} {2016})}\BibitemShut
  {NoStop}%
\bibitem [{\citenamefont {Kucsko}\ \emph {et~al.}(2018)\citenamefont {Kucsko},
  \citenamefont {Choi}, \citenamefont {Choi}, \citenamefont {Maurer},
  \citenamefont {Zhou}, \citenamefont {Landig}, \citenamefont {Sumiya},
  \citenamefont {Onoda}, \citenamefont {Isoya}, \citenamefont {Jelezko},
  \citenamefont {Demler}, \citenamefont {Yao},\ and\ \citenamefont
  {Lukin}}]{Kucsko2018Critical}%
  \BibitemOpen
  \bibfield  {author} {\bibinfo {author} {\bibfnamefont {G.}~\bibnamefont
  {Kucsko}}, \bibinfo {author} {\bibfnamefont {S.}~\bibnamefont {Choi}},
  \bibinfo {author} {\bibfnamefont {J.}~\bibnamefont {Choi}}, \bibinfo {author}
  {\bibfnamefont {P.~C.}\ \bibnamefont {Maurer}}, \bibinfo {author}
  {\bibfnamefont {H.}~\bibnamefont {Zhou}}, \bibinfo {author} {\bibfnamefont
  {R.}~\bibnamefont {Landig}}, \bibinfo {author} {\bibfnamefont
  {H.}~\bibnamefont {Sumiya}}, \bibinfo {author} {\bibfnamefont
  {S.}~\bibnamefont {Onoda}}, \bibinfo {author} {\bibfnamefont
  {J.}~\bibnamefont {Isoya}}, \bibinfo {author} {\bibfnamefont
  {F.}~\bibnamefont {Jelezko}}, \bibinfo {author} {\bibfnamefont
  {E.}~\bibnamefont {Demler}}, \bibinfo {author} {\bibfnamefont {N.~Y.}\
  \bibnamefont {Yao}},\ and\ \bibinfo {author} {\bibfnamefont {M.~D.}\
  \bibnamefont {Lukin}},\ }\bibfield  {title} {\bibinfo {title} {{Critical
  Thermalization of a Disordered Dipolar Spin System in Diamond}},\ }\href
  {https://doi.org/10.1103/PhysRevLett.121.023601} {\bibfield  {journal}
  {\bibinfo  {journal} {Phys. Rev. Lett.}\ }\textbf {\bibinfo {volume} {121}},\
  \bibinfo {pages} {023601} (\bibinfo {year} {2018})}\BibitemShut {NoStop}%
\bibitem [{\citenamefont {Gornyi}\ \emph {et~al.}(2005)\citenamefont {Gornyi},
  \citenamefont {Mirlin},\ and\ \citenamefont {Polyakov}}]{GMP}%
  \BibitemOpen
  \bibfield  {author} {\bibinfo {author} {\bibfnamefont {I.~V.}\ \bibnamefont
  {Gornyi}}, \bibinfo {author} {\bibfnamefont {A.~D.}\ \bibnamefont {Mirlin}},\
  and\ \bibinfo {author} {\bibfnamefont {D.~G.}\ \bibnamefont {Polyakov}},\
  }\bibfield  {title} {\bibinfo {title} {{Interacting Electrons in Disordered
  Wires: Anderson Localization and Low-$T$ Transport}},\ }\href
  {https://doi.org/10.1103/PhysRevLett.95.206603} {\bibfield  {journal}
  {\bibinfo  {journal} {Phys. Rev. Lett.}\ }\textbf {\bibinfo {volume} {95}},\
  \bibinfo {pages} {206603} (\bibinfo {year} {2005})}\BibitemShut {NoStop}%
\bibitem [{\citenamefont {Basko}\ \emph {et~al.}(2006)\citenamefont {Basko},
  \citenamefont {Aleiner},\ and\ \citenamefont {Altshuler}}]{Basko2006Metal}%
  \BibitemOpen
  \bibfield  {author} {\bibinfo {author} {\bibfnamefont {D.}~\bibnamefont
  {Basko}}, \bibinfo {author} {\bibfnamefont {I.}~\bibnamefont {Aleiner}},\
  and\ \bibinfo {author} {\bibfnamefont {B.}~\bibnamefont {Altshuler}},\
  }\bibfield  {title} {\bibinfo {title} {Metal–insulator transition in a
  weakly interacting many-electron system with localized single-particle
  states},\ }\href {https://doi.org/https://doi.org/10.1016/j.aop.2005.11.014}
  {\bibfield  {journal} {\bibinfo  {journal} {Annals of Physics}\ }\textbf
  {\bibinfo {volume} {321}},\ \bibinfo {pages} {1126} (\bibinfo {year}
  {2006})}\BibitemShut {NoStop}%
\bibitem [{\citenamefont {Nandkishore}\ and\ \citenamefont
  {Huse}(2015)}]{Nandkishore2015MBL}%
  \BibitemOpen
  \bibfield  {author} {\bibinfo {author} {\bibfnamefont {R.}~\bibnamefont
  {Nandkishore}}\ and\ \bibinfo {author} {\bibfnamefont {D.~A.}\ \bibnamefont
  {Huse}},\ }\bibfield  {title} {\bibinfo {title} {{Many-Body Localization and
  Thermalization in Quantum Statistical Mechanics}},\ }\href
  {https://doi.org/10.1146/annurev-conmatphys-031214-014726} {\bibfield
  {journal} {\bibinfo  {journal} {Annual Review of Condensed Matter Physics}\
  }\textbf {\bibinfo {volume} {6}},\ \bibinfo {pages} {15} (\bibinfo {year}
  {2015})}\BibitemShut {NoStop}%
\bibitem [{\citenamefont {Abanin}\ \emph {et~al.}(2019)\citenamefont {Abanin},
  \citenamefont {Altman}, \citenamefont {Bloch},\ and\ \citenamefont
  {Serbyn}}]{Abanin2019RevModPhys}%
  \BibitemOpen
  \bibfield  {author} {\bibinfo {author} {\bibfnamefont {D.~A.}\ \bibnamefont
  {Abanin}}, \bibinfo {author} {\bibfnamefont {E.}~\bibnamefont {Altman}},
  \bibinfo {author} {\bibfnamefont {I.}~\bibnamefont {Bloch}},\ and\ \bibinfo
  {author} {\bibfnamefont {M.}~\bibnamefont {Serbyn}},\ }\bibfield  {title}
  {\bibinfo {title} {{Colloquium: Many-body localization, thermalization, and
  entanglement}},\ }\href {https://doi.org/10.1103/RevModPhys.91.021001}
  {\bibfield  {journal} {\bibinfo  {journal} {Rev. Mod. Phys.}\ }\textbf
  {\bibinfo {volume} {91}},\ \bibinfo {pages} {021001} (\bibinfo {year}
  {2019})}\BibitemShut {NoStop}%
\bibitem [{\citenamefont {Gopalakrishnan}\ and\ \citenamefont
  {Parameswaran}(2020)}]{Gopalakrishnan2020Dynamics}%
  \BibitemOpen
  \bibfield  {author} {\bibinfo {author} {\bibfnamefont {S.}~\bibnamefont
  {Gopalakrishnan}}\ and\ \bibinfo {author} {\bibfnamefont {S.}~\bibnamefont
  {Parameswaran}},\ }\bibfield  {title} {\bibinfo {title} {Dynamics and
  transport at the threshold of many-body localization},\ }\href
  {https://doi.org/https://doi.org/10.1016/j.physrep.2020.03.003} {\bibfield
  {journal} {\bibinfo  {journal} {Physics Reports}\ }\textbf {\bibinfo {volume}
  {862}},\ \bibinfo {pages} {1} (\bibinfo {year} {2020})},\ \bibinfo {note}
  {dynamics and transport at the threshold of many-body
  localization}\BibitemShut {NoStop}%
\bibitem [{\citenamefont {Imbrie}(2016)}]{Imbrie2016many}%
  \BibitemOpen
  \bibfield  {author} {\bibinfo {author} {\bibfnamefont {J.~Z.}\ \bibnamefont
  {Imbrie}},\ }\bibfield  {title} {\bibinfo {title} {On many-body localization
  for quantum spin chains},\ }\href {https://doi.org/10.1007/s10955-016-1508-x}
  {\bibfield  {journal} {\bibinfo  {journal} {Journal of Statistical Physics}\
  }\textbf {\bibinfo {volume} {163}},\ \bibinfo {pages} {998} (\bibinfo {year}
  {2016})}\BibitemShut {NoStop}%
\bibitem [{\citenamefont {Huse}\ \emph {et~al.}(2014)\citenamefont {Huse},
  \citenamefont {Nandkishore},\ and\ \citenamefont
  {Oganesyan}}]{Huse2014Phenomenology}%
  \BibitemOpen
  \bibfield  {author} {\bibinfo {author} {\bibfnamefont {D.~A.}\ \bibnamefont
  {Huse}}, \bibinfo {author} {\bibfnamefont {R.}~\bibnamefont {Nandkishore}},\
  and\ \bibinfo {author} {\bibfnamefont {V.}~\bibnamefont {Oganesyan}},\
  }\bibfield  {title} {\bibinfo {title} {Phenomenology of fully
  many-body-localized systems},\ }\href
  {https://doi.org/10.1103/PhysRevB.90.174202} {\bibfield  {journal} {\bibinfo
  {journal} {Phys. Rev. B}\ }\textbf {\bibinfo {volume} {90}},\ \bibinfo
  {pages} {174202} (\bibinfo {year} {2014})}\BibitemShut {NoStop}%
\bibitem [{\citenamefont {Serbyn}\ \emph {et~al.}(2013)\citenamefont {Serbyn},
  \citenamefont {Papi\ifmmode~\acute{c}\else \'{c}\fi{}},\ and\ \citenamefont
  {Abanin}}]{Serbyn2013Local}%
  \BibitemOpen
  \bibfield  {author} {\bibinfo {author} {\bibfnamefont {M.}~\bibnamefont
  {Serbyn}}, \bibinfo {author} {\bibfnamefont {Z.}~\bibnamefont
  {Papi\ifmmode~\acute{c}\else \'{c}\fi{}}},\ and\ \bibinfo {author}
  {\bibfnamefont {D.~A.}\ \bibnamefont {Abanin}},\ }\bibfield  {title}
  {\bibinfo {title} {{Local Conservation Laws and the Structure of the
  Many-Body Localized States}},\ }\href
  {https://doi.org/10.1103/PhysRevLett.111.127201} {\bibfield  {journal}
  {\bibinfo  {journal} {Phys. Rev. Lett.}\ }\textbf {\bibinfo {volume} {111}},\
  \bibinfo {pages} {127201} (\bibinfo {year} {2013})}\BibitemShut {NoStop}%
\bibitem [{\citenamefont {Shiraishi}\ and\ \citenamefont
  {Mori}(2017)}]{ShiraishiMori}%
  \BibitemOpen
  \bibfield  {author} {\bibinfo {author} {\bibfnamefont {N.}~\bibnamefont
  {Shiraishi}}\ and\ \bibinfo {author} {\bibfnamefont {T.}~\bibnamefont
  {Mori}},\ }\bibfield  {title} {\bibinfo {title} {{Systematic Construction of
  Counterexamples to the Eigenstate Thermalization Hypothesis}},\ }\href
  {https://doi.org/10.1103/PhysRevLett.119.030601} {\bibfield  {journal}
  {\bibinfo  {journal} {Phys. Rev. Lett.}\ }\textbf {\bibinfo {volume} {119}},\
  \bibinfo {pages} {030601} (\bibinfo {year} {2017})}\BibitemShut {NoStop}%
\bibitem [{\citenamefont {Moudgalya}\ \emph {et~al.}(2018)\citenamefont
  {Moudgalya}, \citenamefont {Rachel}, \citenamefont {Bernevig},\ and\
  \citenamefont {Regnault}}]{Moudgalya2018}%
  \BibitemOpen
  \bibfield  {author} {\bibinfo {author} {\bibfnamefont {S.}~\bibnamefont
  {Moudgalya}}, \bibinfo {author} {\bibfnamefont {S.}~\bibnamefont {Rachel}},
  \bibinfo {author} {\bibfnamefont {B.~A.}\ \bibnamefont {Bernevig}},\ and\
  \bibinfo {author} {\bibfnamefont {N.}~\bibnamefont {Regnault}},\ }\bibfield
  {title} {\bibinfo {title} {Exact excited states of nonintegrable models},\
  }\href {https://doi.org/10.1103/PhysRevB.98.235155} {\bibfield  {journal}
  {\bibinfo  {journal} {Phys. Rev. B}\ }\textbf {\bibinfo {volume} {98}},\
  \bibinfo {pages} {235155} (\bibinfo {year} {2018})}\BibitemShut {NoStop}%
\bibitem [{\citenamefont {Turner}\ \emph {et~al.}(2018)\citenamefont {Turner},
  \citenamefont {Michailidis}, \citenamefont {Abanin}, \citenamefont {Serbyn},\
  and\ \citenamefont {Papi{\'c}}}]{turner2018weak}%
  \BibitemOpen
  \bibfield  {author} {\bibinfo {author} {\bibfnamefont {C.~J.}\ \bibnamefont
  {Turner}}, \bibinfo {author} {\bibfnamefont {A.~A.}\ \bibnamefont
  {Michailidis}}, \bibinfo {author} {\bibfnamefont {D.~A.}\ \bibnamefont
  {Abanin}}, \bibinfo {author} {\bibfnamefont {M.}~\bibnamefont {Serbyn}},\
  and\ \bibinfo {author} {\bibfnamefont {Z.}~\bibnamefont {Papi{\'c}}},\
  }\bibfield  {title} {\bibinfo {title} {Weak ergodicity breaking from quantum
  many-body scars},\ }\href {https://doi.org/10.1038/s41567-018-0137-5}
  {\bibfield  {journal} {\bibinfo  {journal} {Nature Physics}\ }\textbf
  {\bibinfo {volume} {14}},\ \bibinfo {pages} {745} (\bibinfo {year}
  {2018})}\BibitemShut {NoStop}%
\bibitem [{\citenamefont {Serbyn}\ \emph {et~al.}(2021)\citenamefont {Serbyn},
  \citenamefont {Abanin},\ and\ \citenamefont {Papi{\'c}}}]{serbyn2021quantum}%
  \BibitemOpen
  \bibfield  {author} {\bibinfo {author} {\bibfnamefont {M.}~\bibnamefont
  {Serbyn}}, \bibinfo {author} {\bibfnamefont {D.~A.}\ \bibnamefont {Abanin}},\
  and\ \bibinfo {author} {\bibfnamefont {Z.}~\bibnamefont {Papi{\'c}}},\
  }\bibfield  {title} {\bibinfo {title} {Quantum many-body scars and weak
  breaking of ergodicity},\ }\href {https://doi.org/10.1038/s41567-021-01230-2}
  {\bibfield  {journal} {\bibinfo  {journal} {Nature Physics}\ }\textbf
  {\bibinfo {volume} {17}},\ \bibinfo {pages} {675} (\bibinfo {year}
  {2021})}\BibitemShut {NoStop}%
\bibitem [{\citenamefont {Moudgalya}\ \emph
  {et~al.}(2021{\natexlab{a}})\citenamefont {Moudgalya}, \citenamefont
  {Bernevig},\ and\ \citenamefont {Regnault}}]{moudgalya2021quantum}%
  \BibitemOpen
  \bibfield  {author} {\bibinfo {author} {\bibfnamefont {S.}~\bibnamefont
  {Moudgalya}}, \bibinfo {author} {\bibfnamefont {B.~A.}\ \bibnamefont
  {Bernevig}},\ and\ \bibinfo {author} {\bibfnamefont {N.}~\bibnamefont
  {Regnault}},\ }\href@noop {} {\bibinfo {title} {Quantum many-body scars and
  hilbert space fragmentation: A review of exact results}} (\bibinfo {year}
  {2021}{\natexlab{a}}),\ \Eprint {https://arxiv.org/abs/2109.00548}
  {arXiv:2109.00548 [cond-mat.str-el]} \BibitemShut {NoStop}%
\bibitem [{\citenamefont {Smith}\ \emph
  {et~al.}(2017{\natexlab{a}})\citenamefont {Smith}, \citenamefont {Knolle},
  \citenamefont {Kovrizhin},\ and\ \citenamefont {Moessner}}]{knolle1}%
  \BibitemOpen
  \bibfield  {author} {\bibinfo {author} {\bibfnamefont {A.}~\bibnamefont
  {Smith}}, \bibinfo {author} {\bibfnamefont {J.}~\bibnamefont {Knolle}},
  \bibinfo {author} {\bibfnamefont {D.~L.}\ \bibnamefont {Kovrizhin}},\ and\
  \bibinfo {author} {\bibfnamefont {R.}~\bibnamefont {Moessner}},\ }\bibfield
  {title} {\bibinfo {title} {{Disorder-Free Localization}},\ }\href
  {https://doi.org/10.1103/PhysRevLett.118.266601} {\bibfield  {journal}
  {\bibinfo  {journal} {Phys. Rev. Lett.}\ }\textbf {\bibinfo {volume} {118}},\
  \bibinfo {pages} {266601} (\bibinfo {year} {2017}{\natexlab{a}})}\BibitemShut
  {NoStop}%
\bibitem [{\citenamefont {Smith}\ \emph
  {et~al.}(2017{\natexlab{b}})\citenamefont {Smith}, \citenamefont {Knolle},
  \citenamefont {Moessner},\ and\ \citenamefont {Kovrizhin}}]{knolle2}%
  \BibitemOpen
  \bibfield  {author} {\bibinfo {author} {\bibfnamefont {A.}~\bibnamefont
  {Smith}}, \bibinfo {author} {\bibfnamefont {J.}~\bibnamefont {Knolle}},
  \bibinfo {author} {\bibfnamefont {R.}~\bibnamefont {Moessner}},\ and\
  \bibinfo {author} {\bibfnamefont {D.~L.}\ \bibnamefont {Kovrizhin}},\
  }\bibfield  {title} {\bibinfo {title} {{Absence of Ergodicity without
  Quenched Disorder: From Quantum Disentangled Liquids to Many-Body
  Localization}},\ }\href {https://doi.org/10.1103/PhysRevLett.119.176601}
  {\bibfield  {journal} {\bibinfo  {journal} {Phys. Rev. Lett.}\ }\textbf
  {\bibinfo {volume} {119}},\ \bibinfo {pages} {176601} (\bibinfo {year}
  {2017}{\natexlab{b}})}\BibitemShut {NoStop}%
\bibitem [{\citenamefont {Smith}\ \emph {et~al.}(2018)\citenamefont {Smith},
  \citenamefont {Knolle}, \citenamefont {Moessner},\ and\ \citenamefont
  {Kovrizhin}}]{knolle3}%
  \BibitemOpen
  \bibfield  {author} {\bibinfo {author} {\bibfnamefont {A.}~\bibnamefont
  {Smith}}, \bibinfo {author} {\bibfnamefont {J.}~\bibnamefont {Knolle}},
  \bibinfo {author} {\bibfnamefont {R.}~\bibnamefont {Moessner}},\ and\
  \bibinfo {author} {\bibfnamefont {D.~L.}\ \bibnamefont {Kovrizhin}},\
  }\bibfield  {title} {\bibinfo {title} {Dynamical localization in
  {${\ensuremath{\mathbb{Z}}}_{2}$} lattice gauge theories},\ }\href
  {https://doi.org/10.1103/PhysRevB.97.245137} {\bibfield  {journal} {\bibinfo
  {journal} {Phys. Rev. B}\ }\textbf {\bibinfo {volume} {97}},\ \bibinfo
  {pages} {245137} (\bibinfo {year} {2018})}\BibitemShut {NoStop}%
\bibitem [{\citenamefont {Brenes}\ \emph {et~al.}(2018)\citenamefont {Brenes},
  \citenamefont {Dalmonte}, \citenamefont {Heyl},\ and\ \citenamefont
  {Scardicchio}}]{Brenes2018}%
  \BibitemOpen
  \bibfield  {author} {\bibinfo {author} {\bibfnamefont {M.}~\bibnamefont
  {Brenes}}, \bibinfo {author} {\bibfnamefont {M.}~\bibnamefont {Dalmonte}},
  \bibinfo {author} {\bibfnamefont {M.}~\bibnamefont {Heyl}},\ and\ \bibinfo
  {author} {\bibfnamefont {A.}~\bibnamefont {Scardicchio}},\ }\bibfield
  {title} {\bibinfo {title} {{Many-Body Localization Dynamics from Gauge
  Invariance}},\ }\href {https://doi.org/10.1103/PhysRevLett.120.030601}
  {\bibfield  {journal} {\bibinfo  {journal} {Phys. Rev. Lett.}\ }\textbf
  {\bibinfo {volume} {120}},\ \bibinfo {pages} {030601} (\bibinfo {year}
  {2018})}\BibitemShut {NoStop}%
\bibitem [{\citenamefont {Parameswaran}\ and\ \citenamefont
  {Gopalakrishnan}(2017)}]{nonfermiglasses}%
  \BibitemOpen
  \bibfield  {author} {\bibinfo {author} {\bibfnamefont {S.~A.}\ \bibnamefont
  {Parameswaran}}\ and\ \bibinfo {author} {\bibfnamefont {S.}~\bibnamefont
  {Gopalakrishnan}},\ }\bibfield  {title} {\bibinfo {title} {{Non-Fermi
  Glasses: Localized Descendants of Fractionalized Metals}},\ }\href
  {https://doi.org/10.1103/PhysRevLett.119.146601} {\bibfield  {journal}
  {\bibinfo  {journal} {Phys. Rev. Lett.}\ }\textbf {\bibinfo {volume} {119}},\
  \bibinfo {pages} {146601} (\bibinfo {year} {2017})}\BibitemShut {NoStop}%
\bibitem [{\citenamefont {Smith}\ \emph {et~al.}(2019)\citenamefont {Smith},
  \citenamefont {Knolle}, \citenamefont {Moessner},\ and\ \citenamefont
  {Kovrizhin}}]{Smith2019}%
  \BibitemOpen
  \bibfield  {author} {\bibinfo {author} {\bibfnamefont {A.}~\bibnamefont
  {Smith}}, \bibinfo {author} {\bibfnamefont {J.}~\bibnamefont {Knolle}},
  \bibinfo {author} {\bibfnamefont {R.}~\bibnamefont {Moessner}},\ and\
  \bibinfo {author} {\bibfnamefont {D.~L.}\ \bibnamefont {Kovrizhin}},\
  }\bibfield  {title} {\bibinfo {title} {{Logarithmic Spreading of
  Out-of-Time-Ordered Correlators without Many-Body Localization}},\ }\href
  {https://doi.org/10.1103/PhysRevLett.123.086602} {\bibfield  {journal}
  {\bibinfo  {journal} {Phys. Rev. Lett.}\ }\textbf {\bibinfo {volume} {123}},\
  \bibinfo {pages} {086602} (\bibinfo {year} {2019})}\BibitemShut {NoStop}%
\bibitem [{\citenamefont {Russomanno}\ \emph {et~al.}(2020)\citenamefont
  {Russomanno}, \citenamefont {Notarnicola}, \citenamefont {Surace},
  \citenamefont {Fazio}, \citenamefont {Dalmonte},\ and\ \citenamefont
  {Heyl}}]{Russomanno2020}%
  \BibitemOpen
  \bibfield  {author} {\bibinfo {author} {\bibfnamefont {A.}~\bibnamefont
  {Russomanno}}, \bibinfo {author} {\bibfnamefont {S.}~\bibnamefont
  {Notarnicola}}, \bibinfo {author} {\bibfnamefont {F.~M.}\ \bibnamefont
  {Surace}}, \bibinfo {author} {\bibfnamefont {R.}~\bibnamefont {Fazio}},
  \bibinfo {author} {\bibfnamefont {M.}~\bibnamefont {Dalmonte}},\ and\
  \bibinfo {author} {\bibfnamefont {M.}~\bibnamefont {Heyl}},\ }\bibfield
  {title} {\bibinfo {title} {{Homogeneous Floquet time crystal protected by
  gauge invariance}},\ }\href
  {https://doi.org/10.1103/PhysRevResearch.2.012003} {\bibfield  {journal}
  {\bibinfo  {journal} {Phys. Rev. Research}\ }\textbf {\bibinfo {volume}
  {2}},\ \bibinfo {pages} {012003} (\bibinfo {year} {2020})}\BibitemShut
  {NoStop}%
\bibitem [{\citenamefont {Karpov}\ \emph {et~al.}(2021)\citenamefont {Karpov},
  \citenamefont {Verdel}, \citenamefont {Huang}, \citenamefont {Schmitt},\ and\
  \citenamefont {Heyl}}]{karpov2020disorderfree}%
  \BibitemOpen
  \bibfield  {author} {\bibinfo {author} {\bibfnamefont {P.}~\bibnamefont
  {Karpov}}, \bibinfo {author} {\bibfnamefont {R.}~\bibnamefont {Verdel}},
  \bibinfo {author} {\bibfnamefont {Y.-P.}\ \bibnamefont {Huang}}, \bibinfo
  {author} {\bibfnamefont {M.}~\bibnamefont {Schmitt}},\ and\ \bibinfo {author}
  {\bibfnamefont {M.}~\bibnamefont {Heyl}},\ }\bibfield  {title} {\bibinfo
  {title} {{Disorder-Free Localization in an Interacting 2D Lattice Gauge
  Theory}},\ }\href {https://doi.org/10.1103/PhysRevLett.126.130401} {\bibfield
   {journal} {\bibinfo  {journal} {Phys. Rev. Lett.}\ }\textbf {\bibinfo
  {volume} {126}},\ \bibinfo {pages} {130401} (\bibinfo {year}
  {2021})}\BibitemShut {NoStop}%
\bibitem [{\citenamefont {Hart}\ \emph {et~al.}(2021)\citenamefont {Hart},
  \citenamefont {Gopalakrishnan},\ and\ \citenamefont
  {Castelnovo}}]{Hart2021Logarithmic}%
  \BibitemOpen
  \bibfield  {author} {\bibinfo {author} {\bibfnamefont {O.}~\bibnamefont
  {Hart}}, \bibinfo {author} {\bibfnamefont {S.}~\bibnamefont
  {Gopalakrishnan}},\ and\ \bibinfo {author} {\bibfnamefont {C.}~\bibnamefont
  {Castelnovo}},\ }\bibfield  {title} {\bibinfo {title} {{Logarithmic
  Entanglement Growth from Disorder-Free Localization in the Two-Leg Compass
  Ladder}},\ }\href {https://doi.org/10.1103/PhysRevLett.126.227202} {\bibfield
   {journal} {\bibinfo  {journal} {Phys. Rev. Lett.}\ }\textbf {\bibinfo
  {volume} {126}},\ \bibinfo {pages} {227202} (\bibinfo {year}
  {2021})}\BibitemShut {NoStop}%
\bibitem [{\citenamefont {Khemani}\ \emph {et~al.}(2020)\citenamefont
  {Khemani}, \citenamefont {Hermele},\ and\ \citenamefont
  {Nandkishore}}]{Khemani2020}%
  \BibitemOpen
  \bibfield  {author} {\bibinfo {author} {\bibfnamefont {V.}~\bibnamefont
  {Khemani}}, \bibinfo {author} {\bibfnamefont {M.}~\bibnamefont {Hermele}},\
  and\ \bibinfo {author} {\bibfnamefont {R.}~\bibnamefont {Nandkishore}},\
  }\bibfield  {title} {\bibinfo {title} {Localization from hilbert space
  shattering: {F}rom theory to physical realizations},\ }\href
  {https://doi.org/10.1103/PhysRevB.101.174204} {\bibfield  {journal} {\bibinfo
   {journal} {Phys. Rev. B}\ }\textbf {\bibinfo {volume} {101}},\ \bibinfo
  {pages} {174204} (\bibinfo {year} {2020})}\BibitemShut {NoStop}%
\bibitem [{\citenamefont {Sala}\ \emph {et~al.}(2020)\citenamefont {Sala},
  \citenamefont {Rakovszky}, \citenamefont {Verresen}, \citenamefont {Knap},\
  and\ \citenamefont {Pollmann}}]{SalaFragmentation2020}%
  \BibitemOpen
  \bibfield  {author} {\bibinfo {author} {\bibfnamefont {P.}~\bibnamefont
  {Sala}}, \bibinfo {author} {\bibfnamefont {T.}~\bibnamefont {Rakovszky}},
  \bibinfo {author} {\bibfnamefont {R.}~\bibnamefont {Verresen}}, \bibinfo
  {author} {\bibfnamefont {M.}~\bibnamefont {Knap}},\ and\ \bibinfo {author}
  {\bibfnamefont {F.}~\bibnamefont {Pollmann}},\ }\bibfield  {title} {\bibinfo
  {title} {{Ergodicity Breaking Arising from Hilbert Space Fragmentation in
  Dipole-Conserving Hamiltonians}},\ }\href
  {https://doi.org/10.1103/PhysRevX.10.011047} {\bibfield  {journal} {\bibinfo
  {journal} {Phys. Rev. X}\ }\textbf {\bibinfo {volume} {10}},\ \bibinfo
  {pages} {011047} (\bibinfo {year} {2020})}\BibitemShut {NoStop}%
\bibitem [{\citenamefont {Chamon}(2005)}]{ChamonQuantumGlassiness}%
  \BibitemOpen
  \bibfield  {author} {\bibinfo {author} {\bibfnamefont {C.}~\bibnamefont
  {Chamon}},\ }\bibfield  {title} {\bibinfo {title} {{Quantum Glassiness in
  Strongly Correlated Clean Systems: An Example of Topological
  Overprotection}},\ }\href {https://doi.org/10.1103/PhysRevLett.94.040402}
  {\bibfield  {journal} {\bibinfo  {journal} {Phys. Rev. Lett.}\ }\textbf
  {\bibinfo {volume} {94}},\ \bibinfo {pages} {040402} (\bibinfo {year}
  {2005})}\BibitemShut {NoStop}%
\bibitem [{\citenamefont {Haah}(2011)}]{Haah2011}%
  \BibitemOpen
  \bibfield  {author} {\bibinfo {author} {\bibfnamefont {J.}~\bibnamefont
  {Haah}},\ }\bibfield  {title} {\bibinfo {title} {Local stabilizer codes in
  three dimensions without string logical operators},\ }\href
  {https://doi.org/10.1103/PhysRevA.83.042330} {\bibfield  {journal} {\bibinfo
  {journal} {Phys. Rev. A}\ }\textbf {\bibinfo {volume} {83}},\ \bibinfo
  {pages} {042330} (\bibinfo {year} {2011})}\BibitemShut {NoStop}%
\bibitem [{\citenamefont {Castelnovo}\ and\ \citenamefont
  {Chamon}(2012)}]{CastelnovoGlassiness2012}%
  \BibitemOpen
  \bibfield  {author} {\bibinfo {author} {\bibfnamefont {C.}~\bibnamefont
  {Castelnovo}}\ and\ \bibinfo {author} {\bibfnamefont {C.}~\bibnamefont
  {Chamon}},\ }\bibfield  {title} {\bibinfo {title} {Topological quantum
  glassiness},\ }\href {https://doi.org/10.1080/14786435.2011.609152}
  {\bibfield  {journal} {\bibinfo  {journal} {Philosophical Magazine}\ }\textbf
  {\bibinfo {volume} {92}},\ \bibinfo {pages} {304} (\bibinfo {year}
  {2012})}\BibitemShut {NoStop}%
\bibitem [{\citenamefont {Vijay}\ \emph {et~al.}(2015)\citenamefont {Vijay},
  \citenamefont {Haah},\ and\ \citenamefont {Fu}}]{VijayTopoOrder2015}%
  \BibitemOpen
  \bibfield  {author} {\bibinfo {author} {\bibfnamefont {S.}~\bibnamefont
  {Vijay}}, \bibinfo {author} {\bibfnamefont {J.}~\bibnamefont {Haah}},\ and\
  \bibinfo {author} {\bibfnamefont {L.}~\bibnamefont {Fu}},\ }\bibfield
  {title} {\bibinfo {title} {A new kind of topological quantum order: {A}
  dimensional hierarchy of quasiparticles built from stationary excitations},\
  }\href {https://doi.org/10.1103/PhysRevB.92.235136} {\bibfield  {journal}
  {\bibinfo  {journal} {Phys. Rev. B}\ }\textbf {\bibinfo {volume} {92}},\
  \bibinfo {pages} {235136} (\bibinfo {year} {2015})}\BibitemShut {NoStop}%
\bibitem [{\citenamefont {Vijay}\ \emph {et~al.}(2016)\citenamefont {Vijay},
  \citenamefont {Haah},\ and\ \citenamefont {Fu}}]{VijayFractonTopo2016}%
  \BibitemOpen
  \bibfield  {author} {\bibinfo {author} {\bibfnamefont {S.}~\bibnamefont
  {Vijay}}, \bibinfo {author} {\bibfnamefont {J.}~\bibnamefont {Haah}},\ and\
  \bibinfo {author} {\bibfnamefont {L.}~\bibnamefont {Fu}},\ }\bibfield
  {title} {\bibinfo {title} {Fracton topological order, generalized lattice
  gauge theory, and duality},\ }\href
  {https://doi.org/10.1103/PhysRevB.94.235157} {\bibfield  {journal} {\bibinfo
  {journal} {Phys. Rev. B}\ }\textbf {\bibinfo {volume} {94}},\ \bibinfo
  {pages} {235157} (\bibinfo {year} {2016})}\BibitemShut {NoStop}%
\bibitem [{\citenamefont {Pretko}(2017)}]{PretkoSubdimensional2017}%
  \BibitemOpen
  \bibfield  {author} {\bibinfo {author} {\bibfnamefont {M.}~\bibnamefont
  {Pretko}},\ }\bibfield  {title} {\bibinfo {title} {Subdimensional particle
  structure of higher rank {$U(1)$} spin liquids},\ }\href
  {https://doi.org/10.1103/PhysRevB.95.115139} {\bibfield  {journal} {\bibinfo
  {journal} {Phys. Rev. B}\ }\textbf {\bibinfo {volume} {95}},\ \bibinfo
  {pages} {115139} (\bibinfo {year} {2017})}\BibitemShut {NoStop}%
\bibitem [{\citenamefont {Gromov}(2019)}]{GromovMultipole2019}%
  \BibitemOpen
  \bibfield  {author} {\bibinfo {author} {\bibfnamefont {A.}~\bibnamefont
  {Gromov}},\ }\bibfield  {title} {\bibinfo {title} {{Towards Classification of
  Fracton Phases: The Multipole Algebra}},\ }\href
  {https://doi.org/10.1103/PhysRevX.9.031035} {\bibfield  {journal} {\bibinfo
  {journal} {Phys. Rev. X}\ }\textbf {\bibinfo {volume} {9}},\ \bibinfo {pages}
  {031035} (\bibinfo {year} {2019})}\BibitemShut {NoStop}%
\bibitem [{\citenamefont {Nandkishore}\ and\ \citenamefont
  {Hermele}(2019)}]{NandkishoreFractons2019}%
  \BibitemOpen
  \bibfield  {author} {\bibinfo {author} {\bibfnamefont {R.~M.}\ \bibnamefont
  {Nandkishore}}\ and\ \bibinfo {author} {\bibfnamefont {M.}~\bibnamefont
  {Hermele}},\ }\bibfield  {title} {\bibinfo {title} {Fractons},\ }\href
  {https://doi.org/10.1146/annurev-conmatphys-031218-013604} {\bibfield
  {journal} {\bibinfo  {journal} {Annual Review of Condensed Matter Physics}\
  }\textbf {\bibinfo {volume} {10}},\ \bibinfo {pages} {295} (\bibinfo {year}
  {2019})}\BibitemShut {NoStop}%
\bibitem [{\citenamefont {Pai}\ \emph {et~al.}(2019)\citenamefont {Pai},
  \citenamefont {Pretko},\ and\ \citenamefont {Nandkishore}}]{PPN}%
  \BibitemOpen
  \bibfield  {author} {\bibinfo {author} {\bibfnamefont {S.}~\bibnamefont
  {Pai}}, \bibinfo {author} {\bibfnamefont {M.}~\bibnamefont {Pretko}},\ and\
  \bibinfo {author} {\bibfnamefont {R.~M.}\ \bibnamefont {Nandkishore}},\
  }\bibfield  {title} {\bibinfo {title} {{Localization in Fractonic Random
  Circuits}},\ }\href {https://doi.org/10.1103/PhysRevX.9.021003} {\bibfield
  {journal} {\bibinfo  {journal} {Phys. Rev. X}\ }\textbf {\bibinfo {volume}
  {9}},\ \bibinfo {pages} {021003} (\bibinfo {year} {2019})}\BibitemShut
  {NoStop}%
\bibitem [{\citenamefont {Rakovszky}\ \emph {et~al.}(2020)\citenamefont
  {Rakovszky}, \citenamefont {Sala}, \citenamefont {Verresen}, \citenamefont
  {Knap},\ and\ \citenamefont {Pollmann}}]{RakovszkySLIOM2020}%
  \BibitemOpen
  \bibfield  {author} {\bibinfo {author} {\bibfnamefont {T.}~\bibnamefont
  {Rakovszky}}, \bibinfo {author} {\bibfnamefont {P.}~\bibnamefont {Sala}},
  \bibinfo {author} {\bibfnamefont {R.}~\bibnamefont {Verresen}}, \bibinfo
  {author} {\bibfnamefont {M.}~\bibnamefont {Knap}},\ and\ \bibinfo {author}
  {\bibfnamefont {F.}~\bibnamefont {Pollmann}},\ }\bibfield  {title} {\bibinfo
  {title} {Statistical localization: {F}rom strong fragmentation to strong edge
  modes},\ }\href {https://doi.org/10.1103/PhysRevB.101.125126} {\bibfield
  {journal} {\bibinfo  {journal} {Phys. Rev. B}\ }\textbf {\bibinfo {volume}
  {101}},\ \bibinfo {pages} {125126} (\bibinfo {year} {2020})}\BibitemShut
  {NoStop}%
\bibitem [{\citenamefont {Morningstar}\ \emph {et~al.}(2020)\citenamefont
  {Morningstar}, \citenamefont {Khemani},\ and\ \citenamefont
  {Huse}}]{MorningstarFreezing2020}%
  \BibitemOpen
  \bibfield  {author} {\bibinfo {author} {\bibfnamefont {A.}~\bibnamefont
  {Morningstar}}, \bibinfo {author} {\bibfnamefont {V.}~\bibnamefont
  {Khemani}},\ and\ \bibinfo {author} {\bibfnamefont {D.~A.}\ \bibnamefont
  {Huse}},\ }\bibfield  {title} {\bibinfo {title} {Kinetically constrained
  freezing transition in a dipole-conserving system},\ }\href
  {https://doi.org/10.1103/PhysRevB.101.214205} {\bibfield  {journal} {\bibinfo
   {journal} {Phys. Rev. B}\ }\textbf {\bibinfo {volume} {101}},\ \bibinfo
  {pages} {214205} (\bibinfo {year} {2020})}\BibitemShut {NoStop}%
\bibitem [{\citenamefont {De~Tomasi}\ \emph {et~al.}(2019)\citenamefont
  {De~Tomasi}, \citenamefont {Hetterich}, \citenamefont {Sala},\ and\
  \citenamefont {Pollmann}}]{DeTomasiDynamics2019}%
  \BibitemOpen
  \bibfield  {author} {\bibinfo {author} {\bibfnamefont {G.}~\bibnamefont
  {De~Tomasi}}, \bibinfo {author} {\bibfnamefont {D.}~\bibnamefont
  {Hetterich}}, \bibinfo {author} {\bibfnamefont {P.}~\bibnamefont {Sala}},\
  and\ \bibinfo {author} {\bibfnamefont {F.}~\bibnamefont {Pollmann}},\
  }\bibfield  {title} {\bibinfo {title} {Dynamics of strongly interacting
  systems: From fock-space fragmentation to many-body localization},\ }\href
  {https://doi.org/10.1103/PhysRevB.100.214313} {\bibfield  {journal} {\bibinfo
   {journal} {Phys. Rev. B}\ }\textbf {\bibinfo {volume} {100}},\ \bibinfo
  {pages} {214313} (\bibinfo {year} {2019})}\BibitemShut {NoStop}%
\bibitem [{\citenamefont {Moudgalya}\ \emph
  {et~al.}(2021{\natexlab{b}})\citenamefont {Moudgalya}, \citenamefont {Prem},
  \citenamefont {Nandkishore}, \citenamefont {Regnault},\ and\ \citenamefont
  {Bernevig}}]{Moudgalya2021}%
  \BibitemOpen
  \bibfield  {author} {\bibinfo {author} {\bibfnamefont {S.}~\bibnamefont
  {Moudgalya}}, \bibinfo {author} {\bibfnamefont {A.}~\bibnamefont {Prem}},
  \bibinfo {author} {\bibfnamefont {R.}~\bibnamefont {Nandkishore}}, \bibinfo
  {author} {\bibfnamefont {N.}~\bibnamefont {Regnault}},\ and\ \bibinfo
  {author} {\bibfnamefont {B.~A.}\ \bibnamefont {Bernevig}},\ }\bibinfo {title}
  {{Thermalization and Its Absence within Krylov Subspaces of a Constrained
  Hamiltonian}},\ in\ \href {https://doi.org/10.1142/9789811231711_0009} {\emph
  {\bibinfo {booktitle} {Memorial Volume for Shoucheng Zhang}}}\ (\bibinfo
  {publisher} {World Scientific},\ \bibinfo {year} {2021})\ Chap.\ \bibinfo
  {chapter} {Chapter 7}, pp.\ \bibinfo {pages} {147--209}\BibitemShut {NoStop}%
\bibitem [{\citenamefont {Moudgalya}\ and\ \citenamefont
  {Motrunich}(2021)}]{Moudgalya2021Commutant}%
  \BibitemOpen
  \bibfield  {author} {\bibinfo {author} {\bibfnamefont {S.}~\bibnamefont
  {Moudgalya}}\ and\ \bibinfo {author} {\bibfnamefont {O.~I.}\ \bibnamefont
  {Motrunich}},\ }\href@noop {} {\bibinfo {title} {Hilbert space fragmentation
  and commutant algebras}} (\bibinfo {year} {2021}),\ \Eprint
  {https://arxiv.org/abs/2108.10324} {arXiv:2108.10324 [cond-mat.stat-mech]}
  \BibitemShut {NoStop}%
\bibitem [{\citenamefont {Khudorozhkov}\ \emph {et~al.}(2021)\citenamefont
  {Khudorozhkov}, \citenamefont {Tiwari}, \citenamefont {Chamon},\ and\
  \citenamefont {Neupert}}]{tiwari}%
  \BibitemOpen
  \bibfield  {author} {\bibinfo {author} {\bibfnamefont {A.}~\bibnamefont
  {Khudorozhkov}}, \bibinfo {author} {\bibfnamefont {A.}~\bibnamefont
  {Tiwari}}, \bibinfo {author} {\bibfnamefont {C.}~\bibnamefont {Chamon}},\
  and\ \bibinfo {author} {\bibfnamefont {T.}~\bibnamefont {Neupert}},\
  }\bibfield  {title} {\bibinfo {title} {Hilbert space fragmentation in a 2d
  quantum spin system with subsystem symmetries},\ }\href@noop {} {\bibfield
  {journal} {\bibinfo  {journal} {arXiv preprint arXiv:2107.09690}\ } (\bibinfo
  {year} {2021})}\BibitemShut {NoStop}%
\bibitem [{\citenamefont {Mukherjee}\ \emph {et~al.}(2021)\citenamefont
  {Mukherjee}, \citenamefont {Banerjee}, \citenamefont {Sengupta},\ and\
  \citenamefont {Sen}}]{Sen}%
  \BibitemOpen
  \bibfield  {author} {\bibinfo {author} {\bibfnamefont {B.}~\bibnamefont
  {Mukherjee}}, \bibinfo {author} {\bibfnamefont {D.}~\bibnamefont {Banerjee}},
  \bibinfo {author} {\bibfnamefont {K.}~\bibnamefont {Sengupta}},\ and\
  \bibinfo {author} {\bibfnamefont {A.}~\bibnamefont {Sen}},\ }\bibfield
  {title} {\bibinfo {title} {Minimal model for hilbert space fragmentation with
  local constraints},\ }\href {https://doi.org/10.1103/PhysRevB.104.155117}
  {\bibfield  {journal} {\bibinfo  {journal} {Phys. Rev. B}\ }\textbf {\bibinfo
  {volume} {104}},\ \bibinfo {pages} {155117} (\bibinfo {year}
  {2021})}\BibitemShut {NoStop}%
\bibitem [{\citenamefont {van Nieuwenburg}\ \emph {et~al.}(2019)\citenamefont
  {van Nieuwenburg}, \citenamefont {Baum},\ and\ \citenamefont
  {Refael}}]{vanNieuwenburg2019}%
  \BibitemOpen
  \bibfield  {author} {\bibinfo {author} {\bibfnamefont {E.}~\bibnamefont {van
  Nieuwenburg}}, \bibinfo {author} {\bibfnamefont {Y.}~\bibnamefont {Baum}},\
  and\ \bibinfo {author} {\bibfnamefont {G.}~\bibnamefont {Refael}},\
  }\bibfield  {title} {\bibinfo {title} {From {B}loch oscillations to many-body
  localization in clean interacting systems},\ }\href
  {https://doi.org/10.1073/pnas.1819316116} {\bibfield  {journal} {\bibinfo
  {journal} {Proceedings of the National Academy of Sciences}\ }\textbf
  {\bibinfo {volume} {116}},\ \bibinfo {pages} {9269} (\bibinfo {year}
  {2019})}\BibitemShut {NoStop}%
\bibitem [{\citenamefont {Taylor}\ \emph {et~al.}(2020)\citenamefont {Taylor},
  \citenamefont {Schulz}, \citenamefont {Pollmann},\ and\ \citenamefont
  {Moessner}}]{Taylor2020Experimental}%
  \BibitemOpen
  \bibfield  {author} {\bibinfo {author} {\bibfnamefont {S.~R.}\ \bibnamefont
  {Taylor}}, \bibinfo {author} {\bibfnamefont {M.}~\bibnamefont {Schulz}},
  \bibinfo {author} {\bibfnamefont {F.}~\bibnamefont {Pollmann}},\ and\
  \bibinfo {author} {\bibfnamefont {R.}~\bibnamefont {Moessner}},\ }\bibfield
  {title} {\bibinfo {title} {Experimental probes of {S}tark many-body
  localization},\ }\href {https://doi.org/10.1103/PhysRevB.102.054206}
  {\bibfield  {journal} {\bibinfo  {journal} {Phys. Rev. B}\ }\textbf {\bibinfo
  {volume} {102}},\ \bibinfo {pages} {054206} (\bibinfo {year}
  {2020})}\BibitemShut {NoStop}%
\bibitem [{\citenamefont {Yoshinaga}\ \emph {et~al.}(2021)\citenamefont
  {Yoshinaga}, \citenamefont {Hakoshima}, \citenamefont {Imoto}, \citenamefont
  {Matsuzaki},\ and\ \citenamefont {Hamazaki}}]{yoshinaga2021emergence}%
  \BibitemOpen
  \bibfield  {author} {\bibinfo {author} {\bibfnamefont {A.}~\bibnamefont
  {Yoshinaga}}, \bibinfo {author} {\bibfnamefont {H.}~\bibnamefont
  {Hakoshima}}, \bibinfo {author} {\bibfnamefont {T.}~\bibnamefont {Imoto}},
  \bibinfo {author} {\bibfnamefont {Y.}~\bibnamefont {Matsuzaki}},\ and\
  \bibinfo {author} {\bibfnamefont {R.}~\bibnamefont {Hamazaki}},\ }\href@noop
  {} {\bibinfo {title} {Emergence of hilbert space fragmentation in ising
  models with a weak transverse field}} (\bibinfo {year} {2021}),\ \Eprint
  {https://arxiv.org/abs/2111.05586} {arXiv:2111.05586 [cond-mat.stat-mech]}
  \BibitemShut {NoStop}%
\bibitem [{\citenamefont {Yang}\ \emph {et~al.}(2020)\citenamefont {Yang},
  \citenamefont {Liu}, \citenamefont {Gorshkov},\ and\ \citenamefont
  {Iadecola}}]{YangIadecola2020}%
  \BibitemOpen
  \bibfield  {author} {\bibinfo {author} {\bibfnamefont {Z.-C.}\ \bibnamefont
  {Yang}}, \bibinfo {author} {\bibfnamefont {F.}~\bibnamefont {Liu}}, \bibinfo
  {author} {\bibfnamefont {A.~V.}\ \bibnamefont {Gorshkov}},\ and\ \bibinfo
  {author} {\bibfnamefont {T.}~\bibnamefont {Iadecola}},\ }\bibfield  {title}
  {\bibinfo {title} {{Hilbert-Space Fragmentation from Strict Confinement}},\
  }\href {https://doi.org/10.1103/PhysRevLett.124.207602} {\bibfield  {journal}
  {\bibinfo  {journal} {Phys. Rev. Lett.}\ }\textbf {\bibinfo {volume} {124}},\
  \bibinfo {pages} {207602} (\bibinfo {year} {2020})}\BibitemShut {NoStop}%
\bibitem [{\citenamefont {Chen}\ and\ \citenamefont
  {Iadecola}(2021)}]{Chen2021}%
  \BibitemOpen
  \bibfield  {author} {\bibinfo {author} {\bibfnamefont {I.-C.}\ \bibnamefont
  {Chen}}\ and\ \bibinfo {author} {\bibfnamefont {T.}~\bibnamefont
  {Iadecola}},\ }\bibfield  {title} {\bibinfo {title} {Emergent symmetries and
  slow quantum dynamics in a rydberg-atom chain with confinement},\ }\href
  {https://doi.org/10.1103/PhysRevB.103.214304} {\bibfield  {journal} {\bibinfo
   {journal} {Phys. Rev. B}\ }\textbf {\bibinfo {volume} {103}},\ \bibinfo
  {pages} {214304} (\bibinfo {year} {2021})}\BibitemShut {NoStop}%
\bibitem [{\citenamefont {Ritort}\ and\ \citenamefont
  {Sollich}(2003)}]{Ritort2003Glassy}%
  \BibitemOpen
  \bibfield  {author} {\bibinfo {author} {\bibfnamefont {F.}~\bibnamefont
  {Ritort}}\ and\ \bibinfo {author} {\bibfnamefont {P.}~\bibnamefont
  {Sollich}},\ }\bibfield  {title} {\bibinfo {title} {Glassy dynamics of
  kinetically constrained models},\ }\href
  {https://doi.org/10.1080/0001873031000093582} {\bibfield  {journal} {\bibinfo
   {journal} {Advances in Physics}\ }\textbf {\bibinfo {volume} {52}},\
  \bibinfo {pages} {219} (\bibinfo {year} {2003})}\BibitemShut {NoStop}%
\bibitem [{\citenamefont {Berthier}\ \emph {et~al.}(2011)\citenamefont
  {Berthier}, \citenamefont {Biroli}, \citenamefont {Bouchaud}, \citenamefont
  {Cipelletti},\ and\ \citenamefont {van Saarloos}}]{berthier2011dynamical}%
  \BibitemOpen
  \bibfield  {author} {\bibinfo {author} {\bibfnamefont {L.}~\bibnamefont
  {Berthier}}, \bibinfo {author} {\bibfnamefont {G.}~\bibnamefont {Biroli}},
  \bibinfo {author} {\bibfnamefont {J.-P.}\ \bibnamefont {Bouchaud}}, \bibinfo
  {author} {\bibfnamefont {L.}~\bibnamefont {Cipelletti}},\ and\ \bibinfo
  {author} {\bibfnamefont {W.}~\bibnamefont {van Saarloos}},\ }\href
  {https://doi.org/10.1093/acprof:oso/9780199691470.001.0001} {\emph {\bibinfo
  {title} {Dynamical heterogeneities in glasses, colloids, and granular
  media}}},\ Vol.\ \bibinfo {volume} {150}\ (\bibinfo  {publisher} {OUP
  Oxford},\ \bibinfo {year} {2011})\BibitemShut {NoStop}%
\bibitem [{\citenamefont {Iaconis}\ \emph {et~al.}(2019)\citenamefont
  {Iaconis}, \citenamefont {Vijay},\ and\ \citenamefont
  {Nandkishore}}]{IaconisSubsystem}%
  \BibitemOpen
  \bibfield  {author} {\bibinfo {author} {\bibfnamefont {J.}~\bibnamefont
  {Iaconis}}, \bibinfo {author} {\bibfnamefont {S.}~\bibnamefont {Vijay}},\
  and\ \bibinfo {author} {\bibfnamefont {R.}~\bibnamefont {Nandkishore}},\
  }\bibfield  {title} {\bibinfo {title} {Anomalous subdiffusion from subsystem
  symmetries},\ }\href {https://doi.org/10.1103/PhysRevB.100.214301} {\bibfield
   {journal} {\bibinfo  {journal} {Phys. Rev. B}\ }\textbf {\bibinfo {volume}
  {100}},\ \bibinfo {pages} {214301} (\bibinfo {year} {2019})}\BibitemShut
  {NoStop}%
\bibitem [{\citenamefont {Feldmeier}\ \emph {et~al.}(2020)\citenamefont
  {Feldmeier}, \citenamefont {Sala}, \citenamefont {De~Tomasi}, \citenamefont
  {Pollmann},\ and\ \citenamefont {Knap}}]{FeldmeierAnomalousDiffusion}%
  \BibitemOpen
  \bibfield  {author} {\bibinfo {author} {\bibfnamefont {J.}~\bibnamefont
  {Feldmeier}}, \bibinfo {author} {\bibfnamefont {P.}~\bibnamefont {Sala}},
  \bibinfo {author} {\bibfnamefont {G.}~\bibnamefont {De~Tomasi}}, \bibinfo
  {author} {\bibfnamefont {F.}~\bibnamefont {Pollmann}},\ and\ \bibinfo
  {author} {\bibfnamefont {M.}~\bibnamefont {Knap}},\ }\bibfield  {title}
  {\bibinfo {title} {{Anomalous Diffusion in Dipole- and
  Higher-Moment-Conserving Systems}},\ }\href
  {https://doi.org/10.1103/PhysRevLett.125.245303} {\bibfield  {journal}
  {\bibinfo  {journal} {Phys. Rev. Lett.}\ }\textbf {\bibinfo {volume} {125}},\
  \bibinfo {pages} {245303} (\bibinfo {year} {2020})}\BibitemShut {NoStop}%
\bibitem [{\citenamefont {Adler}(1991)}]{adler1991bootstrap}%
  \BibitemOpen
  \bibfield  {author} {\bibinfo {author} {\bibfnamefont {J.}~\bibnamefont
  {Adler}},\ }\bibfield  {title} {\bibinfo {title} {Bootstrap percolation},\
  }\href@noop {} {\bibfield  {journal} {\bibinfo  {journal} {Physica A:
  Statistical Mechanics and its Applications}\ }\textbf {\bibinfo {volume}
  {171}},\ \bibinfo {pages} {453} (\bibinfo {year} {1991})}\BibitemShut
  {NoStop}%
\bibitem [{\citenamefont {De~Gregorio}\ \emph {et~al.}(2016)\citenamefont
  {De~Gregorio}, \citenamefont {Lawlor},\ and\ \citenamefont
  {Dawson}}]{DeGregorio2016}%
  \BibitemOpen
  \bibfield  {author} {\bibinfo {author} {\bibfnamefont {P.}~\bibnamefont
  {De~Gregorio}}, \bibinfo {author} {\bibfnamefont {A.}~\bibnamefont
  {Lawlor}},\ and\ \bibinfo {author} {\bibfnamefont {K.~A.}\ \bibnamefont
  {Dawson}},\ }\bibinfo {title} {Bootstrap percolation},\ in\ \href
  {https://doi.org/10.1007/978-3-642-27737-5_41-3} {\emph {\bibinfo {booktitle}
  {Encyclopedia of Complexity and Systems Science}}},\ \bibinfo {editor}
  {edited by\ \bibinfo {editor} {\bibfnamefont {R.~A.}\ \bibnamefont
  {Meyers}}}\ (\bibinfo  {publisher} {Springer Berlin Heidelberg},\ \bibinfo
  {address} {Berlin, Heidelberg},\ \bibinfo {year} {2016})\ pp.\ \bibinfo
  {pages} {1--26}\BibitemShut {NoStop}%
\bibitem [{\citenamefont {Nandkishore}(2014)}]{mblcontinuum1}%
  \BibitemOpen
  \bibfield  {author} {\bibinfo {author} {\bibfnamefont {R.}~\bibnamefont
  {Nandkishore}},\ }\bibfield  {title} {\bibinfo {title} {Many-body
  localization and delocalization in the two-dimensional continuum},\ }\href
  {https://doi.org/10.1103/PhysRevB.90.184204} {\bibfield  {journal} {\bibinfo
  {journal} {Phys. Rev. B}\ }\textbf {\bibinfo {volume} {90}},\ \bibinfo
  {pages} {184204} (\bibinfo {year} {2014})}\BibitemShut {NoStop}%
\bibitem [{\citenamefont {Gornyi}\ \emph {et~al.}(2017)\citenamefont {Gornyi},
  \citenamefont {Mirlin}, \citenamefont {MÃŒller},\ and\ \citenamefont
  {Polyakov}}]{mblcontinuum2}%
  \BibitemOpen
  \bibfield  {author} {\bibinfo {author} {\bibfnamefont {I.~V.}\ \bibnamefont
  {Gornyi}}, \bibinfo {author} {\bibfnamefont {A.~D.}\ \bibnamefont {Mirlin}},
  \bibinfo {author} {\bibfnamefont {M.}~\bibnamefont {MÃŒller}},\ and\
  \bibinfo {author} {\bibfnamefont {D.~G.}\ \bibnamefont {Polyakov}},\
  }\bibfield  {title} {\bibinfo {title} {Absence of many-body localization in a
  continuum},\ }\href {https://doi.org/https://doi.org/10.1002/andp.201600365}
  {\bibfield  {journal} {\bibinfo  {journal} {Annalen der Physik}\ }\textbf
  {\bibinfo {volume} {529}},\ \bibinfo {pages} {1600365} (\bibinfo {year}
  {2017})}\BibitemShut {NoStop}%
\bibitem [{\citenamefont {Bertoli}\ \emph {et~al.}(2019)\citenamefont
  {Bertoli}, \citenamefont {Altshuler},\ and\ \citenamefont
  {Shlyapnikov}}]{mblcontinuum3}%
  \BibitemOpen
  \bibfield  {author} {\bibinfo {author} {\bibfnamefont {G.}~\bibnamefont
  {Bertoli}}, \bibinfo {author} {\bibfnamefont {B.~L.}\ \bibnamefont
  {Altshuler}},\ and\ \bibinfo {author} {\bibfnamefont {G.~V.}\ \bibnamefont
  {Shlyapnikov}},\ }\bibfield  {title} {\bibinfo {title} {Many-body
  localization in continuum systems: Two-dimensional bosons},\ }\href
  {https://doi.org/10.1103/PhysRevA.100.013628} {\bibfield  {journal} {\bibinfo
   {journal} {Phys. Rev. A}\ }\textbf {\bibinfo {volume} {100}},\ \bibinfo
  {pages} {013628} (\bibinfo {year} {2019})}\BibitemShut {NoStop}%
\bibitem [{\citenamefont {Lieb}\ \emph {et~al.}(1961)\citenamefont {Lieb},
  \citenamefont {Schultz},\ and\ \citenamefont {Mattis}}]{Lieb1961Soluble}%
  \BibitemOpen
  \bibfield  {author} {\bibinfo {author} {\bibfnamefont {E.}~\bibnamefont
  {Lieb}}, \bibinfo {author} {\bibfnamefont {T.}~\bibnamefont {Schultz}},\ and\
  \bibinfo {author} {\bibfnamefont {D.}~\bibnamefont {Mattis}},\ }\bibfield
  {title} {\bibinfo {title} {Two soluble models of an antiferromagnetic
  chain},\ }\href
  {https://doi.org/https://doi.org/10.1016/0003-4916(61)90115-4} {\bibfield
  {journal} {\bibinfo  {journal} {Annals of Physics}\ }\textbf {\bibinfo
  {volume} {16}},\ \bibinfo {pages} {407} (\bibinfo {year} {1961})}\BibitemShut
  {NoStop}%
\bibitem [{\citenamefont {Pfeuty}(1970)}]{Pfeuty1970}%
  \BibitemOpen
  \bibfield  {author} {\bibinfo {author} {\bibfnamefont {P.}~\bibnamefont
  {Pfeuty}},\ }\bibfield  {title} {\bibinfo {title} {The one-dimensional ising
  model with a transverse field},\ }\href
  {https://doi.org/https://doi.org/10.1016/0003-4916(70)90270-8} {\bibfield
  {journal} {\bibinfo  {journal} {Annals of Physics}\ }\textbf {\bibinfo
  {volume} {57}},\ \bibinfo {pages} {79} (\bibinfo {year} {1970})}\BibitemShut
  {NoStop}%
\bibitem [{\citenamefont {Stein}(1997)}]{stein1997flow}%
  \BibitemOpen
  \bibfield  {author} {\bibinfo {author} {\bibfnamefont {J.}~\bibnamefont
  {Stein}},\ }\bibfield  {title} {\bibinfo {title} {Flow equations and the
  strong-coupling expansion for the {H}ubbard model},\ }\href
  {https://doi.org/doi.org/10.1007/BF02508481} {\bibfield  {journal} {\bibinfo
  {journal} {Journal of Statistical Physics}\ }\textbf {\bibinfo {volume}
  {88}},\ \bibinfo {pages} {487} (\bibinfo {year} {1997})}\BibitemShut
  {NoStop}%
\bibitem [{\citenamefont {Vidal}\ \emph
  {et~al.}(2009{\natexlab{a}})\citenamefont {Vidal}, \citenamefont {Dusuel},\
  and\ \citenamefont {Schmidt}}]{VidalParallel2009}%
  \BibitemOpen
  \bibfield  {author} {\bibinfo {author} {\bibfnamefont {J.}~\bibnamefont
  {Vidal}}, \bibinfo {author} {\bibfnamefont {S.}~\bibnamefont {Dusuel}},\ and\
  \bibinfo {author} {\bibfnamefont {K.~P.}\ \bibnamefont {Schmidt}},\
  }\bibfield  {title} {\bibinfo {title} {Low-energy effective theory of the
  toric code model in a parallel magnetic field},\ }\href
  {https://doi.org/10.1103/PhysRevB.79.033109} {\bibfield  {journal} {\bibinfo
  {journal} {Phys. Rev. B}\ }\textbf {\bibinfo {volume} {79}},\ \bibinfo
  {pages} {033109} (\bibinfo {year} {2009}{\natexlab{a}})}\BibitemShut
  {NoStop}%
\bibitem [{\citenamefont {Vidal}\ \emph
  {et~al.}(2009{\natexlab{b}})\citenamefont {Vidal}, \citenamefont {Thomale},
  \citenamefont {Schmidt},\ and\ \citenamefont {Dusuel}}]{VidalTransverse2009}%
  \BibitemOpen
  \bibfield  {author} {\bibinfo {author} {\bibfnamefont {J.}~\bibnamefont
  {Vidal}}, \bibinfo {author} {\bibfnamefont {R.}~\bibnamefont {Thomale}},
  \bibinfo {author} {\bibfnamefont {K.~P.}\ \bibnamefont {Schmidt}},\ and\
  \bibinfo {author} {\bibfnamefont {S.}~\bibnamefont {Dusuel}},\ }\bibfield
  {title} {\bibinfo {title} {Self-duality and bound states of the toric code
  model in a transverse field},\ }\href
  {https://doi.org/10.1103/PhysRevB.80.081104} {\bibfield  {journal} {\bibinfo
  {journal} {Phys. Rev. B}\ }\textbf {\bibinfo {volume} {80}},\ \bibinfo
  {pages} {081104} (\bibinfo {year} {2009}{\natexlab{b}})}\BibitemShut
  {NoStop}%
\bibitem [{\citenamefont {Knetter}\ and\ \citenamefont
  {Uhrig}(2000)}]{knetter2000perturbation}%
  \BibitemOpen
  \bibfield  {author} {\bibinfo {author} {\bibfnamefont {C.}~\bibnamefont
  {Knetter}}\ and\ \bibinfo {author} {\bibfnamefont {G.~S.}\ \bibnamefont
  {Uhrig}},\ }\bibfield  {title} {\bibinfo {title} {Perturbation theory by flow
  equations: dimerized and frustrated {$S = 1/2$} chain},\ }\href
  {https://doi.org/doi.org/10.1007/s100510050026} {\bibfield  {journal}
  {\bibinfo  {journal} {The European Physical Journal B-Condensed Matter and
  Complex Systems}\ }\textbf {\bibinfo {volume} {13}},\ \bibinfo {pages} {209}
  (\bibinfo {year} {2000})}\BibitemShut {NoStop}%
\bibitem [{\citenamefont {Knetter}\ \emph {et~al.}(2003)\citenamefont
  {Knetter}, \citenamefont {Schmidt},\ and\ \citenamefont
  {tz~S~Uhrig}}]{Knetter2003Structure}%
  \BibitemOpen
  \bibfield  {author} {\bibinfo {author} {\bibfnamefont {C.}~\bibnamefont
  {Knetter}}, \bibinfo {author} {\bibfnamefont {K.~P.}\ \bibnamefont
  {Schmidt}},\ and\ \bibinfo {author} {\bibfnamefont {G.}~\bibnamefont
  {tz~S~Uhrig}},\ }\bibfield  {title} {\bibinfo {title} {The structure of
  operators in effective particle-conserving models},\ }\href
  {https://doi.org/10.1088/0305-4470/36/29/302} {\bibfield  {journal} {\bibinfo
   {journal} {Journal of Physics A: Mathematical and General}\ }\textbf
  {\bibinfo {volume} {36}},\ \bibinfo {pages} {7889} (\bibinfo {year}
  {2003})}\BibitemShut {NoStop}%
\bibitem [{\citenamefont {Iadecola}\ and\ \citenamefont
  {Schecter}(2020)}]{Iadecola2020}%
  \BibitemOpen
  \bibfield  {author} {\bibinfo {author} {\bibfnamefont {T.}~\bibnamefont
  {Iadecola}}\ and\ \bibinfo {author} {\bibfnamefont {M.}~\bibnamefont
  {Schecter}},\ }\bibfield  {title} {\bibinfo {title} {Quantum many-body scar
  states with emergent kinetic constraints and finite-entanglement revivals},\
  }\href {https://doi.org/10.1103/PhysRevB.101.024306} {\bibfield  {journal}
  {\bibinfo  {journal} {Phys. Rev. B}\ }\textbf {\bibinfo {volume} {101}},\
  \bibinfo {pages} {024306} (\bibinfo {year} {2020})}\BibitemShut {NoStop}%
\bibitem [{\citenamefont {Abanin}\ \emph {et~al.}(2017)\citenamefont {Abanin},
  \citenamefont {De~Roeck}, \citenamefont {Ho},\ and\ \citenamefont
  {Huveneers}}]{abanin2017rigorous}%
  \BibitemOpen
  \bibfield  {author} {\bibinfo {author} {\bibfnamefont {D.}~\bibnamefont
  {Abanin}}, \bibinfo {author} {\bibfnamefont {W.}~\bibnamefont {De~Roeck}},
  \bibinfo {author} {\bibfnamefont {W.~W.}\ \bibnamefont {Ho}},\ and\ \bibinfo
  {author} {\bibfnamefont {F.}~\bibnamefont {Huveneers}},\ }\bibfield  {title}
  {\bibinfo {title} {A rigorous theory of many-body prethermalization for
  periodically driven and closed quantum systems},\ }\href
  {https://doi.org/10.1007/s00220-017-2930-x} {\bibfield  {journal} {\bibinfo
  {journal} {Communications in Mathematical Physics}\ }\textbf {\bibinfo
  {volume} {354}},\ \bibinfo {pages} {809} (\bibinfo {year}
  {2017})}\BibitemShut {NoStop}%
\bibitem [{\citenamefont {Rokhsar}\ and\ \citenamefont
  {Kivelson}(1988)}]{RokhsarKivelson}%
  \BibitemOpen
  \bibfield  {author} {\bibinfo {author} {\bibfnamefont {D.~S.}\ \bibnamefont
  {Rokhsar}}\ and\ \bibinfo {author} {\bibfnamefont {S.~A.}\ \bibnamefont
  {Kivelson}},\ }\bibfield  {title} {\bibinfo {title} {Superconductivity and
  the {Q}uantum {H}ard-{C}ore {D}imer {G}as},\ }\href
  {https://doi.org/10.1103/PhysRevLett.61.2376} {\bibfield  {journal} {\bibinfo
   {journal} {Phys. Rev. Lett.}\ }\textbf {\bibinfo {volume} {61}},\ \bibinfo
  {pages} {2376} (\bibinfo {year} {1988})}\BibitemShut {NoStop}%
\bibitem [{\citenamefont {Mondaini}\ and\ \citenamefont
  {Rigol}(2017)}]{MondainiRigolTFIM_II}%
  \BibitemOpen
  \bibfield  {author} {\bibinfo {author} {\bibfnamefont {R.}~\bibnamefont
  {Mondaini}}\ and\ \bibinfo {author} {\bibfnamefont {M.}~\bibnamefont
  {Rigol}},\ }\bibfield  {title} {\bibinfo {title} {Eigenstate thermalization
  in the two-dimensional transverse field ising model. ii. off-diagonal matrix
  elements of observables},\ }\href
  {https://doi.org/10.1103/PhysRevE.96.012157} {\bibfield  {journal} {\bibinfo
  {journal} {Phys. Rev. E}\ }\textbf {\bibinfo {volume} {96}},\ \bibinfo
  {pages} {012157} (\bibinfo {year} {2017})}\BibitemShut {NoStop}%
\bibitem [{\citenamefont {Wegner}(1971)}]{WegnerDuality1971}%
  \BibitemOpen
  \bibfield  {author} {\bibinfo {author} {\bibfnamefont {F.~J.}\ \bibnamefont
  {Wegner}},\ }\bibfield  {title} {\bibinfo {title} {{Duality in Generalized
  Ising Models and Phase Transitions without Local Order Parameters}},\ }\href
  {https://doi.org/10.1063/1.1665530} {\bibfield  {journal} {\bibinfo
  {journal} {Journal of Mathematical Physics}\ }\textbf {\bibinfo {volume}
  {12}},\ \bibinfo {pages} {2259} (\bibinfo {year} {1971})}\BibitemShut
  {NoStop}%
\bibitem [{\citenamefont {Kogut}(1979)}]{Kogut_RevModPhys1979}%
  \BibitemOpen
  \bibfield  {author} {\bibinfo {author} {\bibfnamefont {J.~B.}\ \bibnamefont
  {Kogut}},\ }\bibfield  {title} {\bibinfo {title} {An introduction to lattice
  gauge theory and spin systems},\ }\href
  {https://doi.org/10.1103/RevModPhys.51.659} {\bibfield  {journal} {\bibinfo
  {journal} {Rev. Mod. Phys.}\ }\textbf {\bibinfo {volume} {51}},\ \bibinfo
  {pages} {659} (\bibinfo {year} {1979})}\BibitemShut {NoStop}%
\bibitem [{\citenamefont {Fradkin}(2013)}]{fradkin_2013}%
  \BibitemOpen
  \bibfield  {author} {\bibinfo {author} {\bibfnamefont {E.}~\bibnamefont
  {Fradkin}},\ }\href {https://doi.org/10.1017/CBO9781139015509} {\emph
  {\bibinfo {title} {Field Theories of Condensed Matter Physics}}},\ \bibinfo
  {edition} {2nd}\ ed.\ (\bibinfo  {publisher} {Cambridge University Press},\
  \bibinfo {year} {2013})\BibitemShut {NoStop}%
\bibitem [{\citenamefont {Kitaev}(2003)}]{Kitaev2003}%
  \BibitemOpen
  \bibfield  {author} {\bibinfo {author} {\bibfnamefont {A.}~\bibnamefont
  {Kitaev}},\ }\bibfield  {title} {\bibinfo {title} {Fault-tolerant quantum
  computation by anyons},\ }\href
  {https://doi.org/https://doi.org/10.1016/S0003-4916(02)00018-0} {\bibfield
  {journal} {\bibinfo  {journal} {Annals of Physics}\ }\textbf {\bibinfo
  {volume} {303}},\ \bibinfo {pages} {2} (\bibinfo {year} {2003})}\BibitemShut
  {NoStop}%
\bibitem [{\citenamefont {Lacroix}\ \emph {et~al.}(2011)\citenamefont
  {Lacroix}, \citenamefont {Mendels},\ and\ \citenamefont
  {Mila}}]{lacroix2011introduction}%
  \BibitemOpen
  \bibfield  {author} {\bibinfo {author} {\bibfnamefont {C.}~\bibnamefont
  {Lacroix}}, \bibinfo {author} {\bibfnamefont {P.}~\bibnamefont {Mendels}},\
  and\ \bibinfo {author} {\bibfnamefont {F.}~\bibnamefont {Mila}},\ }\href
  {https://doi.org/doi.org/10.1007/978-3-642-10589-0} {\emph {\bibinfo {title}
  {Introduction to frustrated magnetism: materials, experiments, theory}}},\
  Vol.\ \bibinfo {volume} {164}\ (\bibinfo  {publisher} {Springer Science \&
  Business Media},\ \bibinfo {year} {2011})\BibitemShut {NoStop}%
\bibitem [{\citenamefont {Gromov}\ \emph {et~al.}(2020)\citenamefont {Gromov},
  \citenamefont {Lucas},\ and\ \citenamefont {Nandkishore}}]{GLN}%
  \BibitemOpen
  \bibfield  {author} {\bibinfo {author} {\bibfnamefont {A.}~\bibnamefont
  {Gromov}}, \bibinfo {author} {\bibfnamefont {A.}~\bibnamefont {Lucas}},\ and\
  \bibinfo {author} {\bibfnamefont {R.~M.}\ \bibnamefont {Nandkishore}},\
  }\bibfield  {title} {\bibinfo {title} {Fracton hydrodynamics},\ }\href
  {https://doi.org/10.1103/PhysRevResearch.2.033124} {\bibfield  {journal}
  {\bibinfo  {journal} {Phys. Rev. Research}\ }\textbf {\bibinfo {volume}
  {2}},\ \bibinfo {pages} {033124} (\bibinfo {year} {2020})}\BibitemShut
  {NoStop}%
\bibitem [{SM()}]{SM}%
  \BibitemOpen
  \href@noop {} {}\bibinfo {note} {See Supplementary Material (appended) for a
  discussion of corner cases pertaining to the growth of an unstable minority
  cluster.}\BibitemShut {Stop}%
\bibitem [{\citenamefont {J{\"a}ckle}\ \emph {et~al.}(1991)\citenamefont
  {J{\"a}ckle}, \citenamefont {Frob{\"o}se},\ and\ \citenamefont
  {Kn{\"o}dler}}]{jackie1991size}%
  \BibitemOpen
  \bibfield  {author} {\bibinfo {author} {\bibfnamefont {J.}~\bibnamefont
  {J{\"a}ckle}}, \bibinfo {author} {\bibfnamefont {K.}~\bibnamefont
  {Frob{\"o}se}},\ and\ \bibinfo {author} {\bibfnamefont {D.}~\bibnamefont
  {Kn{\"o}dler}},\ }\bibfield  {title} {\bibinfo {title} {Size dependence of
  self-diffusion in the hard-square lattice gas},\ }\href@noop {} {\bibfield
  {journal} {\bibinfo  {journal} {Journal of statistical physics}\ }\textbf
  {\bibinfo {volume} {63}},\ \bibinfo {pages} {249} (\bibinfo {year}
  {1991})}\BibitemShut {NoStop}%
\bibitem [{\citenamefont {Aizenman}\ and\ \citenamefont
  {Lebowitz}(1988)}]{AizenmanLebowitz1988}%
  \BibitemOpen
  \bibfield  {author} {\bibinfo {author} {\bibfnamefont {M.}~\bibnamefont
  {Aizenman}}\ and\ \bibinfo {author} {\bibfnamefont {J.~L.}\ \bibnamefont
  {Lebowitz}},\ }\bibfield  {title} {\bibinfo {title} {Metastability effects in
  bootstrap percolation},\ }\href {https://doi.org/10.1088/0305-4470/21/19/017}
  {\bibfield  {journal} {\bibinfo  {journal} {Journal of Physics A:
  Mathematical and General}\ }\textbf {\bibinfo {volume} {21}},\ \bibinfo
  {pages} {3801} (\bibinfo {year} {1988})}\BibitemShut {NoStop}%
\bibitem [{\citenamefont {Ertel}\ \emph {et~al.}(1988)\citenamefont {Ertel},
  \citenamefont {Froböse},\ and\ \citenamefont {Jäckle}}]{Ertel1988}%
  \BibitemOpen
  \bibfield  {author} {\bibinfo {author} {\bibfnamefont {W.}~\bibnamefont
  {Ertel}}, \bibinfo {author} {\bibfnamefont {K.}~\bibnamefont {Froböse}},\
  and\ \bibinfo {author} {\bibfnamefont {J.}~\bibnamefont {Jäckle}},\
  }\bibfield  {title} {\bibinfo {title} {Constrained diffusion dynamics in the
  hard‐square lattice gas at high density},\ }\href
  {https://doi.org/10.1063/1.454683} {\bibfield  {journal} {\bibinfo  {journal}
  {The Journal of Chemical Physics}\ }\textbf {\bibinfo {volume} {88}},\
  \bibinfo {pages} {5027} (\bibinfo {year} {1988})}\BibitemShut {NoStop}%
\bibitem [{\citenamefont {Holroyd}(2003)}]{holroyd2003sharp}%
  \BibitemOpen
  \bibfield  {author} {\bibinfo {author} {\bibfnamefont {A.~E.}\ \bibnamefont
  {Holroyd}},\ }\bibfield  {title} {\bibinfo {title} {Sharp metastability
  threshold for two-dimensional bootstrap percolation},\ }\href
  {https://doi.org/10.1007/s00440-002-0239-x} {\bibfield  {journal} {\bibinfo
  {journal} {Probability Theory and Related Fields}\ }\textbf {\bibinfo
  {volume} {125}},\ \bibinfo {pages} {195} (\bibinfo {year}
  {2003})}\BibitemShut {NoStop}%
\bibitem [{\citenamefont {De~Gregorio}\ \emph {et~al.}(2005)\citenamefont
  {De~Gregorio}, \citenamefont {Lawlor}, \citenamefont {Bradley},\ and\
  \citenamefont {Dawson}}]{DeGregorio2005exact}%
  \BibitemOpen
  \bibfield  {author} {\bibinfo {author} {\bibfnamefont {P.}~\bibnamefont
  {De~Gregorio}}, \bibinfo {author} {\bibfnamefont {A.}~\bibnamefont {Lawlor}},
  \bibinfo {author} {\bibfnamefont {P.}~\bibnamefont {Bradley}},\ and\ \bibinfo
  {author} {\bibfnamefont {K.~A.}\ \bibnamefont {Dawson}},\ }\bibfield  {title}
  {\bibinfo {title} {Exact solution of a jamming transition: closed equations
  for a bootstrap percolation problem},\ }\href
  {https://doi.org/doi:10.1073/pnas.0408756102} {\bibfield  {journal} {\bibinfo
   {journal} {Proceedings of the National Academy of Sciences}\ }\textbf
  {\bibinfo {volume} {102}},\ \bibinfo {pages} {5669} (\bibinfo {year}
  {2005})}\BibitemShut {NoStop}%
\end{thebibliography}%


\cleardoublepage
\newpage

\onecolumngrid
\begin{center}
\textbf{\large Supplemental Material for ``Hilbert space shattering and dynamical freezing in the quantum Ising model''}
\vskip 1cm
\end{center}
\twocolumngrid

\setcounter{equation}{0}
\setcounter{figure}{0}
\setcounter{table}{0}
\setcounter{page}{1}
\makeatletter
\renewcommand{\theequation}{S\arabic{equation}} 
\renewcommand{\thefigure}{S\arabic{figure}} 

\unappendix 

\setcounter{section}{0}
\setcounter{subsection}{0}

\section{Incorporating towers}

In Sec.~5 of the main text, we described how to incorporate `towers' of height one into the growing cluster.
Here, we describe how the sequence of moves is generalized to incorporate towers of height $h>1$.
\begin{equation*}
    \includegraphics[width=0.95\linewidth,valign=c]{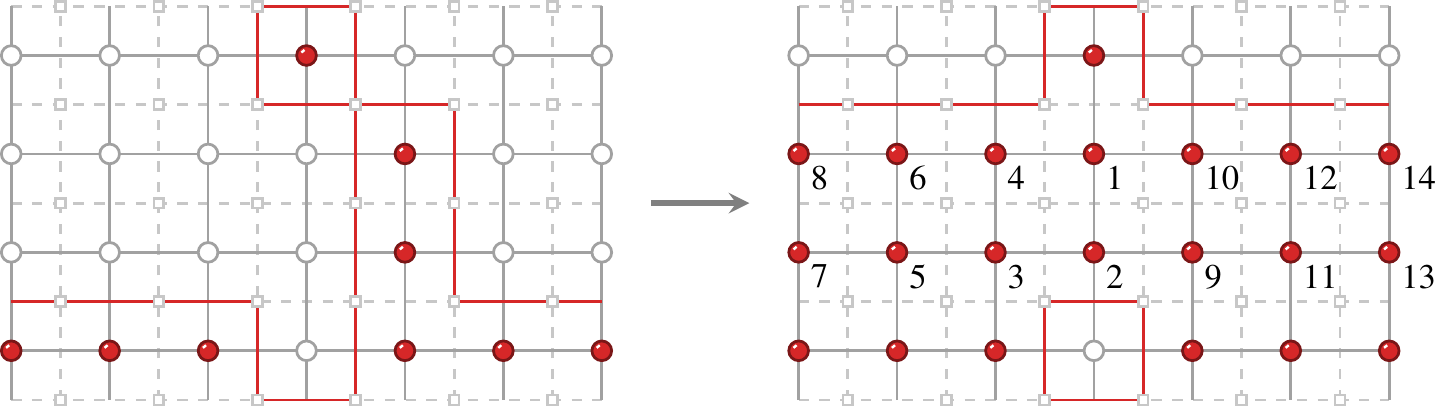}
    \, .
\end{equation*}
The spin is absorbed by the cluster by first propagating a kink down the `tower' of spins (the moves 1, 2 in the above), producing a void 
in the cluster. The baseline can then be increased by two units (moves 3--8 [9--14] propagate a kink to the left [right] corner).
The same procedure allows towers of arbitrary height to be integrated into the cluster. If the spin in the top row instead appears to the right
of the tower, then the kink should be propagated down the right-hand side of the tower. The tower may then be incorporated using the moves
described in (17) in the main text.

\section{Corner cases}

Sec.~5 of the main text was concerned with how kinks interact with spins in adjacent rows. We assumed for simplicity of presentation
that is was always possible to translate the kinks relative to the spin in the outer row. However, this needn't be the case at one of the corners of the square.
Here, we show that these special corner cases still permit the spins in outer rows to be incorporated into the growing cluster.

We begin by considering the special cases that were missed by assuming that the kink structure could be translated into the configuration~(14) in the main text.
The sequence of moves described thereafter cannot take place if the isolated spin in the next row hangs over the edge of the square, thereby preventing it from being connected directly to the cluster:
\begin{equation}
    \includegraphics[width=0.85\linewidth,valign=c]{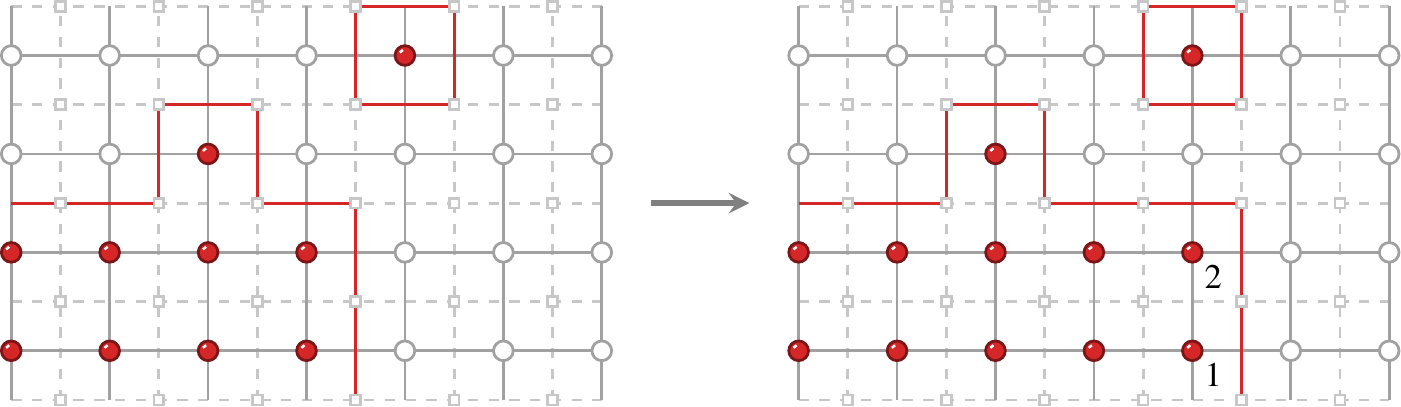}
    \, .
    \label{eqn:overhang}
\end{equation}
In the above we have assumed that the right edge can act as a source of kinks, allowing the square to first grow to the right.
If this is possible then the configuration is reduced to that of (14) in the main text.
We now consider the situation in which bringing in a kink is not possible.
To prevent a proliferation of edge cases, we consider only configurations containing towers of spins of height $h=1$.
This can be enforced by preventing spins in neighboring rows from being nearest neighbors in the initial condition. It should be noted that this restriction is merely for convenience, since it allows us to enumerate all possible interactions between various kink structures at the corner; if the restriction is instead relaxed, then there exist new edge cases that should be examined (in all such cases that we have checked, it is still possible to incorporate the spins into the cluster).
Having made this simplifying assumption, there are just three cases to consider.
The first corresponds to two copies of~\eqref{eqn:overhang} at the corner
\begin{equation*}
    \includegraphics[width=0.7\linewidth,valign=c]{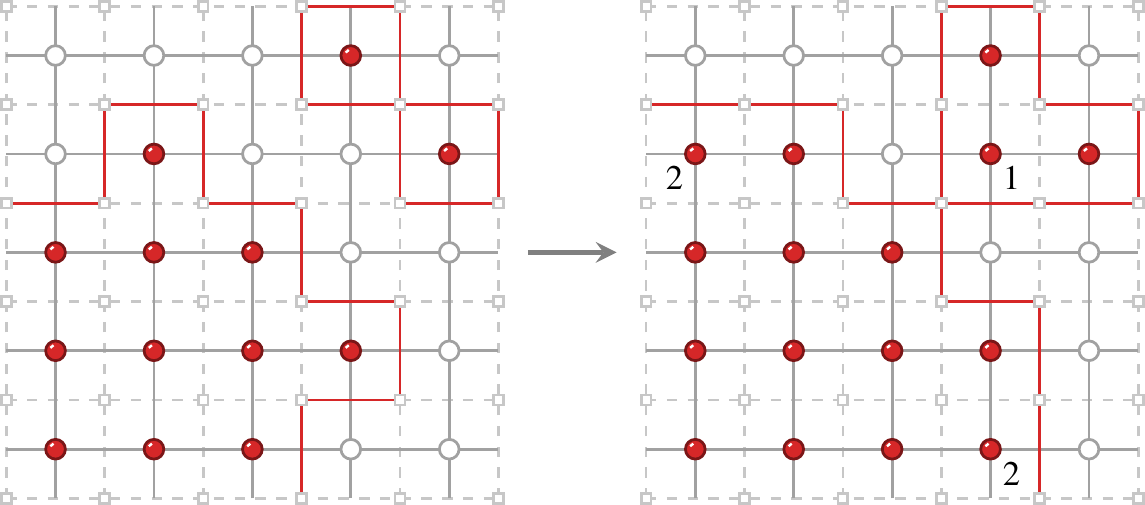}
    \, ,
\end{equation*}
which is therefore reduced to two copies of~(14) in the main text.
The next interaction to consider is between \eqref{eqn:overhang} and (16) at the corner:
\begin{equation}
    \includegraphics[width=0.7\linewidth,valign=c]{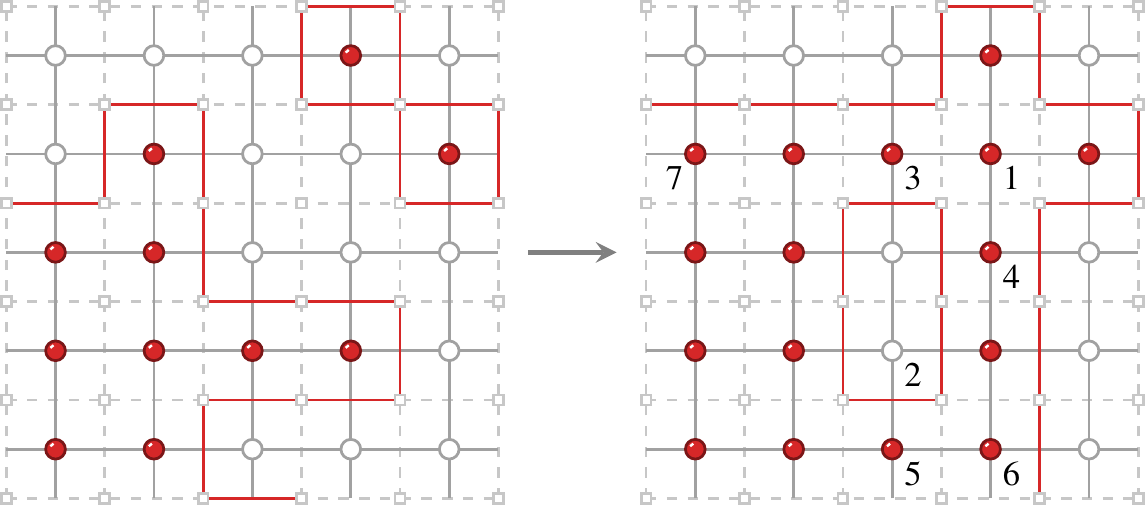}
    \, ,
    \label{eqn:config-AC}
\end{equation}
which is reduced to two copies of (11) in the main text.
The isolated pair of spins is first combined, and the resulting cluster is connected to the main growing cluster by flipping the intervening spins.
The final configuration to consider consists of the interaction between~\eqref{eqn:overhang} and the configuration~(15) in the main text
\begin{equation}
    \includegraphics[width=0.7\linewidth,valign=c]{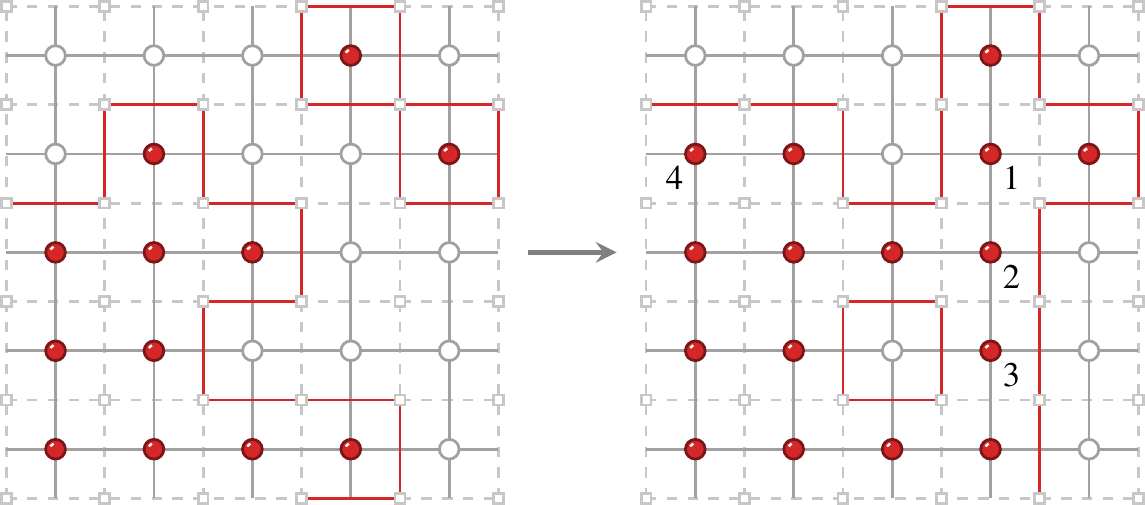}
    \, ,
    \label{eqn:config-AB}
\end{equation}
which leads to the configurations~(14) and (11) on the top and right edges, respectively.

Next, we scrutinize the corner cases associated with the configuration in (15) in the main text (more precisely, its mirror image).
First, we consider the sequence of moves requires to incorporate the isolated spin if the kink structure is prevented from being translated any further to the right due to the presence of the corner 
\begin{equation}
    \includegraphics[width=0.7\linewidth,valign=c]{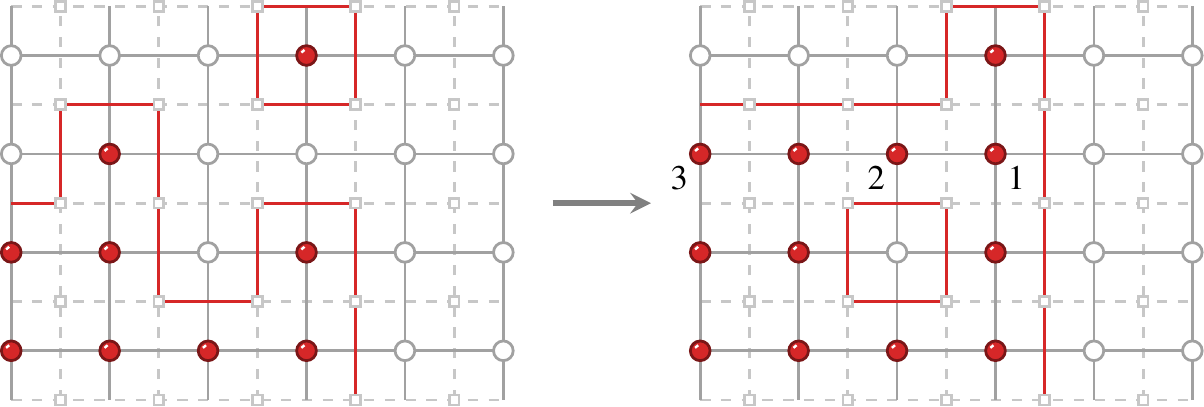}
    \, ,
    \label{eqn:double-kink-corner}
\end{equation}
reducing the configuration to that of (11) in the main text.
If the isolated spin in the top row is positioned one lattice spacing further to the right, a situation analogous to~\eqref{eqn:overhang} arises.
Specifically, if the right edge is able to act as a source of kinks, then this permits us to translate the kink structure right by one unit
\begin{equation}
    \includegraphics[width=0.7\linewidth,valign=c]{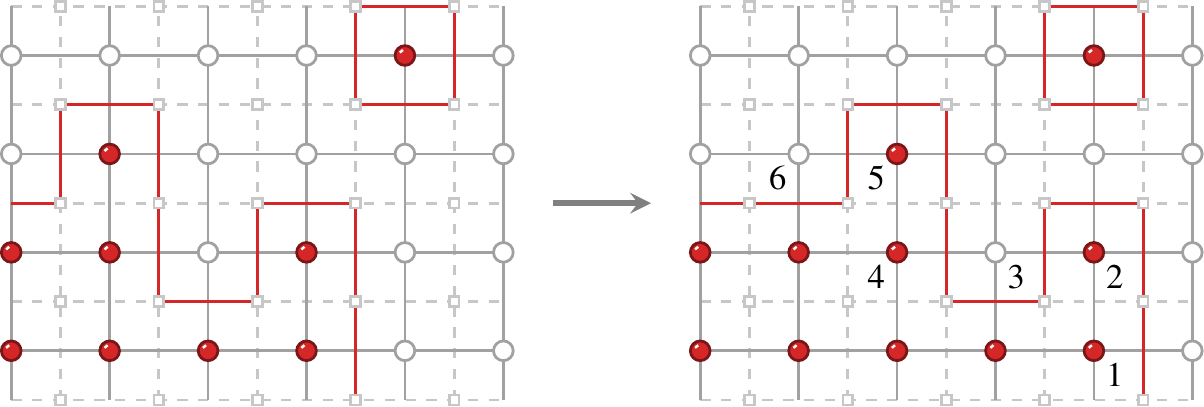}
    \, ,
    \label{eqn:kink-overhang}
\end{equation}
allowing us to proceed via the sequence of moves shown in~\eqref{eqn:double-kink-corner}.
We now look at the cases where the right-hand edge cannot act as a source of kinks. One possibility has already been considered in~\eqref{eqn:config-AB}.
The remaining special cases are the interaction between two copies of \eqref{eqn:kink-overhang}:
\begin{equation}
    \includegraphics[width=0.7\linewidth,valign=c]{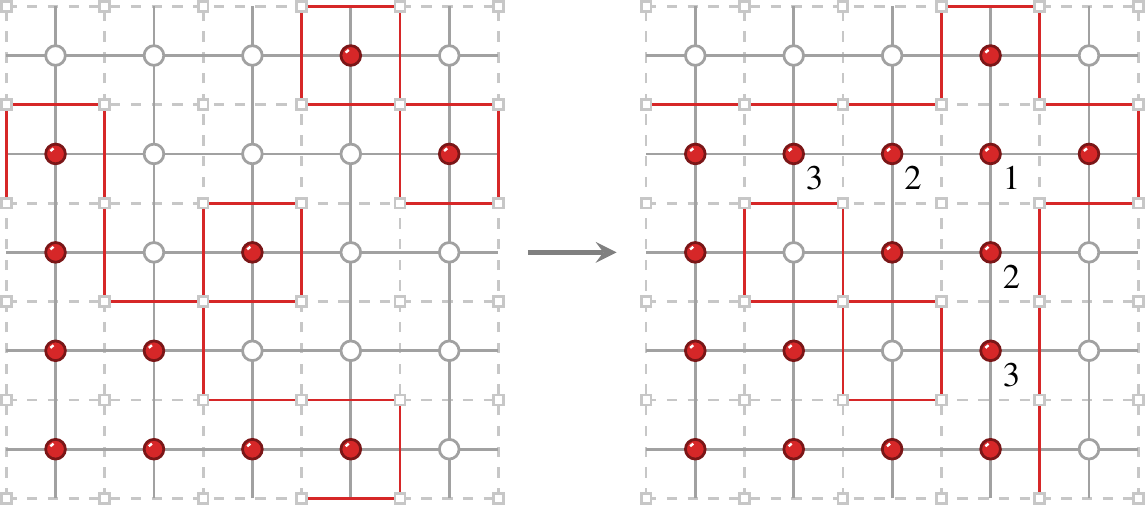}
    \, ,
    \label{eqn:config-CB}
\end{equation}
leading to two copies of (11) in the main text.
In a similar manner to~\eqref{eqn:config-AC} and~\eqref{eqn:config-AB}, the isolated spins are incorporated by first combining them, and then connecting the resulting three-spin cluster to the main growing cluster by flipping the intervening spins.
The final configuration, corresponding to the interaction between~\eqref{eqn:kink-overhang} and (16) in the main text, can also be amalgamated using a similar sequence of moves
\begin{equation}
    \includegraphics[width=0.7\linewidth,valign=c]{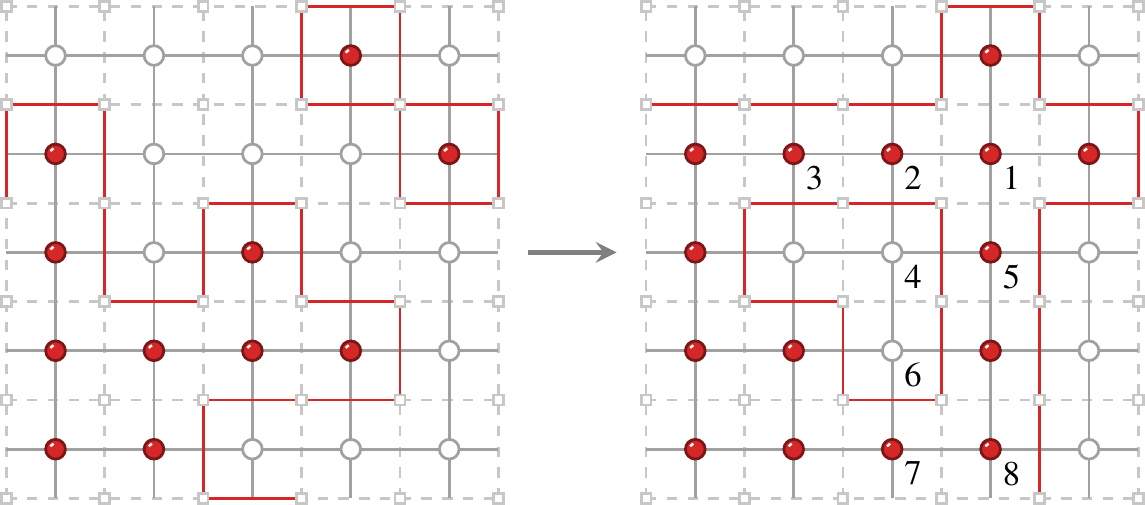}
    \, ,
    \label{eqn:config-CC}
\end{equation}
leading to two copies of (11) in the main text.

The only interaction that remains to be considered is between two copies of~(16) in the main text, which can be combined using sequences of spin flips similar to~\eqref{eqn:config-AC}, \eqref{eqn:config-AB}, \eqref{eqn:config-CB}, and \eqref{eqn:config-CC}
\begin{equation}
    \includegraphics[width=0.7\linewidth,valign=c]{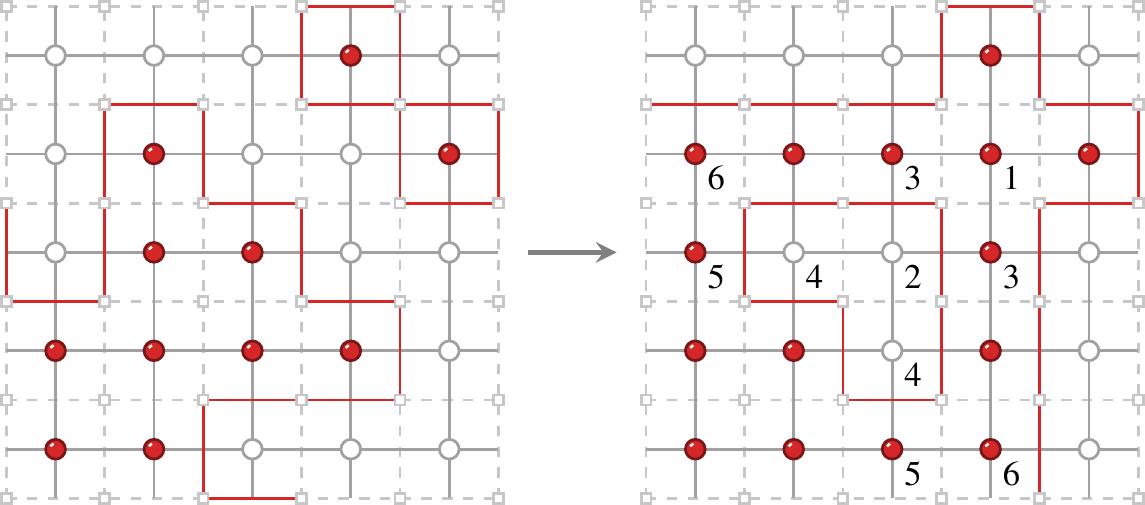}
    \, ,
    \label{eqn:config-BC}
\end{equation}
reducing the configuration to two copies of (11) in the main text with a void.
We have therefore shown that all corner cases involving spin towers of height one can be reduced to kink structures already considered.

\end{document}